\documentclass[]{aa}
\usepackage{graphicx}
\usepackage{txfonts}

\def\lesssim{\,\lower2truept\hbox{${<\atop\hbox{\raise4truept\hbox{$\sim$}}}$}\,}
\def\gtrsim{\,\lower2truept\hbox{${>\atop\hbox{\raise4truept\hbox{$\sim$}}}$}\,}

    \def\smallskip{\vskip 6pt}
    
    \def\M12{${\rm M_{12}}$}

\newcommand{\lsim}{\mathrel{\rlap{\raise -.3ex\hbox{${\scriptstyle\sim}$}}%
                   \raise .6ex\hbox{${\scriptstyle <}$}}}%
\newcommand{\gsim}{\mathrel{\rlap{\raise -.3ex\hbox{${\scriptstyle\sim}$}}%
                   \raise .6ex\hbox{${\scriptstyle >}$}}}%

\begin{document}

\title{Observational tests of the evolution of spheroidal galaxies
and predictions for SIRTF/Spitzer cosmological surveys}
\titlerunning{Evolution of spheroidal galaxies}
\author{L. Silva\inst{1} \and G. De Zotti\inst{2} \and G. L. Granato\inst{2}
\and R. Maiolino\inst{3} \and L. Danese\inst{4}}

\offprints{L. Silva, silva@ts.astro.it}

\institute{INAF-Trieste, Via Tiepolo 11, I-34131 Trieste, Italy
\and INAF-Padova, Vicolo Osservatorio 5, I-35122 Padova, Italy
\and INAF-Arcetri, Largo E. Fermi 5, I-50125 Firenze, Italy \and
SISSA, Via Beirut 4, I-34014 Trieste, Italy}

\date{Received / Accepted }

\abstract{Granato et al. (2004) have elaborated a physically
grounded model exploiting the mutual feedback between star-forming
spheroidal galaxies and the active nuclei growing in their cores
to overcome, in the framework of the hierarchical clustering
scenario for galaxy formation, one of the main challenges facing
such scenario, i.e. the fact that massive spheroidal galaxies
appear to have formed much earlier and faster than predicted by
previous hierarchical models, while the formation process was
slower for less massive objects. Adopting the choice by Granato et
al. (2004) for the parameters governing the history of the star
formation, of chemical abundances and of the gas and dust content
of galaxies, we are left with only two, rather constrained, but
still adjustable, parameters, affecting the time- and
mass-dependent SEDs of spheroidal galaxies. After having
complemented the model with a simple phenomenological description
of evolutionary properties of starburst, normal late--type
galaxies, and of AGNs, we have successfully compared the model
with a broad variety of observational data, including deep
$K$-band, ISOCAM, ISOPHOT, IRAS, SCUBA, radio counts, and the
corresponding redshift distributions, as well as the
1--$1000\,\mu$m background spectrum. Special predictions have been
made for the especially challenging counts and redshift
distributions of EROs. We also present detailed predictions for
the GOODS and SWIRE surveys with the Spitzer Space Telescope. We
find that the GOODS deep survey at $24\,\mu$m and the SWIRE
surveys at 70 and $160\,\mu$m are likely to be severely confusion
limited. The GOODS surveys in the IRAC channels (3.6 to
$8\,\mu$m), reaching flux limits of a few mJy, are expected to
resolve most of the background at these wavelengths, to explore
the full passive evolution phase of spheroidal galaxies and most
of their active star-forming phase, detecting galaxies up to
$z\simeq 4$ and beyond. A substantial number of high $z$
star-forming spheroidal galaxies should also be detected by the
$24\,\mu$m SWIRE and GOODS surveys, while the 70 and $160\,\mu$m
will be particularly useful to study the evolution of such
galaxies in the range $1 \lsim z \lsim 2$. However, starburst
galaxies at $z \lsim 1$--1.5 are expected to be the dominant
population in MIPS channels, except, perhaps, at $160\,\mu$m.


 }

\maketitle

\section{Introduction}

The standard Lambda Cold Dark Matter ($\Lambda$CDM) cosmology is a
well established framework to understand the hierarchical assembly
of dark matter (DM) halos. Indeed, it has been remarkably
successful in matching the observed large-scale structure. However
the complex evolution of the baryonic matter within the potential
wells determined by DM halos is still an open issue, both on
theoretical and on observational grounds.

Full simulations of galaxy formation in a cosmological setting are
far beyond present day computational possibilities. Thus, it is
necessary to introduce at some level rough parametric
prescriptions to deal with the physics of baryons, based on
sometimes debatable assumptions (e.g.\ Binney 2004). A class of
such models, known as semi-analytic models, has been extensively
compared with the available information on galaxy populations at
various redshifts (e.g.\ Lacey et al.\ 1993; Kauffmann, White \&
Guiderdoni, 1993; Cole et al.\ 1994; Kauffmann et al.\ 1999;
Somerville \& Primack 1999; Cole et al.\ 2000; Granato et al.\
2000; Benson et al.\ 2003).

The general strategy consists in using a subset of observations to
calibrate the many model parameters providing a heuristic
description of baryonic processes we don't properly understand.
Besides encouraging successes, current semi-analytic models have
met critical inconsistencies which seems to be deeply linked to
the standard recipes and assumptions. These problems are in
general related to the properties of elliptical galaxies, such as
the color-magnitude and the [$\alpha$/Fe]-M relations (Cole et al.
2000; Thomas 1999; Thomas et al. 2002), and the statistics of
sub-mm and deep IR selected (I- and K-band) samples (Silva 1999;
Chapman et al. 2003; Kaviani et al. 2003; Daddi et al. 2004;
Kashikawa et al. 2003; Poli et al. 2003; Pozzetti et al. 2003;
Somerville et al. 2004; see Cimatti
2003 for a review).

However, the general agreement of a broad variety of observational
data with the hierarchical scenario and the fact that the observed
number of luminous high-redshift galaxies, while substantially
higher than predicted by semi-analytic models, is nevertheless
consistent with the number of sufficiently massive dark matter
halos, indicates that we may not need alternative scenarios, but
just some new ingredients.

Previous work by our group (Granato et al.\ 2001; Romano et al.\
2002; Granato et al.\ 2004) suggests that a crucial ingredient is
the mutual feedback between spheroidal galaxies and active nuclei
at their centers.

Granato et al.\ (2004, henceforth GDS04) presented a detailed
physically motivated model for the early co-evolution of the two
components, in the
framework of the $\Lambda$CDM cosmology.

\begin{figure}[tbp]
\centering
\includegraphics[width=9truecm]{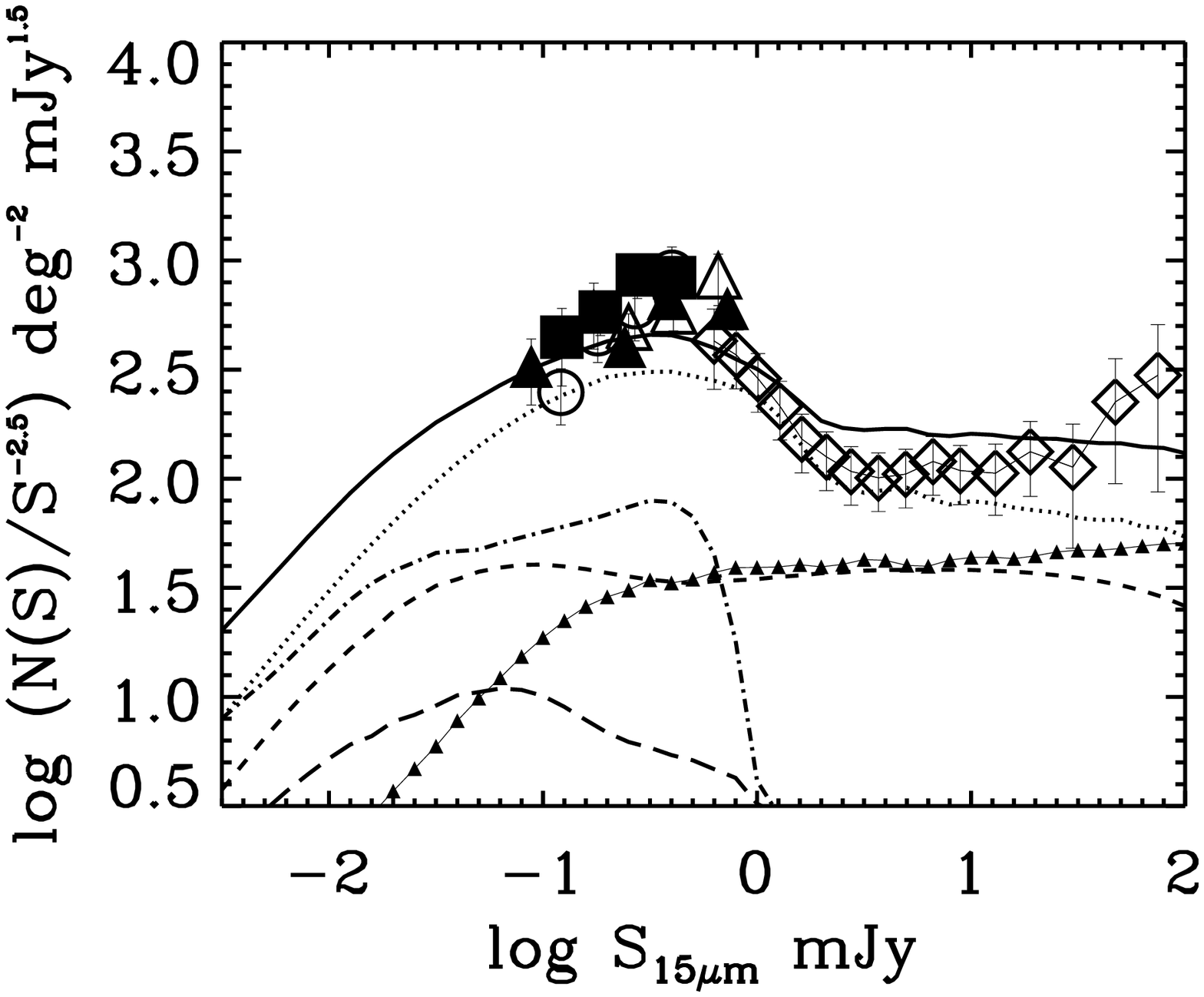}
\includegraphics[width=9truecm]{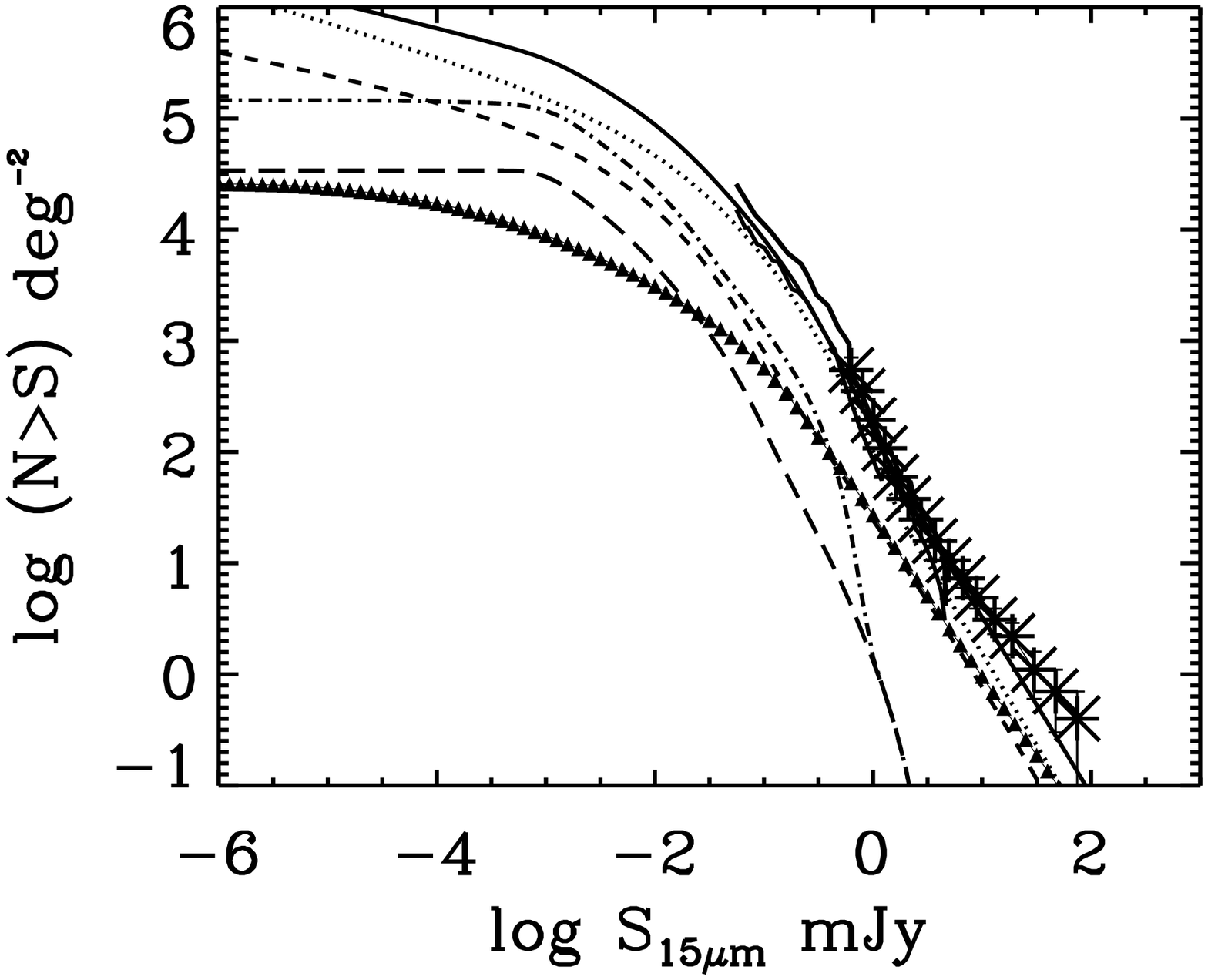}
\caption{Differential (upper panel) and integral (lower panel) $15
\mu$m counts. The solid line is the sum of contributions from
spheroids (dot-dashed line; long dashes single out passively
evolving spheroids), spirals (short dashes), starburst galaxies
(dotted line) and (type 1 + 2) AGN (filled triangles). Data are
from Elbaz et al.\ (1999), Gruppioni et al.\ (2002).}
\label{c15std}
\end{figure}

\begin{figure}[tbp]
\centering
\includegraphics[width=9truecm]{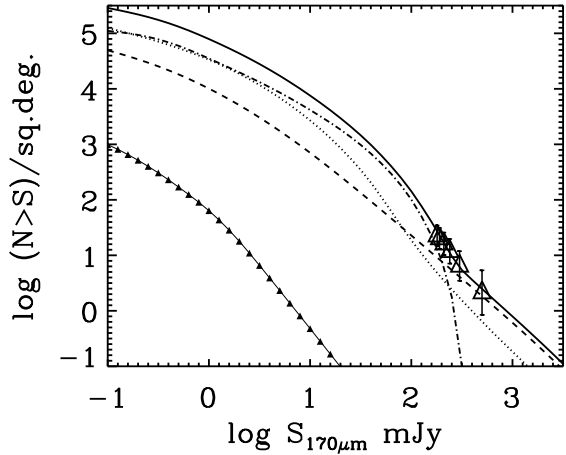}
\includegraphics[width=9truecm]{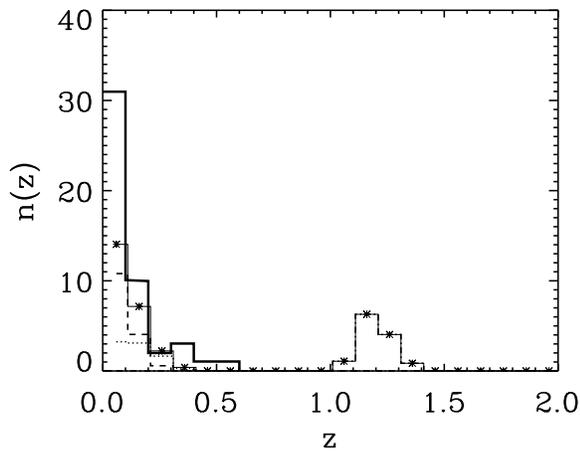}
\caption{$170\,\mu$m counts (upper panel) and redshift
distribution of sources with S$_{170}>223\,$mJy over an area of
$3\,\hbox{deg}^{-2}$. The dotted, dashed, dot-dashed lines and the
filled triangles show the contributions of starburst, spiral,
(star-forming) spheroidal galaxies and AGN, respectively. Data in
the upper panel are by Dole et al.\ (2001). In the lower panel,
the thin continuous line with asterisks is the sum of the various
contributions and the thick continuous histogram shows the data by
Rowan-Robinson et al.\ (2003).} \label{c175std}
\end{figure}

In this paper, we present a comprehensive comparison of the model
with the available data (number counts and redshift distributions)
in near-IR (NIR) to sub-mm bands and extensive predictions
relevant for surveys such as GOODS and SWIRE, which are being
carried out with NASA Spitzer (formerly SIRTF) Observatory. In
Sect.~2 we give a short overview of the GDS04 model for spheroidal
galaxies, a description of the phenomenological approach adopted
to model the evolution of starburst and normal late-type galaxies,
and of active galactic nuclei (AGNs). In Sect.~3 we discuss the
determination of the two main parameters controlling the
time-dependent spectral energy distributions (SEDs) of spheroidal
galaxies. In Sect.~4, the model counts and redshift distributions
are compared with data from the ISO surveys and follow-up, from
the IRAS $60\,\mu$m survey, from the SCUBA $850\,\mu$m surveys and
follow-up, from the radio 1.4 GHz surveys down to sub-mJy flux
densities, and from deep K-band surveys and follow-up. In Sect.~5,
we present our predictions for Spitzer  GOODS and SWIRE surveys.
The main conclusions are summarized in Sect.~6.

We adopt the following cosmological parameters: $\Omega_m=0.3$,
$\Omega_\Lambda=0.7$, $H_0=70$ km s$^{-1}$.

\begin{figure}[tbp]
\centering
\includegraphics[width=9truecm]{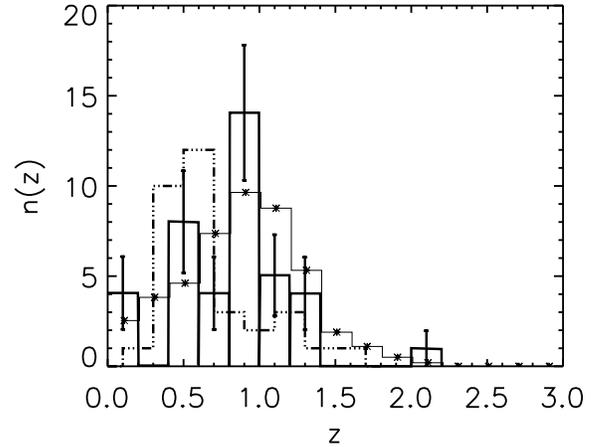}
\includegraphics[width=9truecm]{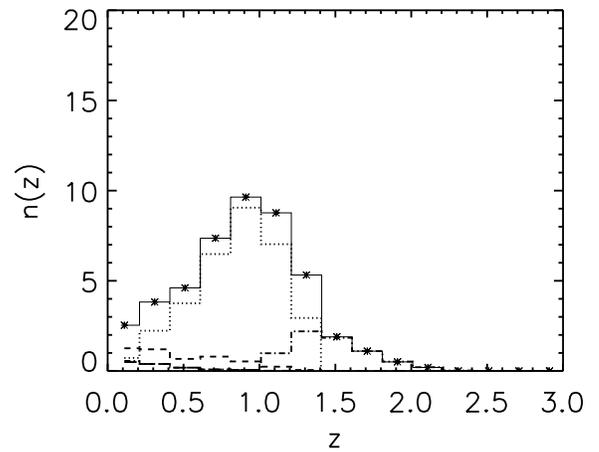}
\caption{Redshift distribution of sources brighter than $0.1\,$mJy
at $15 \mu$m, within an area of $6\cdot 10^{-3}\,\hbox{deg}^{2}$.
In the upper panel the global (spheroids plus spiral and starburst
galaxies) redshift distribution predicted by the model (thin
continuous line with asterisks) is compared with data by Elbaz et
al. (2002, thick solid histogram with error bars) and by
Franceschini et al. (2003, three dot-dashed line). In the lower
panel we show the various contributions to the global model
distribution (again shown by the thin solid histogram with
asterisks): starbursts (dots), spirals (dashes), star-forming
spheroids (dots-dashes), passively evolving spheroids (long
dashes).} \label{nz15std01}
\end{figure}

\begin{figure}[tbp]
\centering
\includegraphics[width=9truecm]{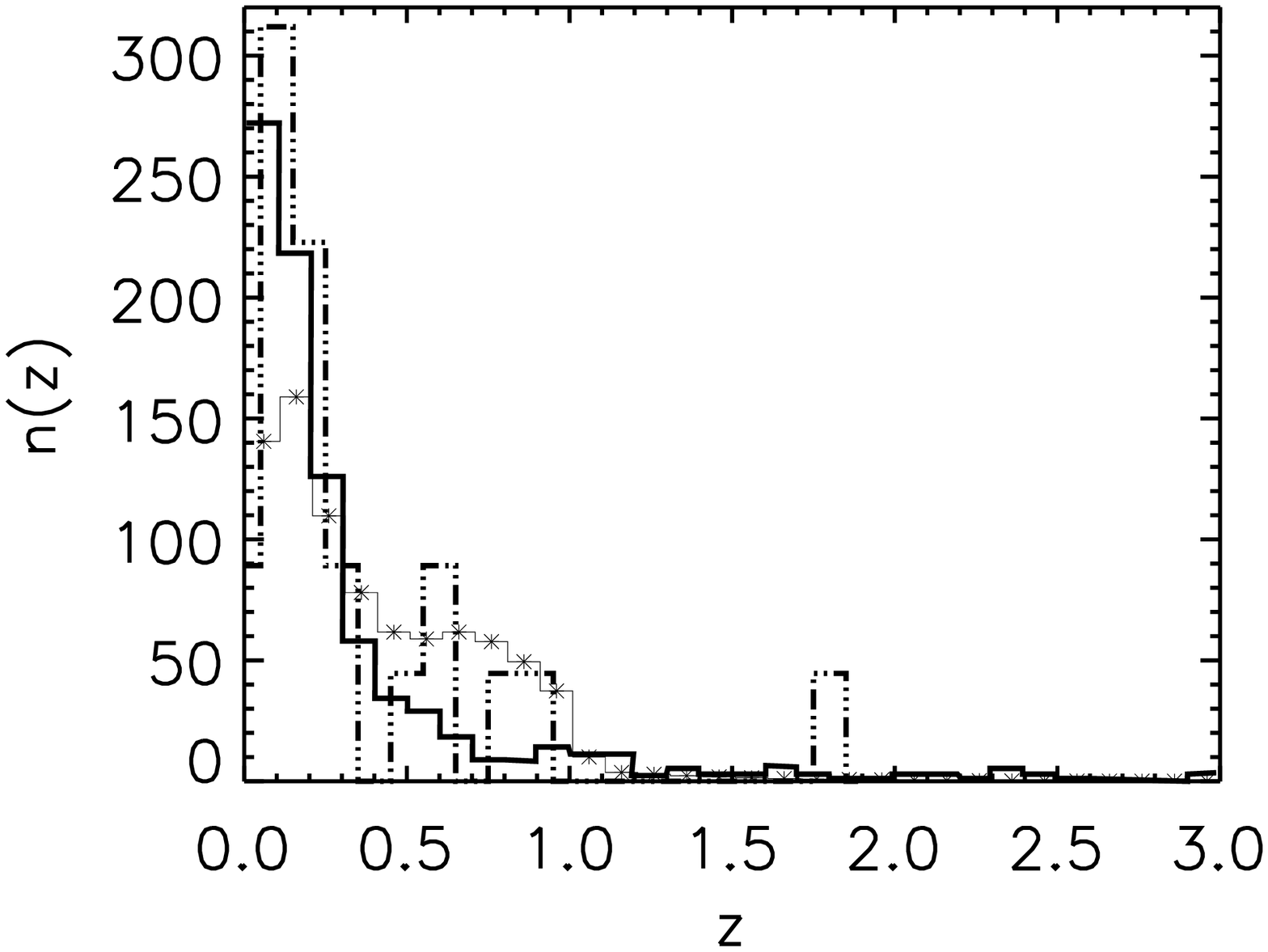}
\includegraphics[width=9truecm]{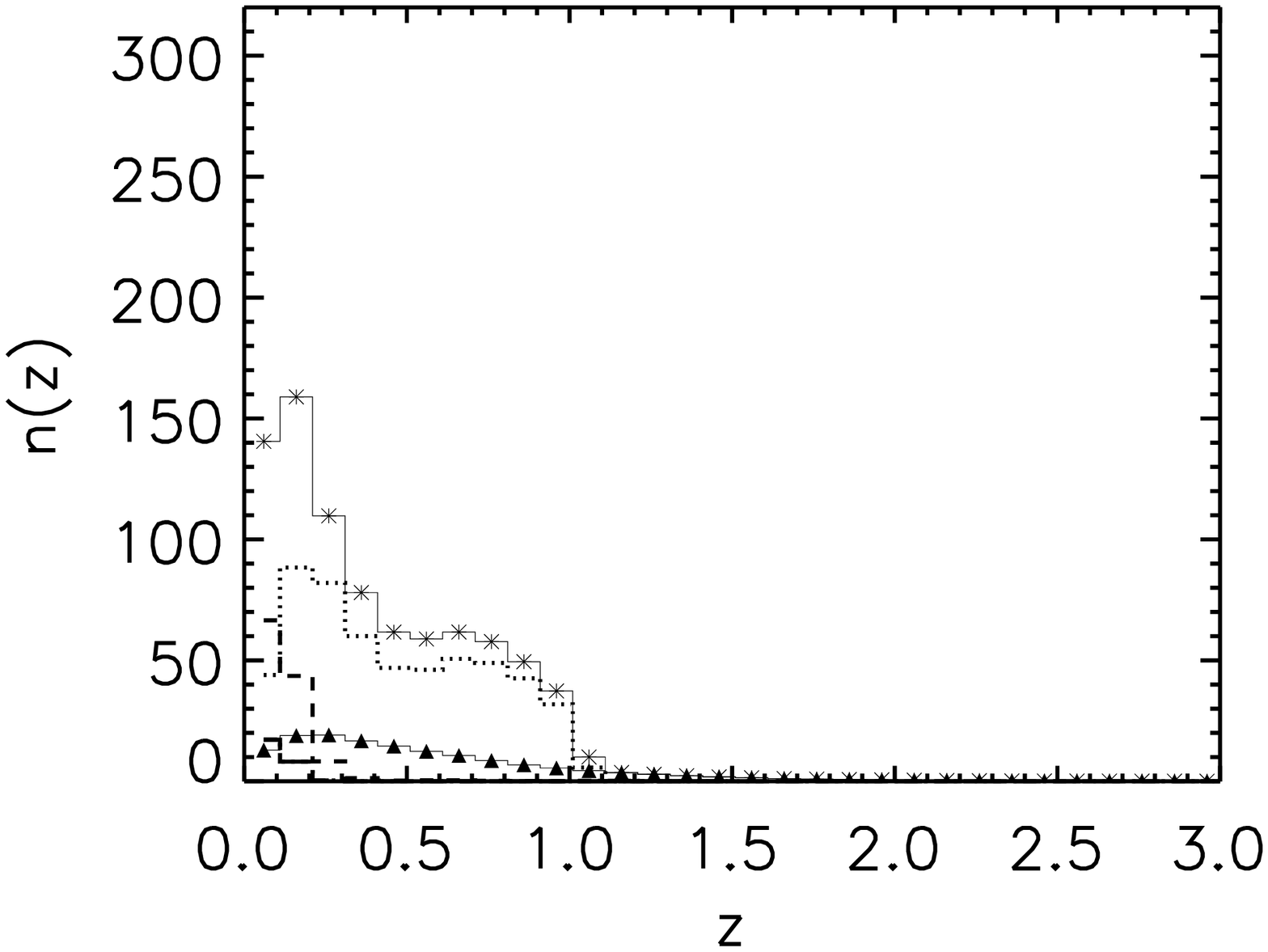}
\caption{Redshift distribution of sources brighter than $1\,$mJy
at $15 \mu$m, within an area of $5.46\,\hbox{deg}^{-2}$. In the
upper panel the global (spheroids plus spirals, starbursts and
AGN) model $z$-distribution (thin continuous line with asterisks)
is compared with data by Rowan-Robinson et al. (2003, thick
continuous line) and Pozzi et al (2003; three dots-dash, scaled to
the same area). In the lower panel, the dotted, short-dashed,
dot-dashed lines and the filled triangles show the contributions
from starburst, spiral, spheroidal galaxies, and AGN respectively,
to the global redshift distribution, represented again by the thin
continuous line with asterisks. The long-dashed line singles out
the contribution of passively evolving spheroids.}
\label{nz15std1}
\end{figure}

\section{Model description}

\subsection{The GDS04 model}
\label{sec:gds04}

The model follows with simple, physically grounded, recipes and a
semi-analytic technique the evolution of the baryonic component of
proto-spheroidal galaxies within massive dark matter (DM) halos
forming at the rate predicted by the standard hierarchical
clustering scenario for a $\Lambda$CDM cosmology. The main
difference with other treatments is in the central role attributed
to the mutual feedback between star formation and growth of a
super massive black hole (SMBH) in the galaxy center.

The kinetic energy fed by supernovae is increasingly effective,
with decreasing halo mass, in slowing down (and eventually
halting) both the star formation and the gas accretion onto the
central black hole. On the contrary, star formation and black hole
growth proceed very effectively in the more massive halos, until
the energy injected by the active nucleus in the surrounding
interstellar gas unbinds it, thus halting both the star formation
and the black hole growth (and establishing the observed
relationship between black hole mass and stellar velocity
dispersion or halo mass). As a result, the physical processes
acting on baryons reverse the order of the formation of spheroidal
galaxies with respect to the hierarchical assembling of DM halos,
in keeping with the previous proposition by Granato et al.\
(2001).

Not only the black hole growth is faster in more massive halos,
but also the feedback of the active nucleus on the interstellar
medium is stronger, to the effect of sweeping out such medium
earlier, thus causing a shorter duration of the active
star-formation phase (for more details, see GDS04).

\begin{figure}[tbp]
\centering
\includegraphics[width=9truecm]{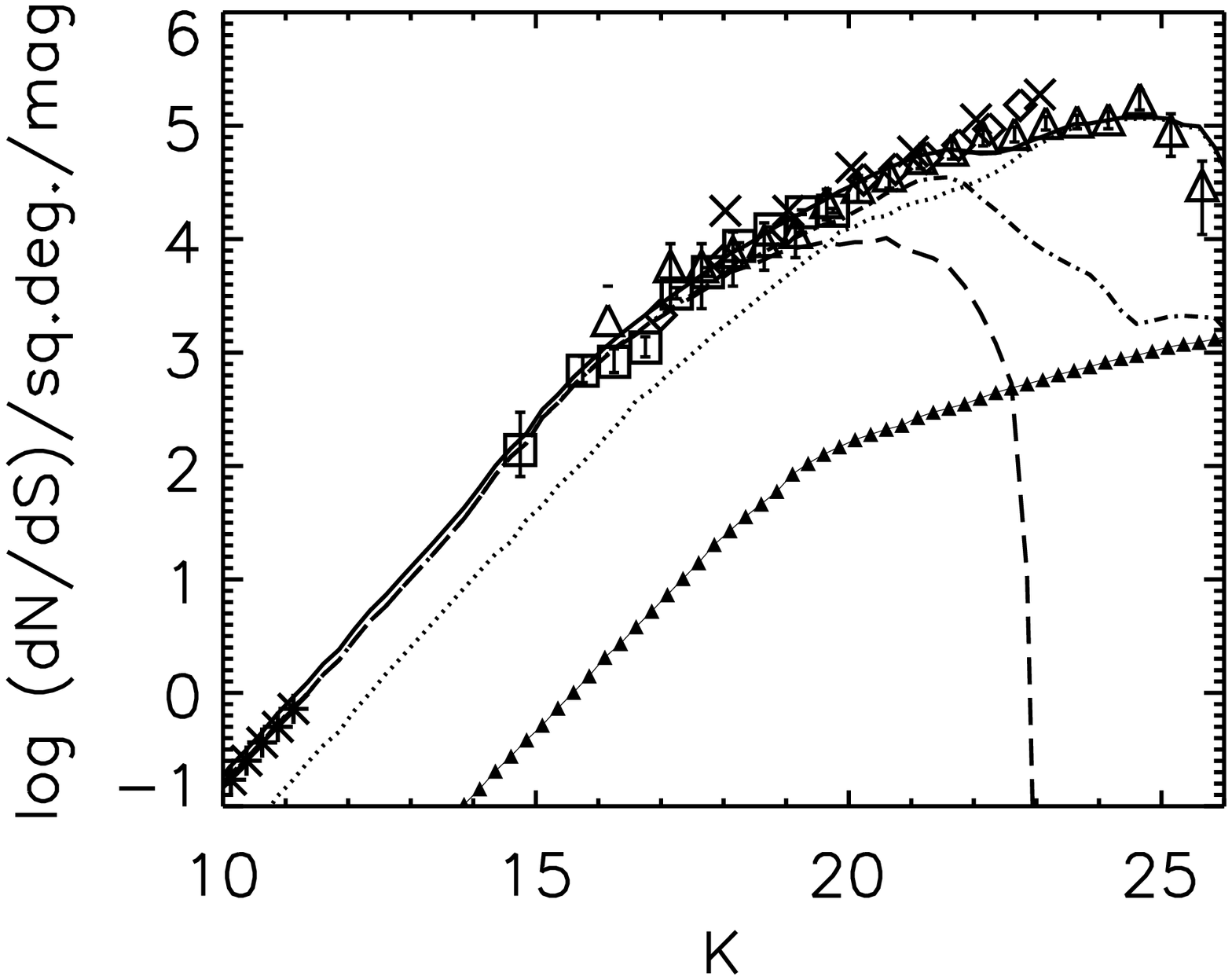}
\caption{K-band counts. Dot-dash: spheroids (the long-dashed line
singles out passively evolving spheroids); dots: late type
galaxies (spirals, irregulars and dwarves), assumed not to evolve
in the near-IR band (see text); filled triangles: AGN. We have
assumed $t_e=0.05\,$Gyr for star-forming spheroids (see text);
with this choice, they show up at $K\gsim 19$ and become
increasingly important with increasing $K$ up to $K\simeq 22$.
Data from Moustakas et al. (1997), Kochanek et al.\ (2001),
Saracco et al.\ (2001), Totani et al.\ (2001), Cimatti et al.\
(2002).} \label{ckstd}
\end{figure}

\begin{figure}[tbp]
\centering
\includegraphics[width=9truecm]{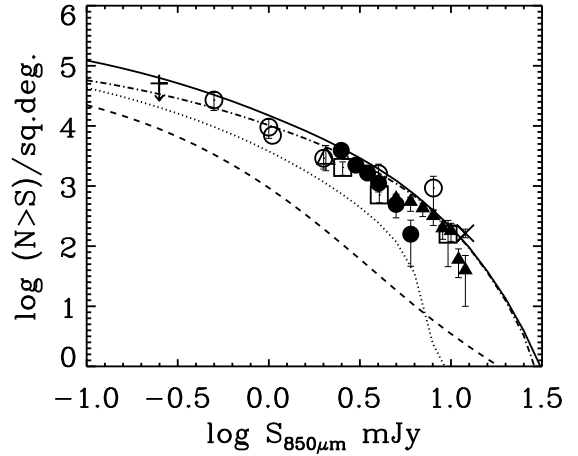}
\caption{$850 \mu$m counts. The dot-dashed, dotted, and dashed
lines refer to (star-forming) spheroids, starburst, and spiral
galaxies, respectively, while the solid line is the sum of the
various contributions. AGN are not visible within the limits of
the plot. Data from Blain et al.\ (1999), Hughes et al.\ (1998),
Barger, Cowie, \& Sanders (1999), Eales et al.\ (2000), Chapman et
al.\ (2002), Borys et al.\ (2002).} \label{c850std}
\end{figure}

\begin{figure}[tbp]
\centering
\includegraphics[width=9truecm]{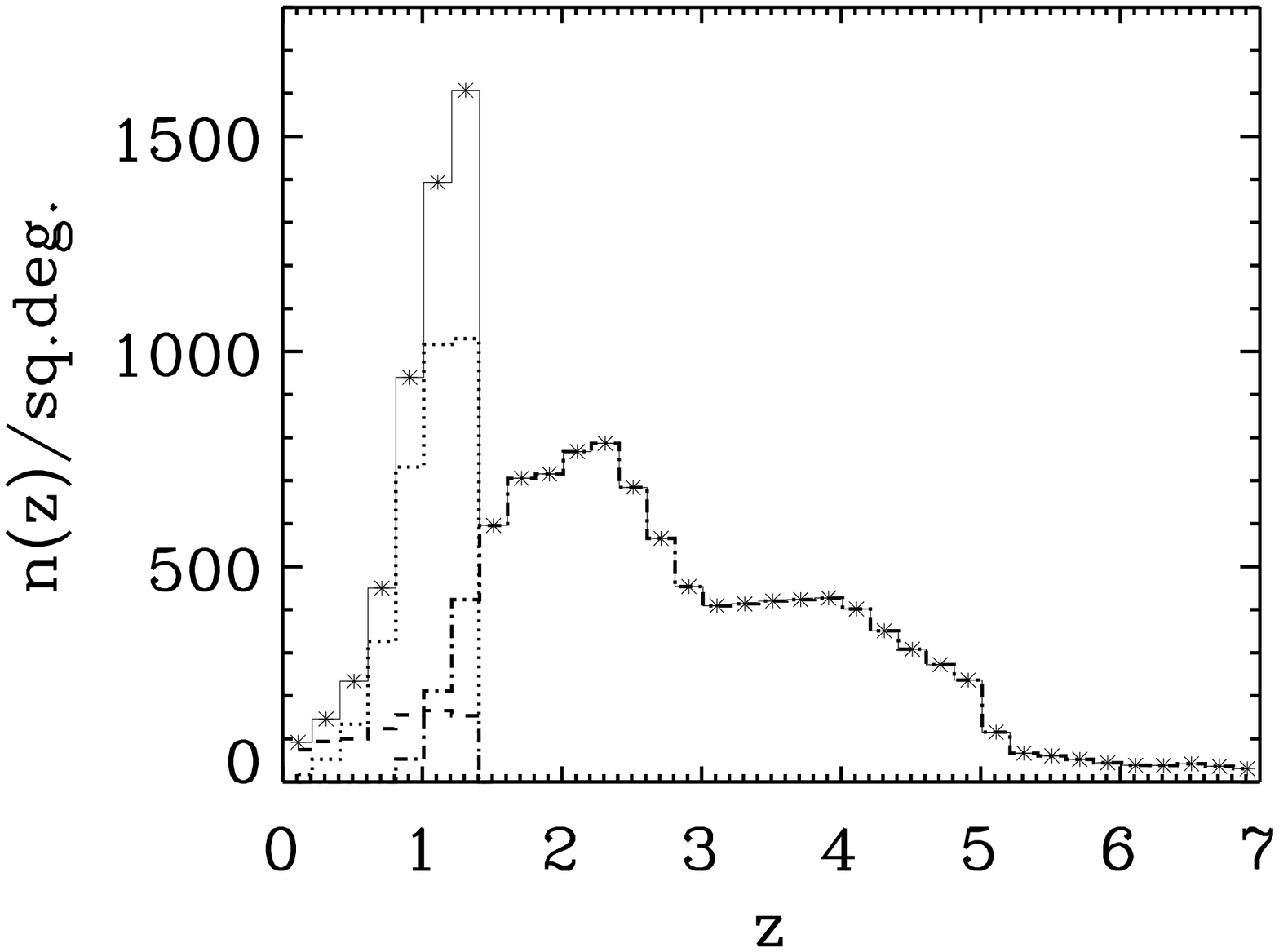}
\caption{Redshift distributions of sources brighter than $1\,$mJy
at $850 \mu$m. The dot-dashed, dotted, and dashed lines refer to
(star-forming) spheroids, starburst, and spiral galaxies,
respectively, while the thin solid line with asterisks is the sum
of the various contributions.} \label{nz850std}
\end{figure}

\subsection{Spectral energy distribution of spheroidal galaxies }
\label{sect:SED}

As in GDS04, we compute the SEDs of spheroidal galaxies using our
code GRASIL, described in Silva et al. (1998; for subsequent
improvements and updates see {\it
http://web.pd.astro.it/granato}). GRASIL computes the
time-dependent UV to radio SED of galaxies, given their star
formation and chemical enrichment histories (derived as described
in the previous section), and with a state of the art treatment of
dust reprocessing. The latter point is fundamental, since during
the phase of intense star-formation, young stars are mixed with a
huge amount of gas, quickly chemically enriched and dust polluted.
The use of local templates (such as M82 or Arp220) may be
inappropriate in these extreme conditions. Broadly speaking,
predictions of fluxes during the active star forming phase become
more and more affected by model uncertainties at shorter and
shorter wavelengths.

\begin{figure}[tbp]
\centering
\includegraphics[width=9truecm]{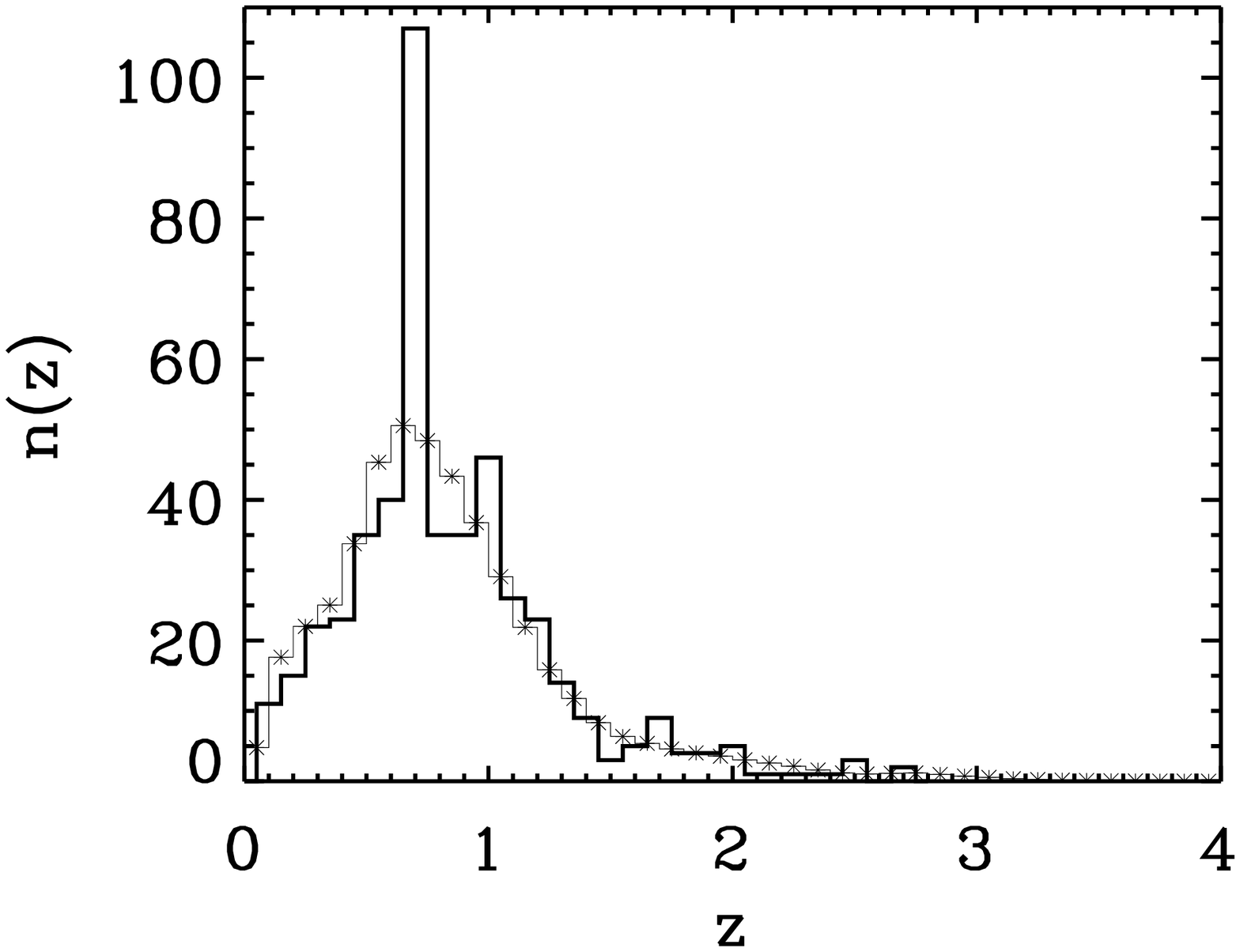}
\includegraphics[width=9truecm]{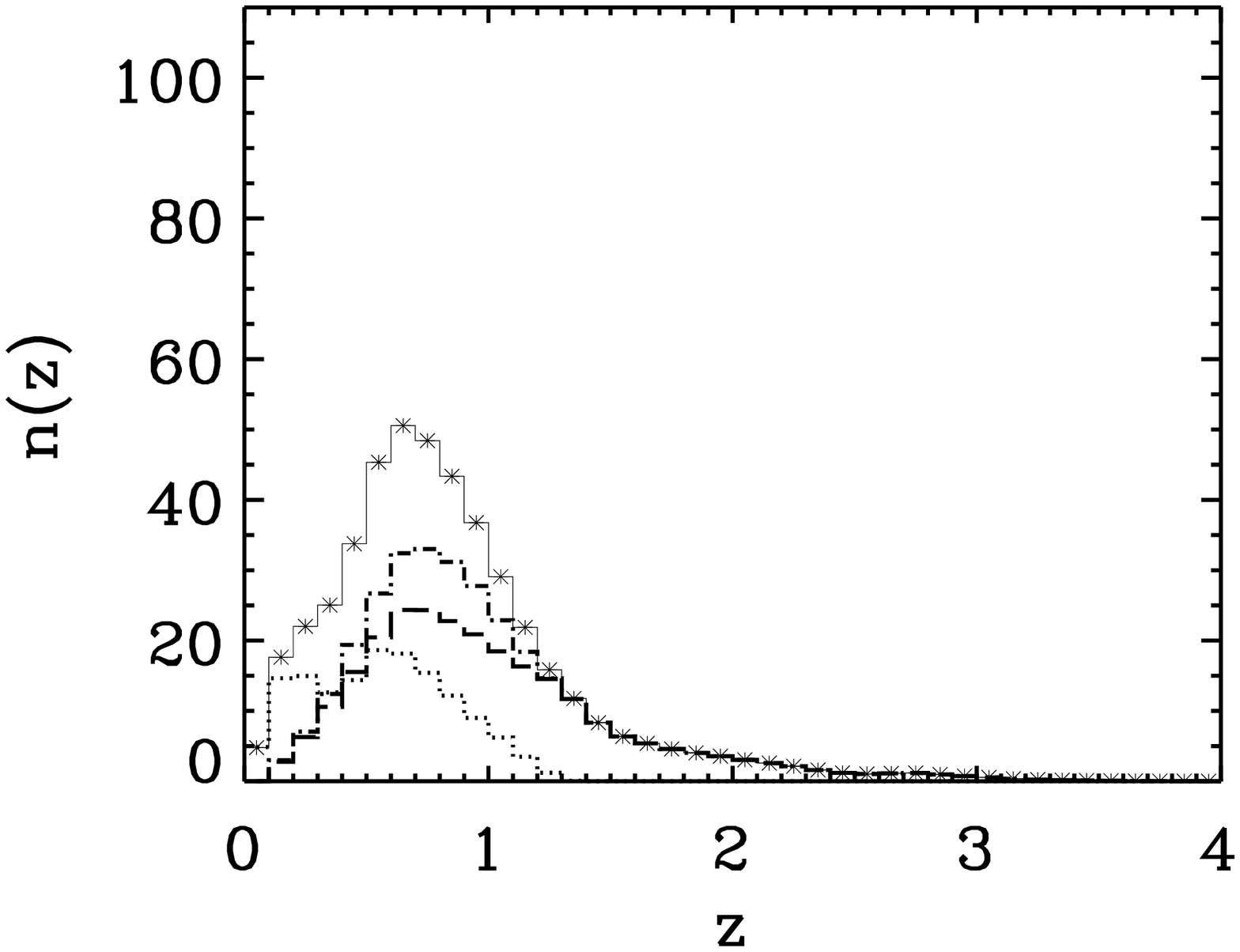}
\caption{Redshift distribution of galaxies brighter than $K=20$.
In the upper panel the total model distribution (thin line with
asterisks) is compared with the, mostly spectroscopic, data by
Cimatti et al. (2002; thick solid line). The lower panel details
the various contributions (dots: non-evolving late type galaxies;
dot-dash: spheroids). The long-dashed line singles out passively
evolving spheroids.} \label{nzk20std}
\end{figure}

\begin{figure}[tbp]
\centering
\includegraphics[width=9truecm]{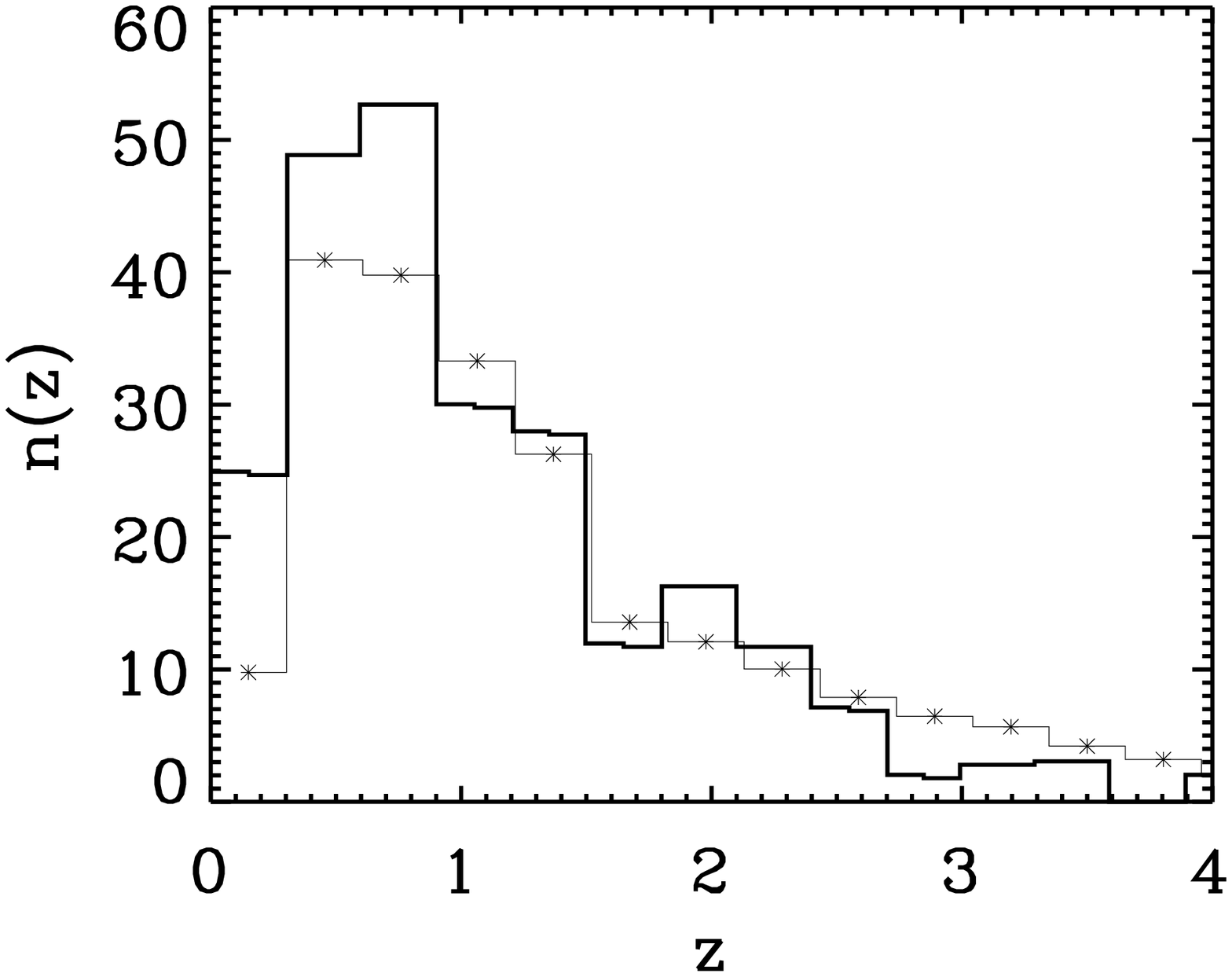}
\includegraphics[width=9truecm]{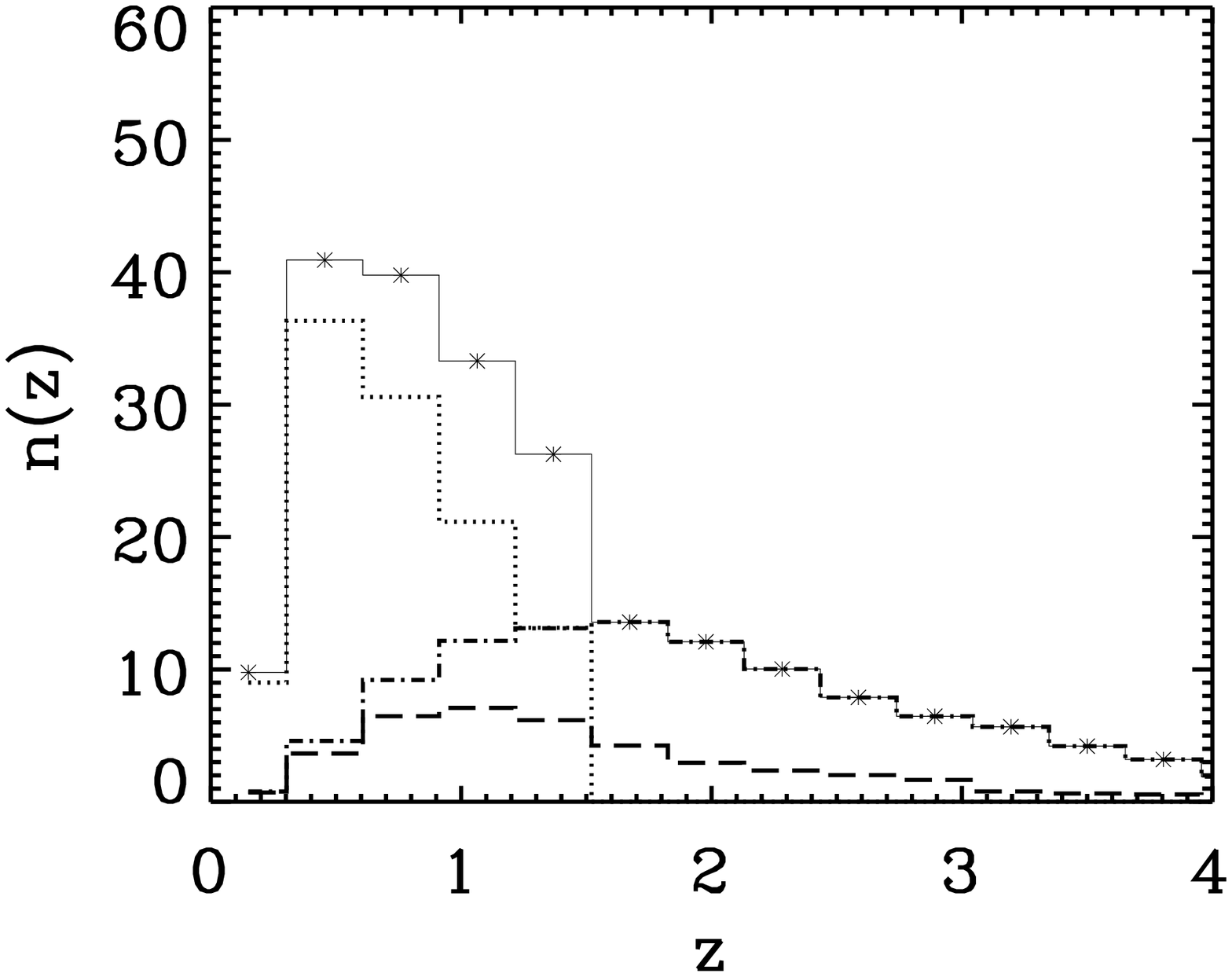}
\caption{Redshift distribution of galaxies brighter than $K=23$.
The meaning of the lines is the same as in
Fig.~\protect{\ref{nzk20std}}. Data (photometric redshifts) from
the Subaru deep survey (Kashikawa et al. 2003). } \label{nzk23std}
\end{figure}

\begin{figure}[tbp]
\centering
\includegraphics[width=9truecm]{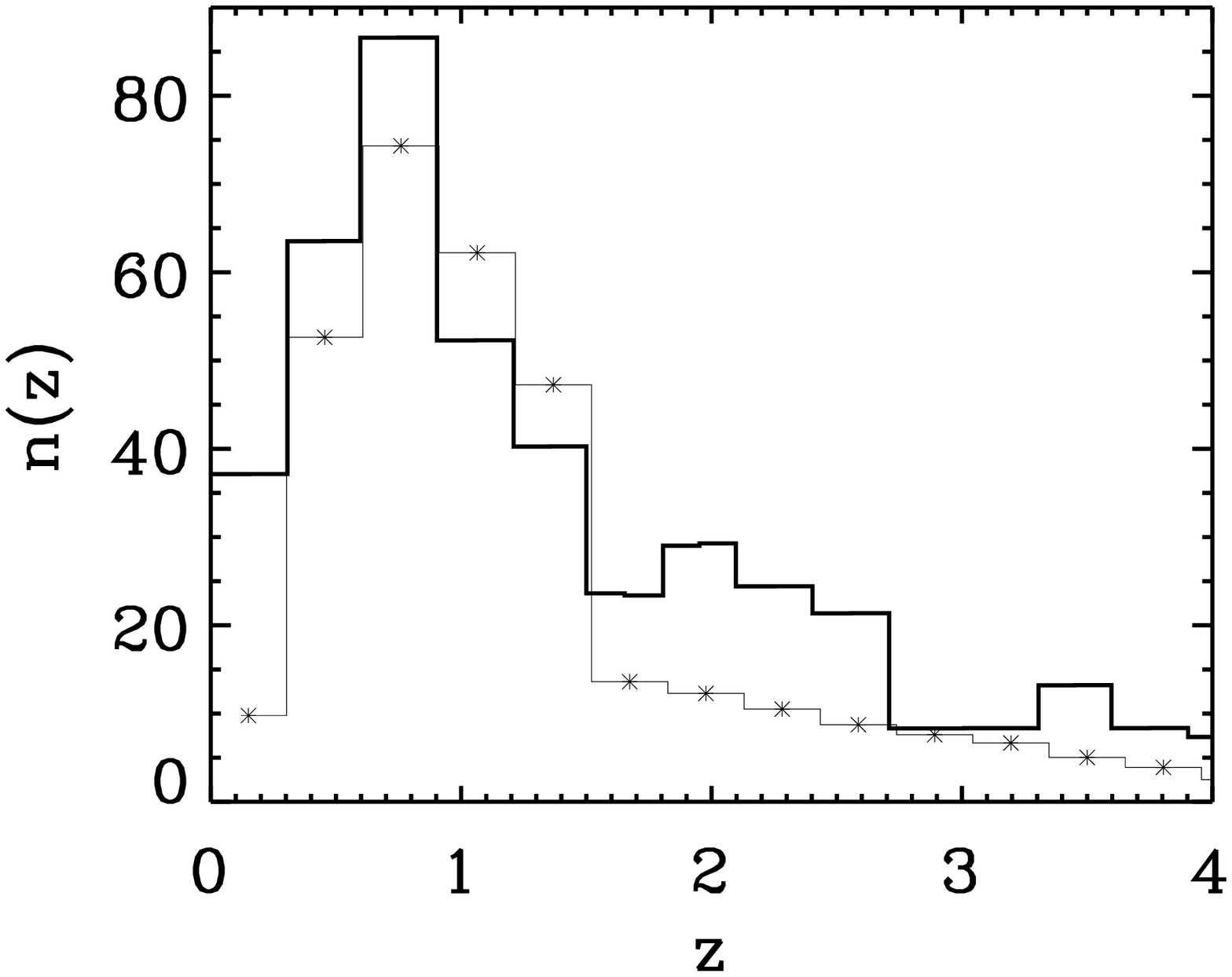}
\includegraphics[width=9truecm]{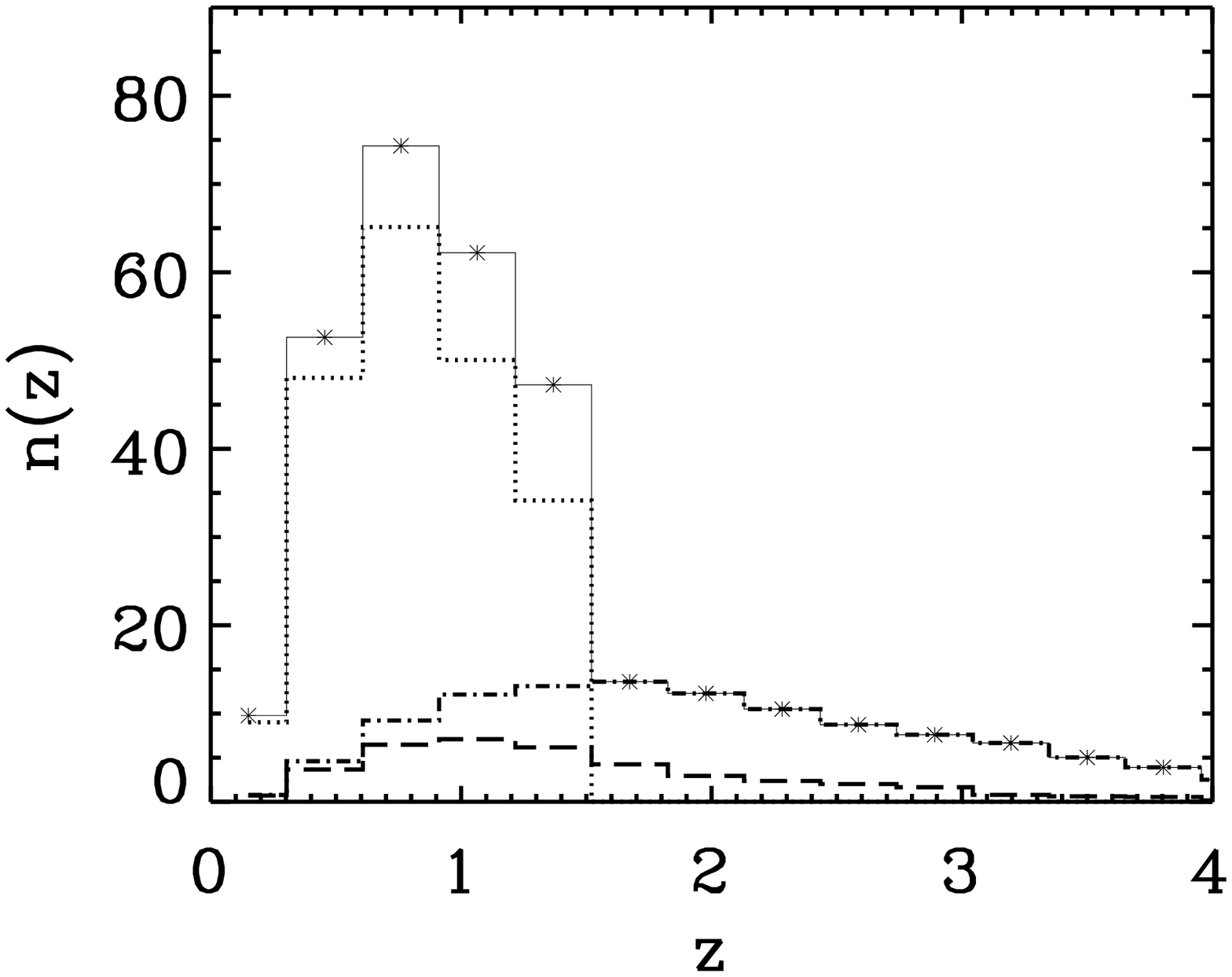}
\caption{Redshift distribution of galaxies brighter than $K=24$.
The meaning of the lines is the same as in
Fig.~\protect{\ref{nzk20std}}. Data (photometric redshifts) from
the Subaru deep survey (Kashikawa et al. 2003).}\label{nzk24std}
\end{figure}

One of the most important distinctive features of GRASIL is that
it has included, for the first time, the effect of {\it
differential extinction} of stellar populations (younger stellar
generations are more affected by dust obscuration), due to the
fact that stars form in a denser than average environment, the
molecular clouds, and progressively get rid of them.

While in principle the GRASIL SEDs depend on several parameters,
in the conditions envisaged here only two of them are really
important: the dust optical depth of molecular clouds ($\tau_{\rm
MC}$), and the escape time scale ($t_e$) of young stars from their
parent molecular clouds. The latter affects more directly the
predictions for the optical-NIR regions and has been determined
fitting the K-band data. The former affects the entire SED, and is
proportional to the dust to gas mass ratio $\delta$ and to $M_{\rm
MC}/r_{\rm MC}^2$, $M_{\rm MC}$ and $r_{\rm MC}$ being the typical
mass and radius of molecular clouds, respectively. We have simply
assumed that $\delta$ increases linearly with the gas metallicity
(e.g., Dwek 1998) and is $\simeq 1/100$ (the Galactic value) for
solar metallicity. The ratio $M_{\rm MC}/r_{\rm MC}^2$ has been
allowed to vary within a range of values compatible with those
found for typical giant molecular clouds in the Milky Way ($M_{\rm
MC} \sim 10^6 M_\odot$, $r_{\rm MC} \sim 15\,$pc).

As for $t_e$, it actually represents the time during which stars
are within an environment of high optical depth ($\tau_{\rm MC}$)
since GRASIL assumes a small optical depth (due to cirrus) outside
molecular clouds. This is likely to be unrealistic in the extreme
conditions envisaged here during the fast star forming phase of
spheroids, particularly in the dense central regions. Indeed in
this case there is a high probability that a light ray coming from
a star outside its parent molecular cloud intersects at least one
molecular cloud over a typical galactic path. Our cure for this
problem is described in Sect.~\ref{sect:param}.

\begin{figure}[tbp]
\centering
\includegraphics[width=9truecm]{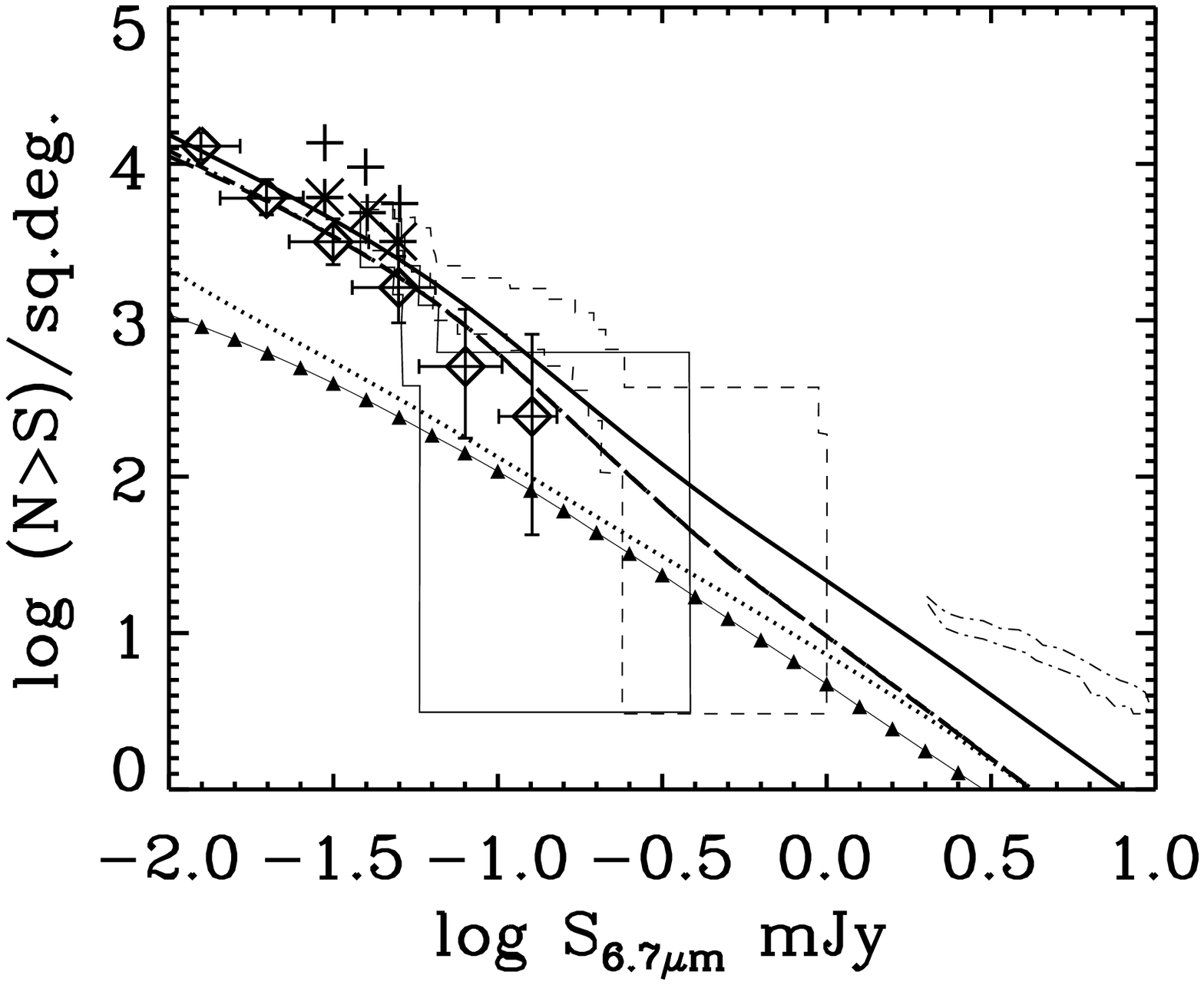}
\includegraphics[width=9truecm]{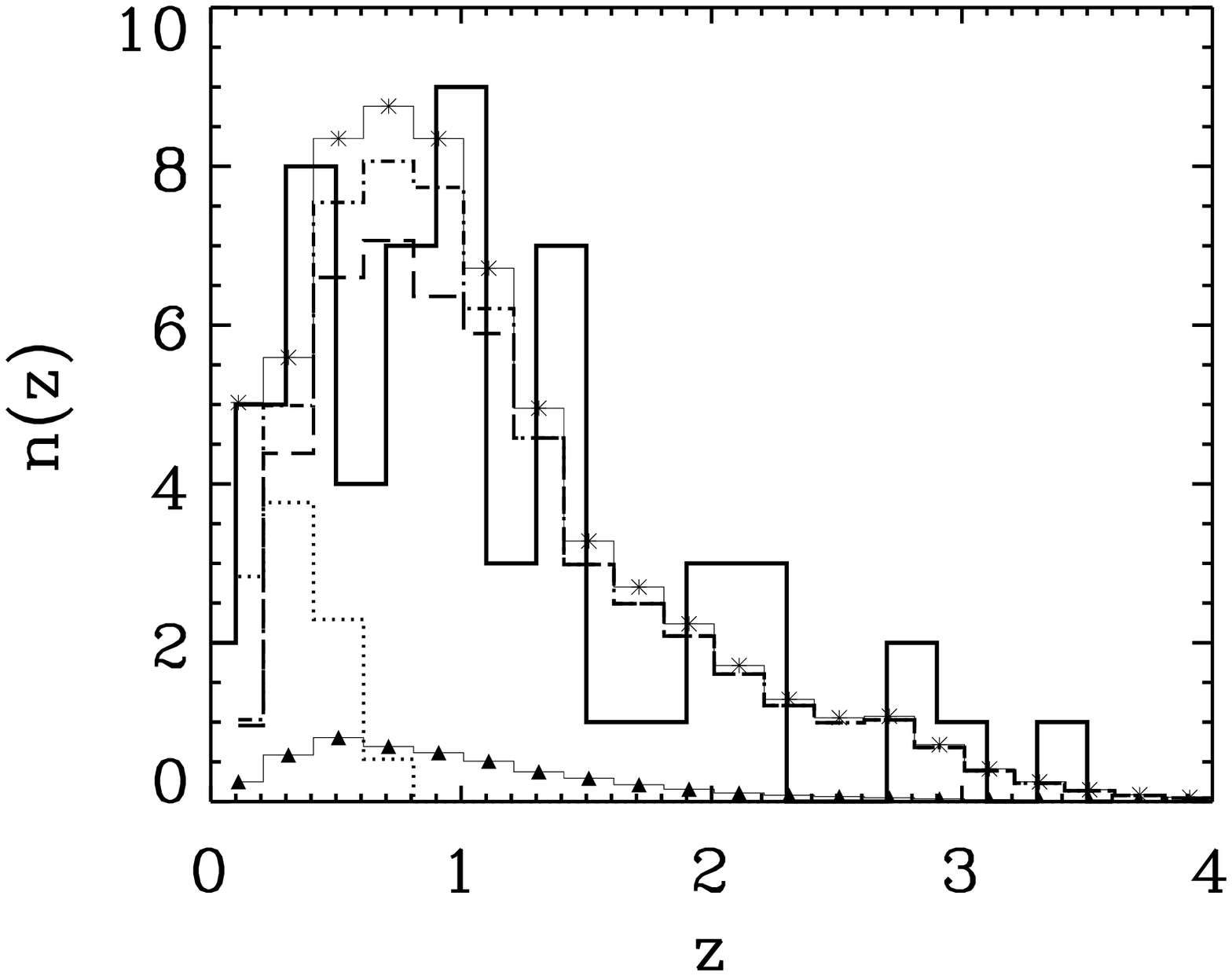}
\caption{$6.7 \mu$m counts and redshift distribution of sources
with $S_{6.7}>10\mu$Jy within $16$ arcmin$^2$. The dot-dashed and
dotted lines refer to spheroids (mostly passively evolving, as
illustrated by the long dashed line, detailing their contribution
to the redshift distribution in the lower panel) and to
non-evolving late-type galaxies, respectively. The filled
triangles refer to AGN. Their total is the thin line with
asterisks. The thick continuous line in the redshift distribution
is from Sato et al. (2004). Observed counts from Taniguchi et al.
(1997), Altieri et al.\ (1999), Serjeant et al.\ (2000), Oliver et
al. (1997, 2002), Sato et al. (2003).} \label{c6.7std}
\end{figure}

\subsection{Prescriptions for other galaxy populations}
\label{sec:other} We use the GDS04 prescriptions to model the star
formation occurring within DM halos (up to a maximum mass of
$\log(M_{\rm vir, max}/M_\odot)=13.2$) virialized at $z_{\rm vir}
\gsim 1.5$ and $M_{\rm vir} \gsim 10^{11.6} M_\odot$. The latter
cuts are meant to crudely filter out galactic halos associated
with a spheroidal galaxy. We envisage disks (and irregulars) as
associated primarily to halos virializing at $z_{\rm vir} \lsim
1.5$, which have incorporated, through merging processes, a large
fraction of halos less massive than $4 \times 10^{11}\,M_\odot$
virializing at earlier times. The latter may become the bulges of
late type galaxies. However, we do not address here the physics of
the formation of disk (and irregular) galaxies. Their
contributions to number counts are estimated following, as usual,
a phenomenological approach, which consists in simple analytic
recipes to evolve their local luminosity functions (LFs), and
appropriate templates for their SEDs to compute K-corrections and
to extrapolate the models to different wavelengths where direct
determinations of the local luminosity functions are not
available.

We adopt the simplest assumptions that are able to describe the
counts and redshift distributions of these populations. The
prescriptions applying in the mid-IR (MIR) to sub-mm spectral
region are different from those in the optical/near-IR (NIR), due
to the different emission mechanisms. Indeed it is well known that
galaxies with very different far-IR (FIR) SEDs can have similar
optical-NIR SEDs. This is clearly witnessed also by the comparison
between the FIR luminosity functions (e.g. Saunders et al. 1990)
and the optical ones. Only the latter is well described by the
Schechter parameterization.

\begin{figure}[tbp]
\centering
\includegraphics[width=9truecm]{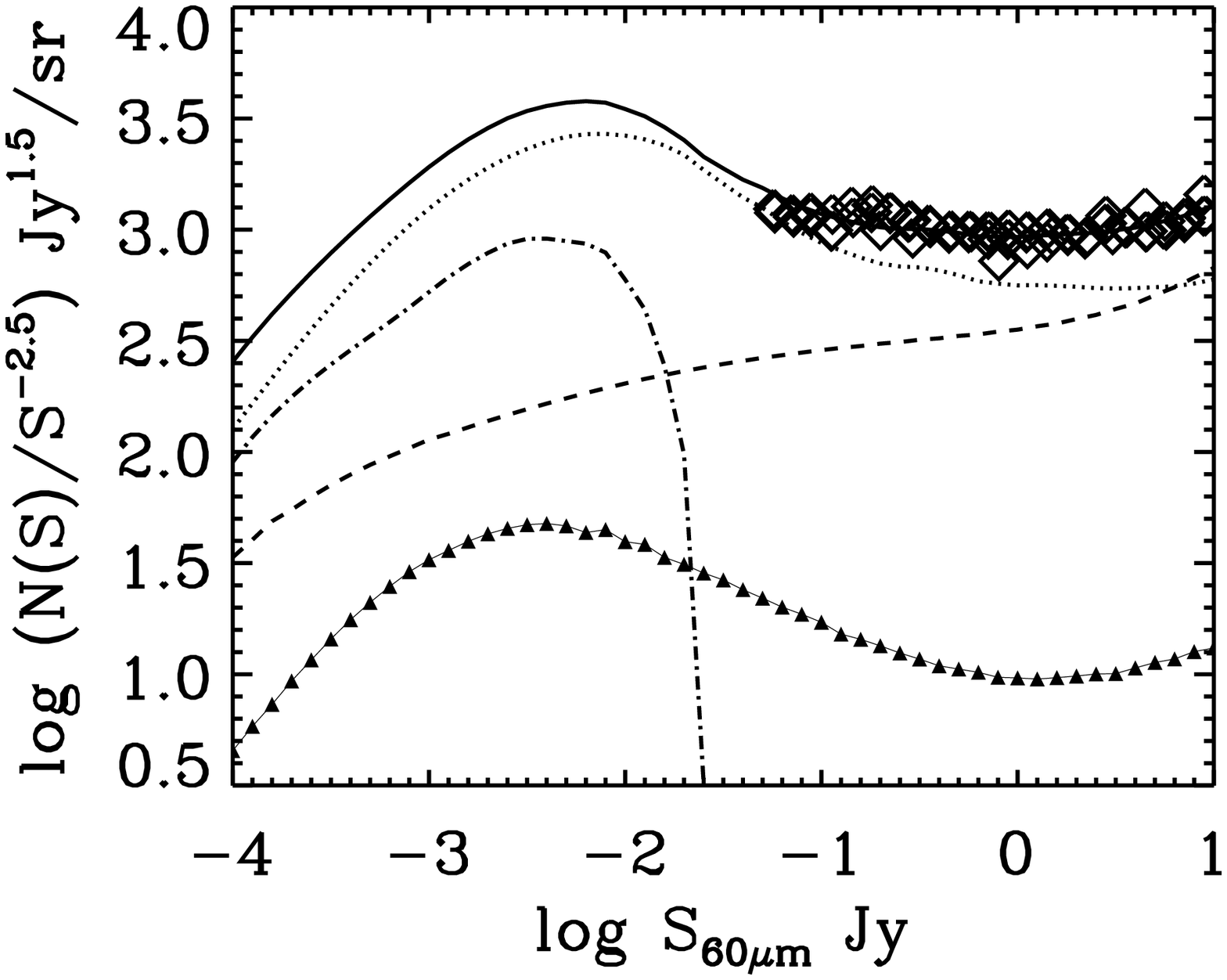}
\caption{IRAS $60 \mu$m differential counts. Dotted line:
starbursts; dashed line: spirals; dot-dashed line: (star-forming)
spheroids; filled triangles: AGN; solid line: total. Data from
Mazzei et al. (2001), Lonsdale et al.\ (1990), Pearson \&
Rowan-Robinson (1996), Bertin et al.\ (1997).} \label{c60std}
\end{figure}

\begin{figure}[tbp]
\centering
\includegraphics[width=9truecm]{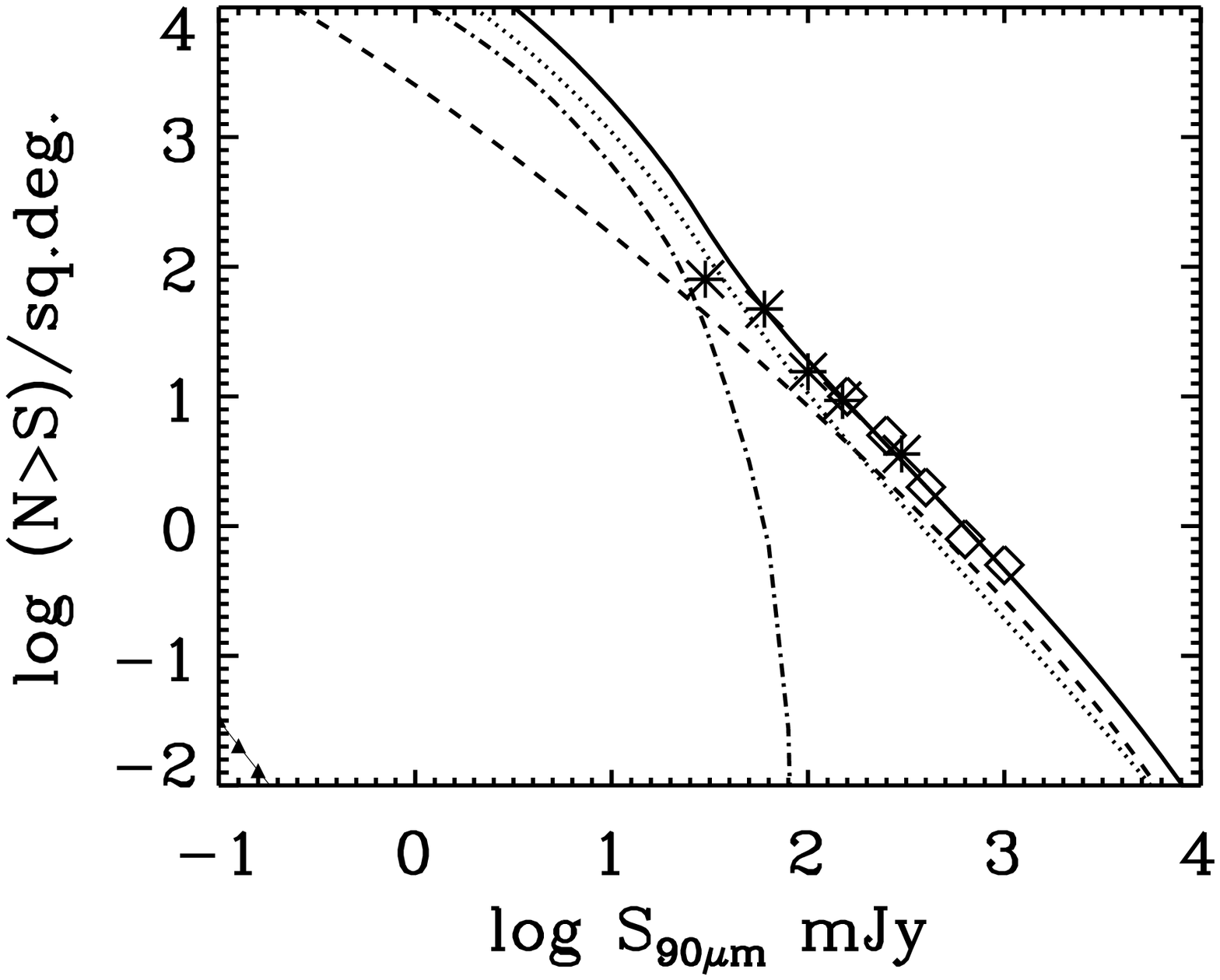}
\includegraphics[width=9truecm]{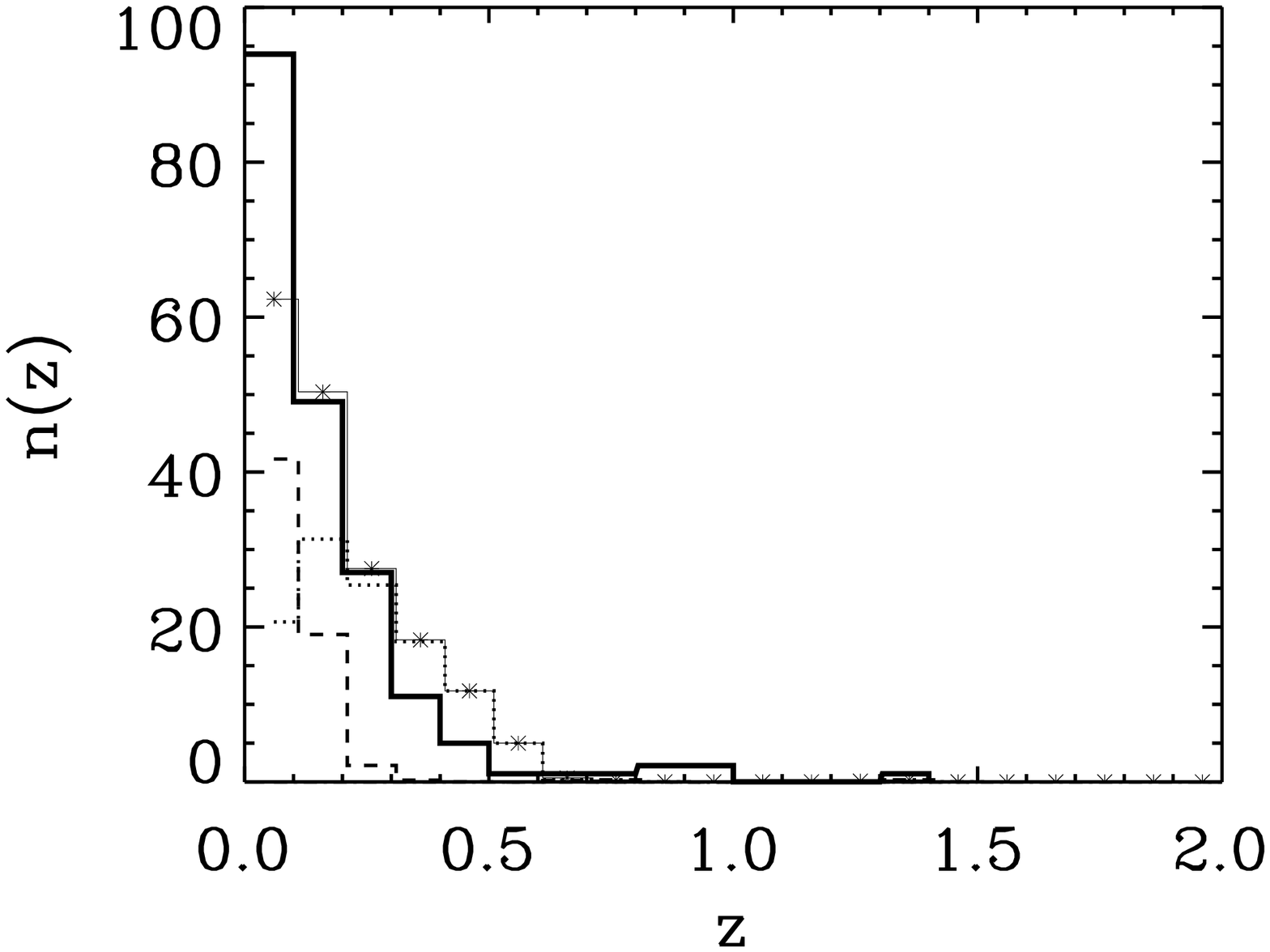}
\caption{90$\,\mu$m counts and redshift distribution of sources
with S$_{90}>$70 mJy within $5.34$ deg$^{-2}$. The SB SED is that
of NGC6090. Dotted line: starbursts; dashed line: spirals;
dot-dashed line: spheroids; filled triangles: AGN. The solid line
in the upper panel and the thin continuous histogram with
asterisks in the lower panel show the sum of all contributions.
The thick continuous histogram displays the observed
$z$-distribution by Rowan-Robinson et al. (2003). Observed counts
from Efstathiou et al.\ (2000) and Rodighiero et al.\ (2003).}
\label{c90std}
\end{figure}

\begin{figure}[tbp]
\centering
\includegraphics[width=9truecm]{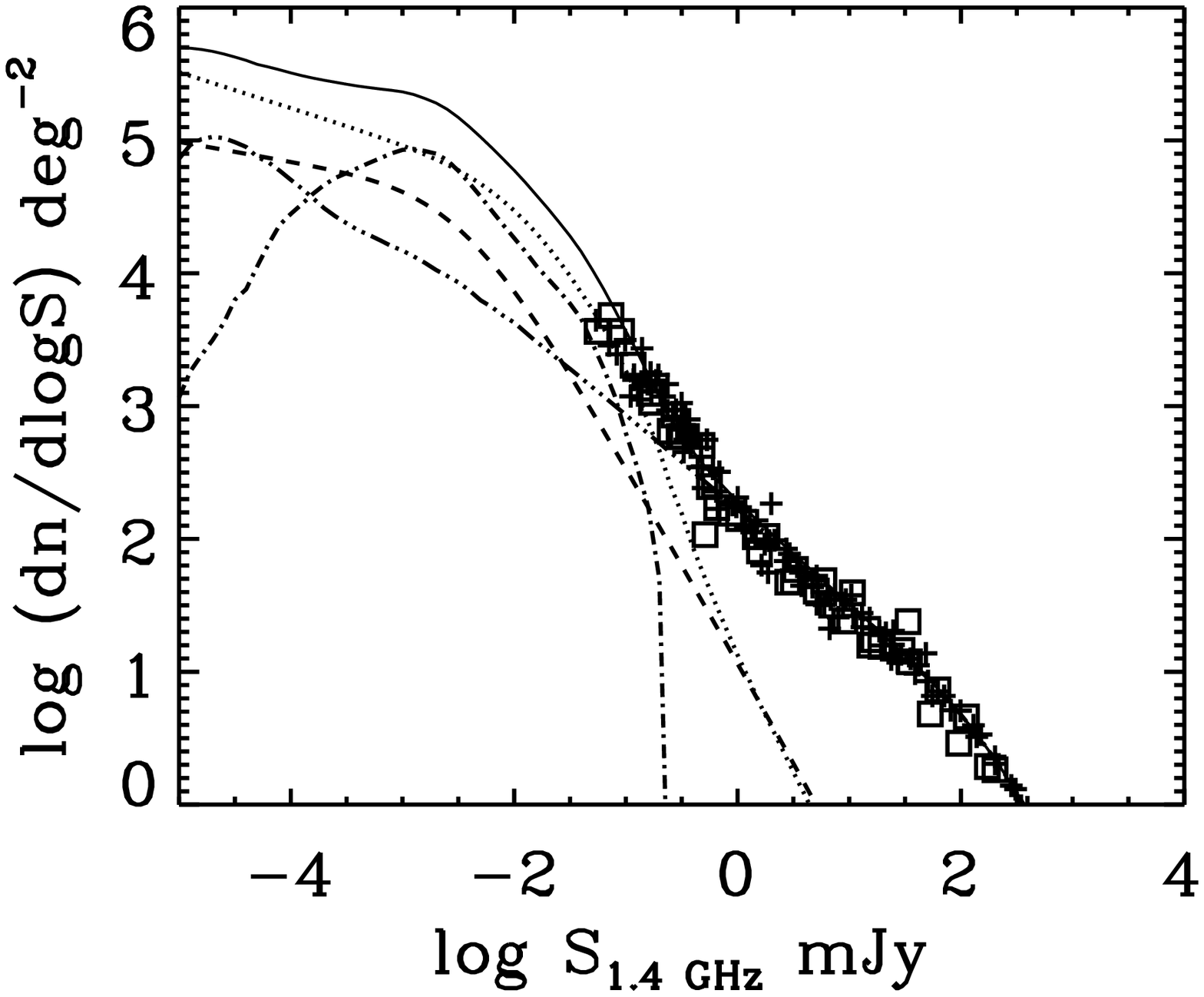}
\caption{$1.4$ GHz counts. Dotted line: starbursts; dashed line:
spirals; dot-dashed line: spheroids; three dots-dash line: sum of
flat- and steep-spectrum radio galaxies according to models by
Dunlop \& Peacock (1990). Data from Windhorst et al.\ (1993),
Gruppioni et al.\ (1997, 1999), Ciliegi et al.\ (1999), Hopkins et
al.\ (1999), Richards (2000).} \label{c1.4std}
\end{figure}

\subsubsection{Mid-IR to sub-mm region}
\label{sec:mirmm}

The IRAS and ISO survey data can be interpreted (e.g.,
Franceschini et al.\ 2001; Gruppioni et al.\ 2002) in terms of a
slowly evolving population of spiral (and irregular) galaxies
(SPs), and a rapidly evolving population of starburst galaxies
(SBs). We adopt the $60\,\mu$m local LFs by Saunders et al (1990),
that have been derived for ``warm'' and ``cold'' galaxies (based
on IRAS colors) that we associate, respectively, to SBs and SPs.

For SBs we used, as templates, the SEDs of NGC6090 and of M82
(Silva et al. 1998). However, both these galaxies have 15 to $60
\mu$m luminosity ratios, $L_{15}/L_{60} \simeq 0.08$, somewhat
higher than the average value for active star forming galaxies
selected at $60\,\mu$m ($<L_{15}/L_{60}> \simeq 0.05$; Mazzei et
al. 2001) and for nearby bright galaxies ($<L_{15}/L_{60}> \simeq
0.04$; Dale et al. 2000). We take $L_{15}/L_{60} = 0.05$ to
extrapolate to 15$\,\mu$m the 60$\,\mu$m SB local LF.

Unless otherwise noted, we have adopted the SED of NGC6090, which
turns out to be closer to the SEDs of starburst galaxies detected
by the ISO FIRBACK survey (Chapman et al. 2002; Sajina et al.
2003; Patris et al. 2003). This is clearly a quite crude
approximation since the starburst SEDs are known to vary
systematically with luminosity (Xu et al. 2001; Chapman et al.
2003; Lagache et al. 2003). A more detailed treatment of the
starburst SEDs is however beyond the scope of the present paper,
which is focussed on spheroidal galaxies. We have therefore chosen
to adopt, for the other populations, the simplest models still
compatible with the data.

The evolutionary parameters of SBs were determined from fits of
the 15$\,\mu$m counts and redshift distributions. Consistent with
earlier analyses, we find that both density and luminosity
evolution is required for the SB population, with $LF[L(z),z] =
LF[L(z)/(1+z)^{2.5}, z=0]\times (1+z)^{3.5}$. Not to exceed the
bright 15$\,\mu$m source counts we had to cut off the local LF at
$L_{60\mu{\rm m}} = 2\times 10^{32}$ erg/s/Hz. A similar
high-luminosity cut-off was introduced by Gruppioni et al.\ (2002)
and Pozzi et al.\ (2003).

As for SPs, we use a pure luminosity evolution model,
$L(z)=L(0)\,(1+z)^{1.5}$ and the SED of the Sc galaxy NGC6946.
Again, a detailed analysis of the contributions of SPs to the
counts and to the redshift distributions at the various
wavelengths, allowing for their colour distribution as well as for
the cold galaxies discovered by the ISOPHOT serendipity survey
(Stickel et al. 2000), is beyond the scope of the present paper.

The LFs of both SBs and SPs are assumed to evolve as described
above up to $z=1$  and to keep constant afterwards, up to a
redshift cut-off $z_{\rm cutoff} =1.5$, above which essentially
all the massive halos are associated, according to GDS04, with
spheroidal galaxies.

\begin{figure}[tbp]
\centering
\includegraphics[width=9truecm]{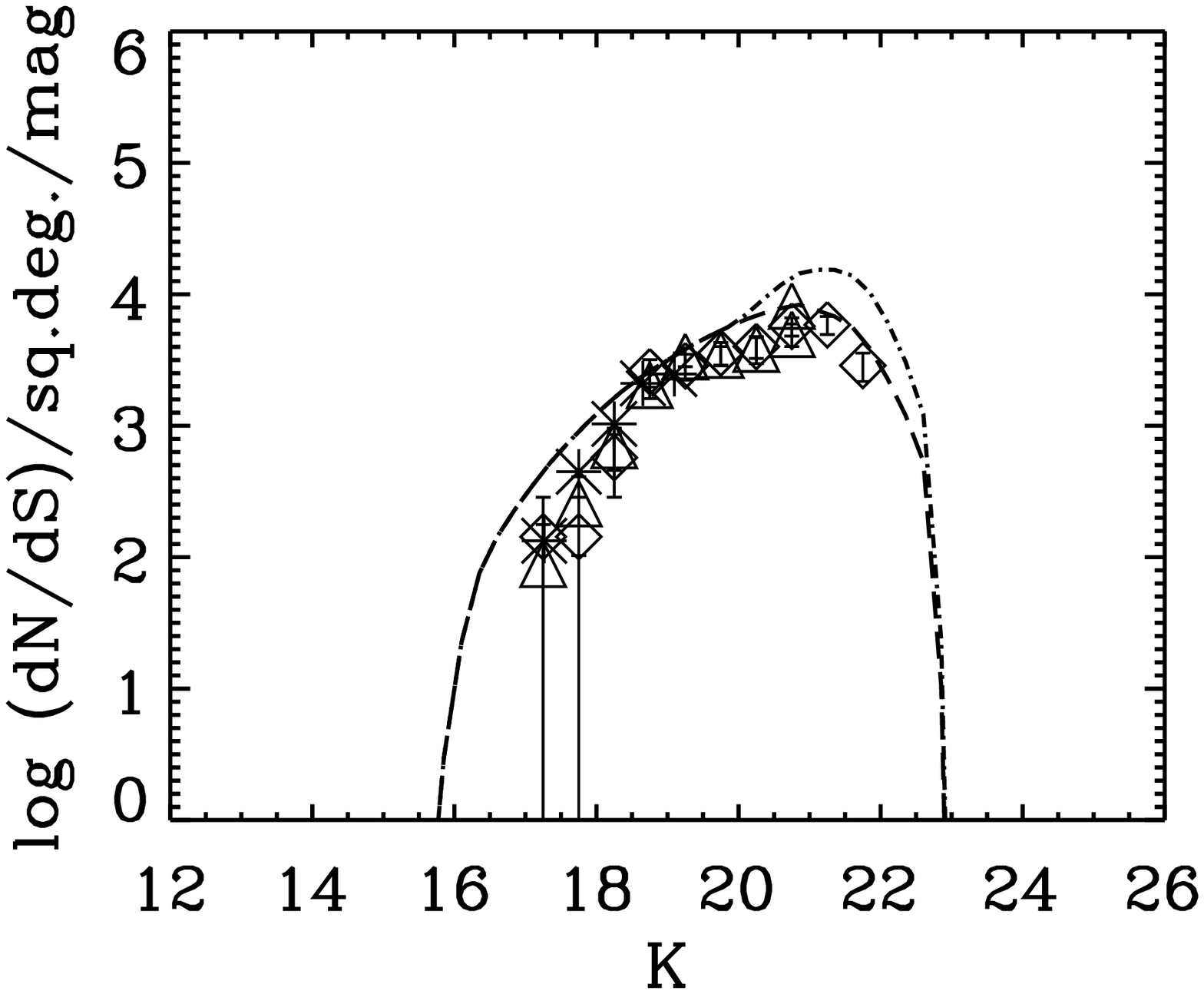}
\includegraphics[width=9truecm]{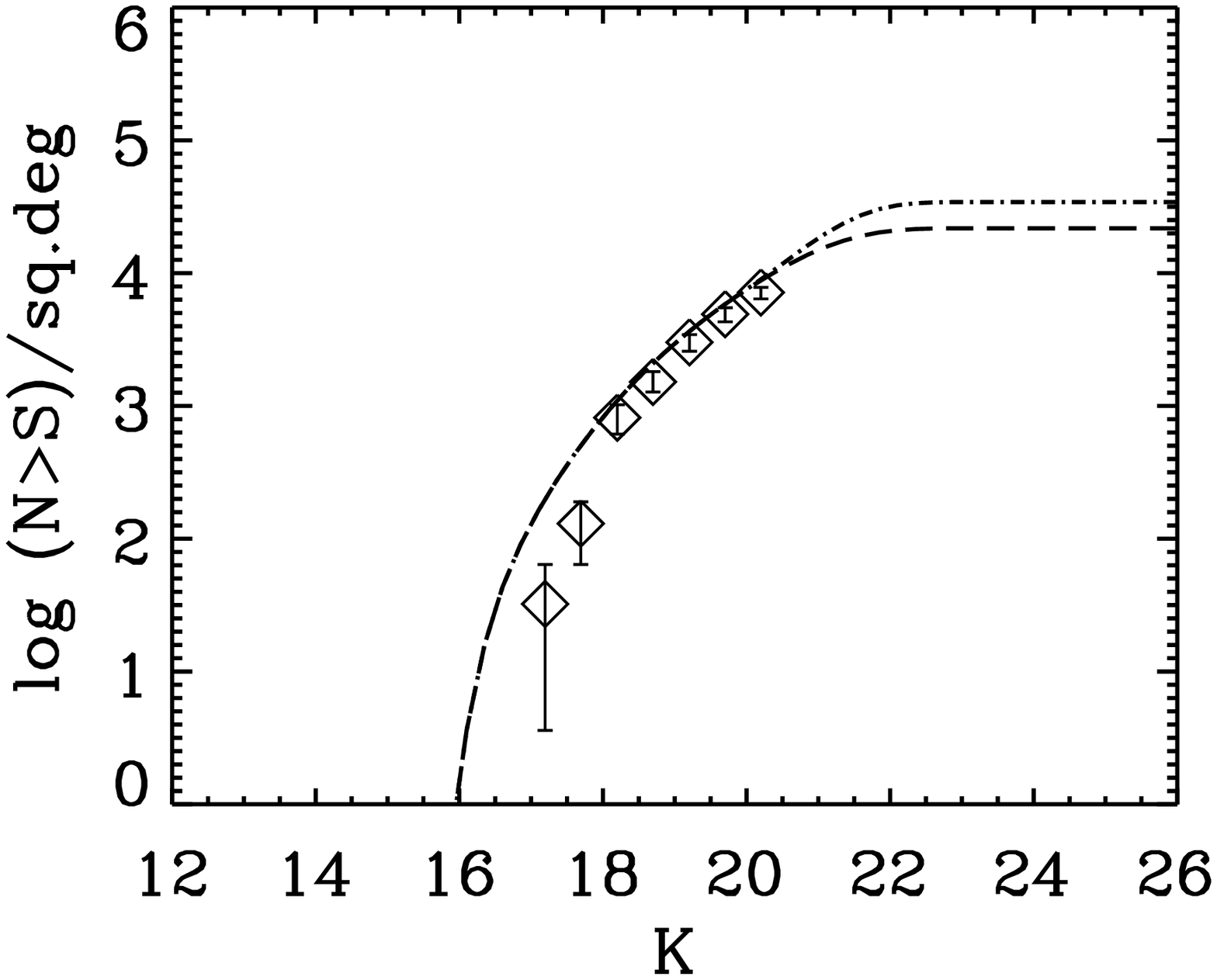}
\caption{Differential and integral K-band counts for EROs
($R-K>5$). The long dashed and dot-dashed lines show the predicted
counts of passively evolving spheroids and of all spheroids
(including the star-forming ones). Data in the upper panel are
from Roche et al.\ (2002, 2003), and Daddi et al.\ (2000). In the
lower panel model predictions are compared with the counts of old
(passively evolving) EROs by Miyazaki et al.\ (2003).}
\label{ckstdero}
\end{figure}

\begin{figure}[tbp]
\centering
\includegraphics[width=9truecm]{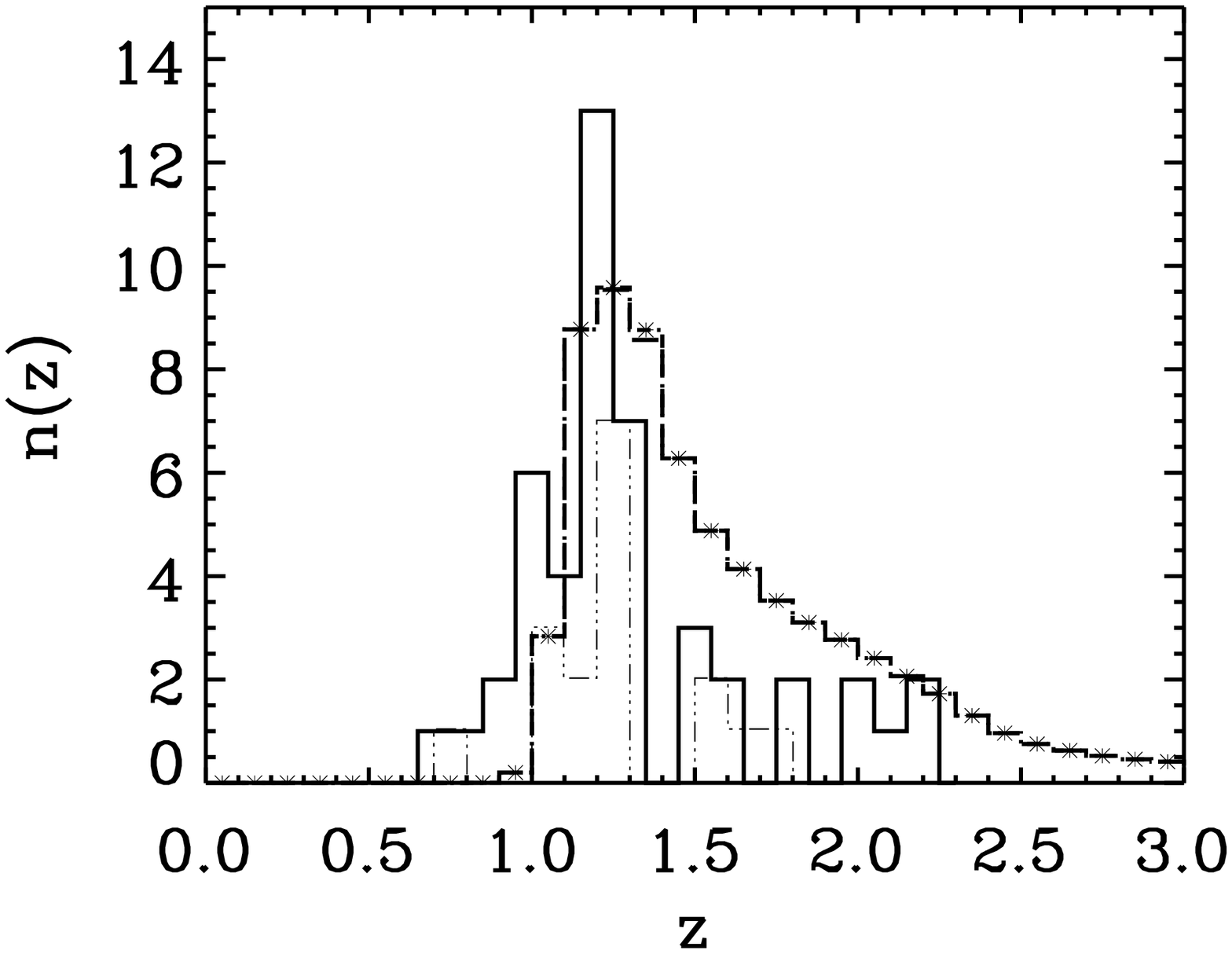}
\includegraphics[width=9truecm]{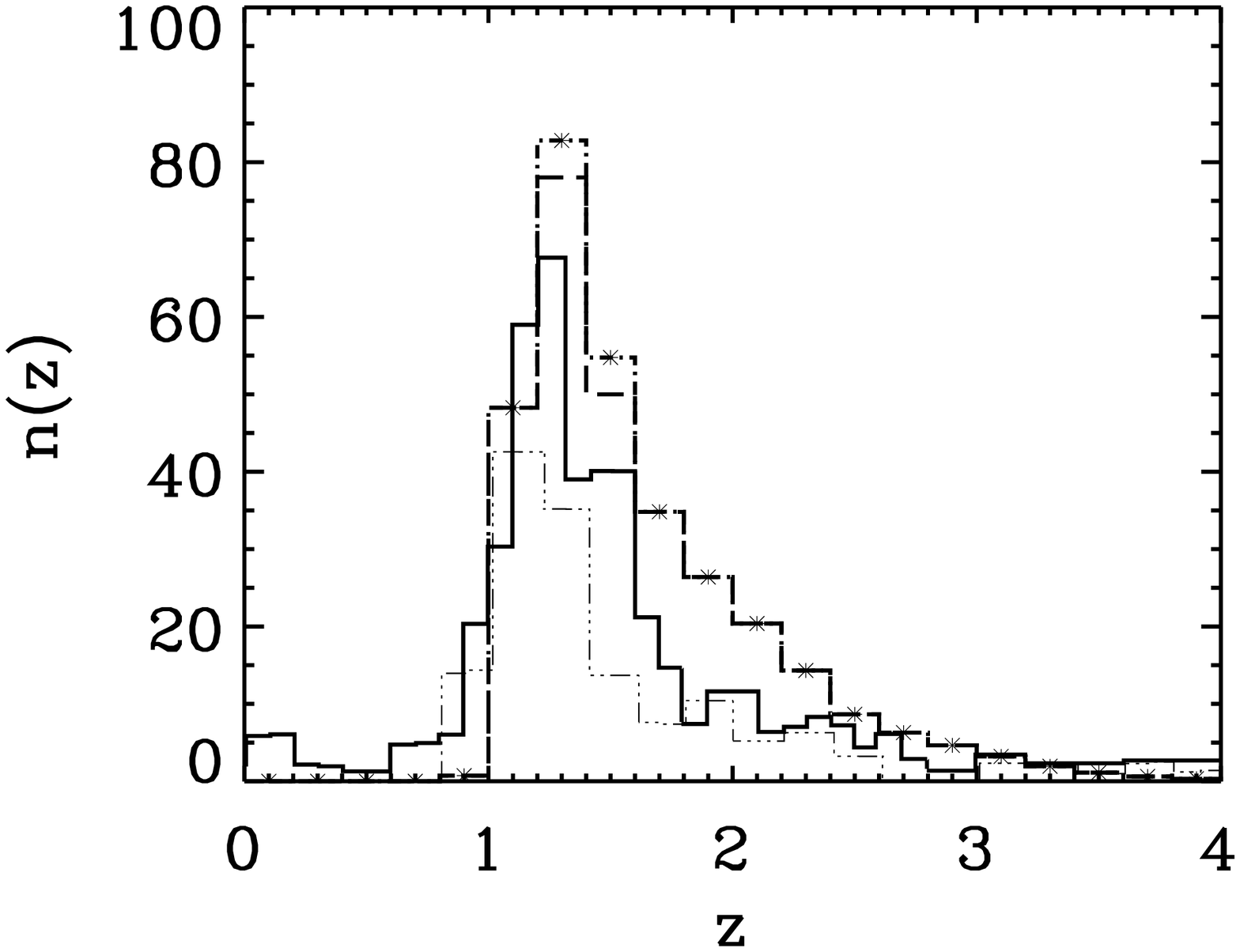}
\includegraphics[width=9truecm]{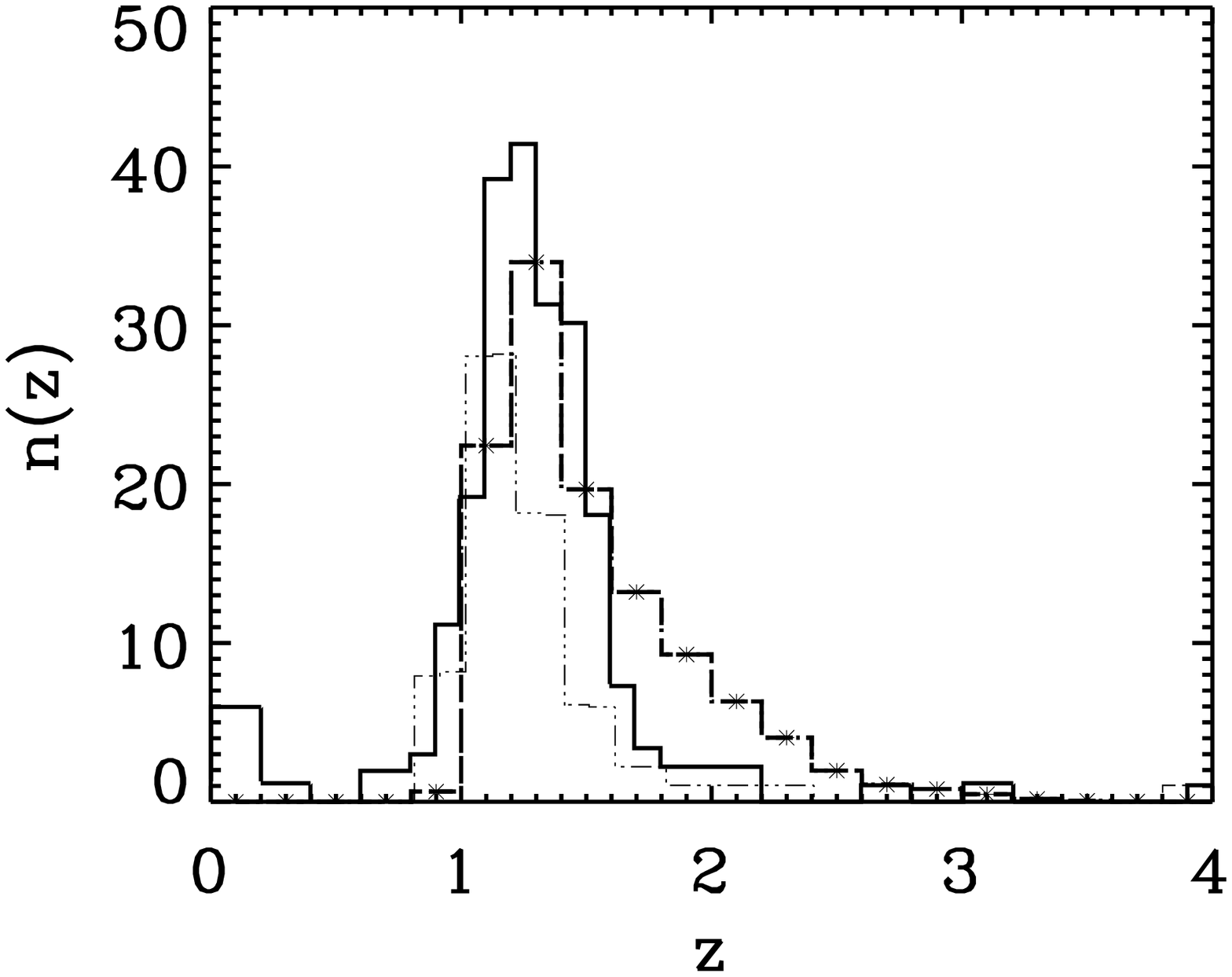}
\caption{From top to bottom: redshift distribution of EROs with
$K<20$ in the K20/CDFS survey area ($32.2$ arcmin$^2$, Cimatti et
al.\ 2003), $K<20.3$ and $K<19.2$ ($114$ arcmin$^2$, Miyazaki et
al.\ 2003). The observed redshift distributions including both
passively evolving and star-forming EROs are shown by the thick
solid lines, while the three dots-dashed lines refer to the
passively evolving only. The model distributions are represented
by the dot-dashed lines with asterisks (all EROs) and by the
long-dashed lines (passively evolving only).} \label{nzkstdero}
\end{figure}

\begin{figure}[tbp]
\centering
\includegraphics[width=7.5truecm,angle=90]{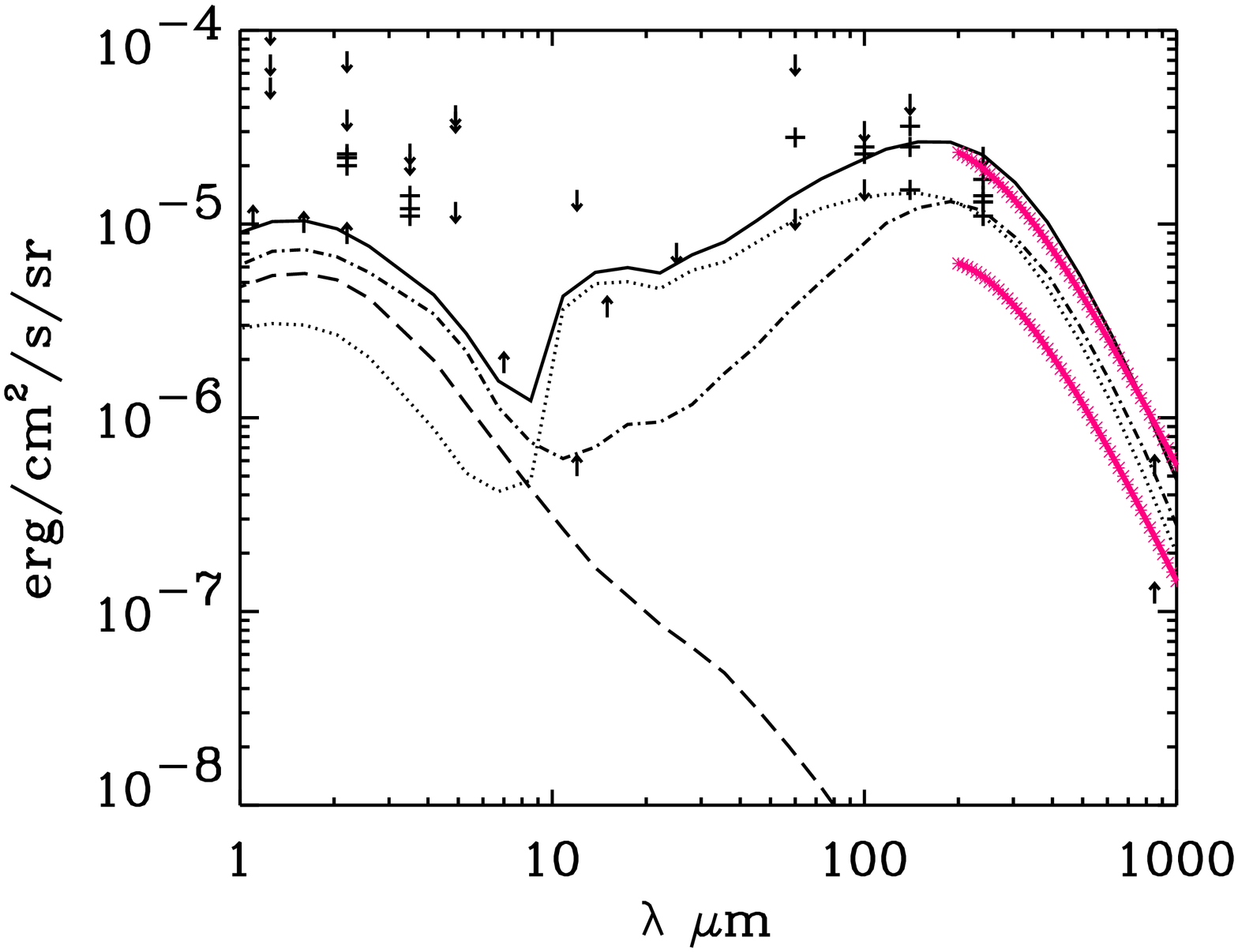}
\includegraphics[width=7.5truecm,angle=90]{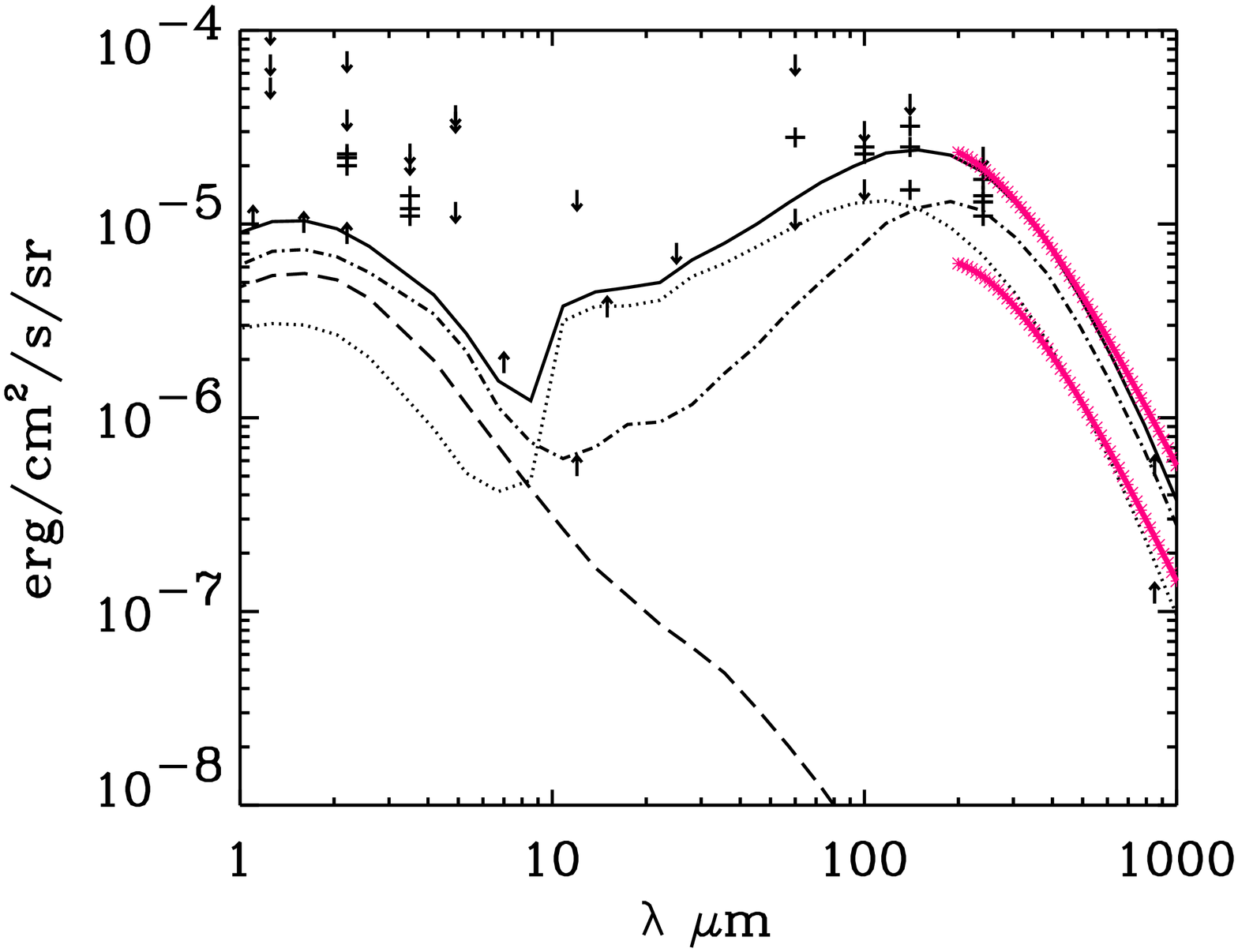}
\caption{Contributions to the $1-1000\mu$m background from
spheroids (dot-dash;  passively evolving spheroids are represented
by long dashes) and late-type galaxies (dots). The leap of the
dotted line at $10\,\mu$m corresponds to the transition between
the two evolutionary regimes for late type galaxies (see text).
The thick solid line shows the sum of all contributions. In the
upper panel we have used the SED of NGC6090 for starburst
galaxies, while in the lower one we have used that of M82. The SED
parameters for forming spheroids are those described in Sect.
\ref{sect:param}, i.e.\ $\tau_{MC, 1\mu m} = 45$, and $t_e=0.05$
Gyr for low mass spheroids, $t_e$ equal to the starburst duration
for the massive ones. Data are from Hauser \& Dwek (2001).}
\label{bgir}
\end{figure}

\subsubsection{Near-IR region}
\label{sec:nir}

The rest-frame NIR SED is generally dominated by old and slowly
evolving low mass stars. To estimate the contribution of all
late--type galaxies (including starburst galaxies) to the counts,
the redshift distributions and the background intensity in the
range $1$--$10\,\mu$m, we have adopted the R-band local LF for
intermediate and late--type galaxies by de~Lapparent et al.
(2003), and the SED of NGC6946. The LF is assumed not to evolve
and, again, a redshift cut-off $z_{\rm cutoff} =1.5$ is
introduced.

\begin{figure}[tbp]
\centering
\includegraphics[width=9truecm]{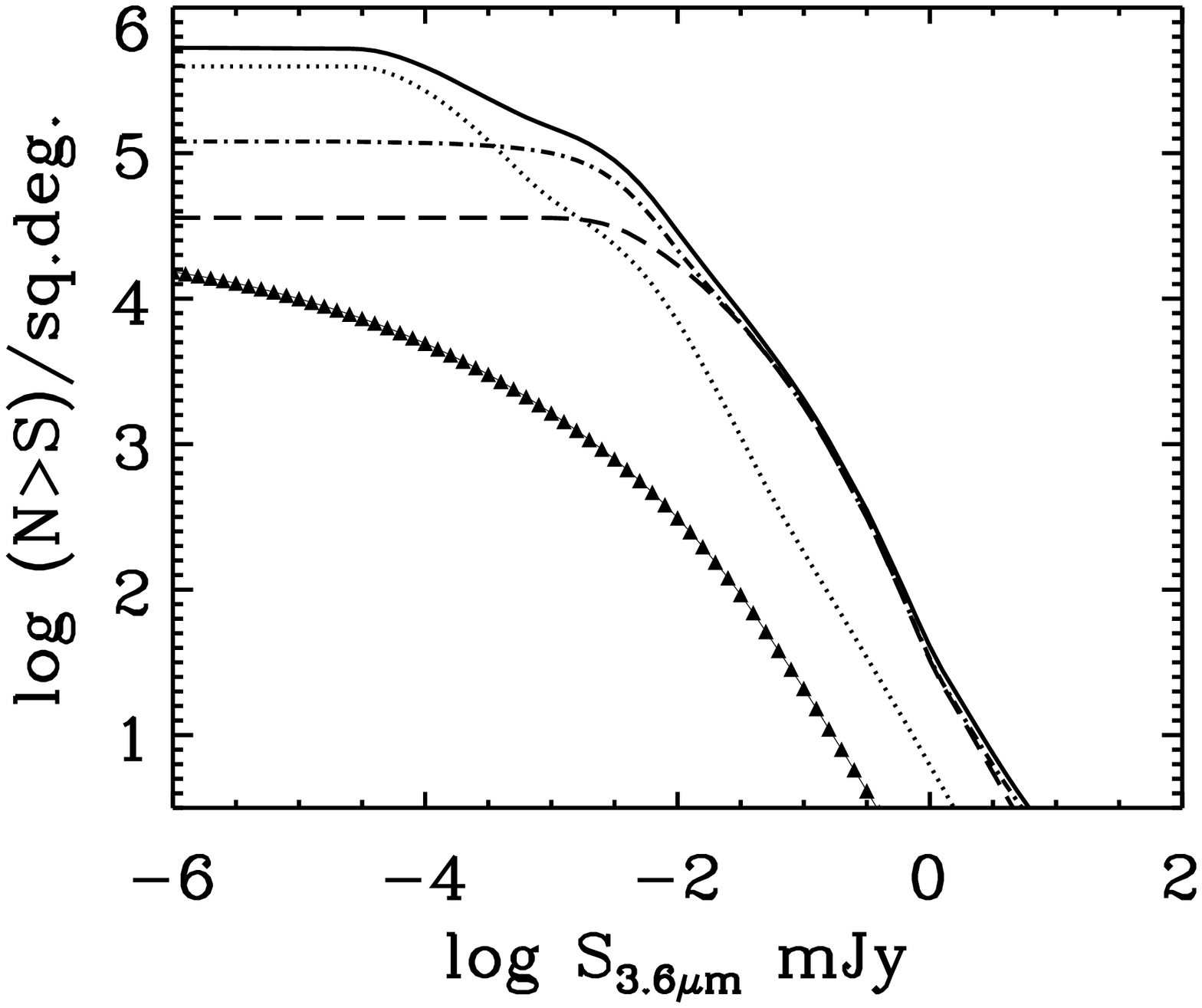}
\caption{Predicted $3.6 \mu$m IRAC source counts. Dot-dashed line:
spheroids (the contribution of passively evolving spheroids is
shown by the long-dashed line); dotted line: late type galaxies;
filled triangles: AGNs. The total counts are shown by the solid
line.} \label{c3.6std}
\end{figure}

\begin{figure}[tbp]
\centering
\includegraphics[width=9truecm]{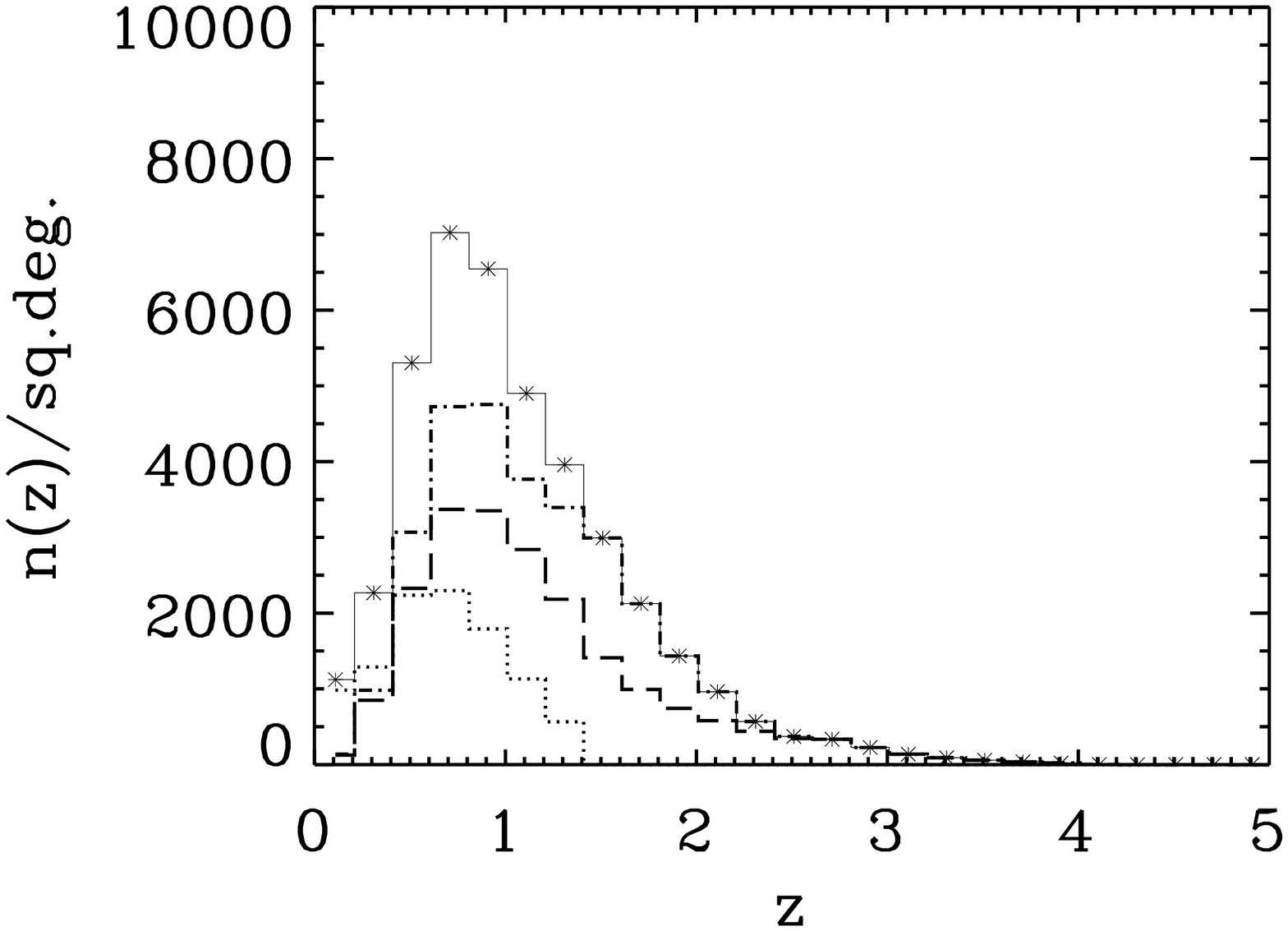}
\includegraphics[width=9truecm]{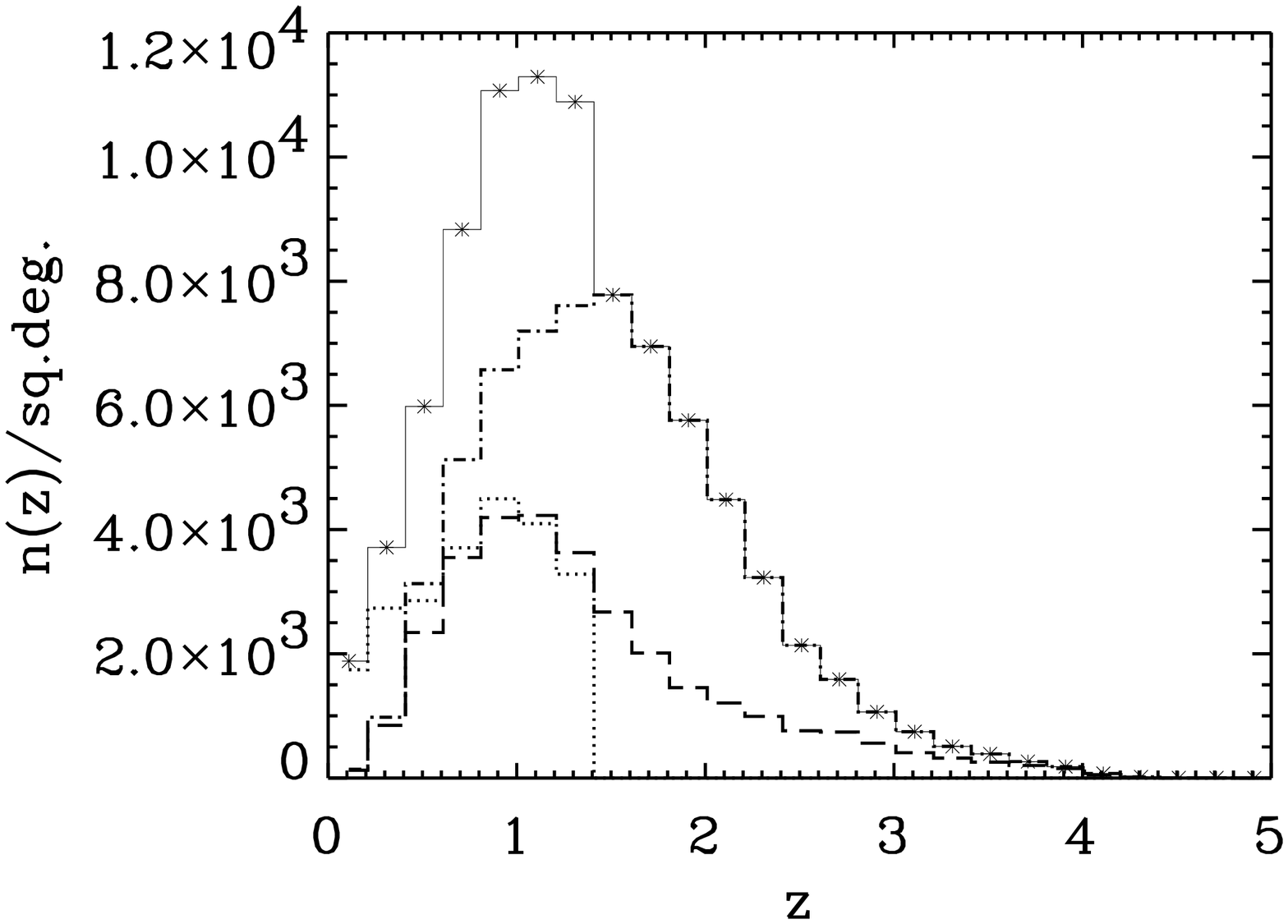}
\caption{Redshift distributions of sources with $S_{3.6\mu{\rm m}}
\ge 7.3\,\mu$Jy (upper panel) and $\ge 3 \mu\,$Jy, the SWIRE and
GOODS limits respectively. Dot-dashed line: spheroids (the
contribution of passively evolving spheroids is shown by the
long-dashed line); dotted line: late type galaxies. The solid
histogram with asterisks is the sum of all contributions. We
expect to find $\simeq 4\times 10^4$ sources per square degree
brighter than $7.3\,\mu$Jy and $\simeq 9\times 10^4$ sources per
square degree brighter than $3\,\mu$Jy. At both flux limits
$\simeq 73\%$ of the detected sources are spheroids. However, the
proportion of those passively evolving decreases from $\simeq
70$\% at $7.3\,\mu$Jy to $\simeq 50\%$ at $3\,\mu$Jy.}
\label{nz3.6std}
\end{figure}

\subsection{Active Galactic Nuclei}
\label{sec:agn}

We have also included the contribution of the nuclear emission by
AGNS (type 1 and 2) to the source counts, computed as in Silva,
Maiolino, \& Granato (2004): the cosmological evolution of AGNS,
derived from X-ray surveys (Ueda et al.\ 2003), has been coupled
to an accurate definition of the average X-ray to sub-mm SEDs of
AGNs, as a function of the absorption column densities N$_H$.

\section{Determination of the GRASIL parameters for spheroidal galaxies}
\label{sect:param} As already mentioned, we need to determine two
GRASIL parameters: the dust optical depth of molecular clouds
($\tau_{\rm MC}$), and the escape time scale ($t_e$) of young
stars from their parent molecular clouds. The former quantity
determines, in particular, the contribution (if any) of
star-forming spheroidal galaxies to the bump in the $15\,\mu$m
counts below a few mJy. Although the observed counts are affected
by a considerable uncertainty, as shown by the spread of results
from different surveys, and the modelling is complicated by the
presence of PAH bands, they can be interpreted in terms of
evolving SBs, with a minor contribution to SPs, particularly at
the brighter flux density levels. Thus, the observed $15\,\mu$m
flux from star-forming spheroids must be substantially suppressed,
implying $\tau_{\rm MC}\gsim 40$ at $1 \mu$m (see
Fig.~\ref{c15std}). On the other hand, a too large value of
$\tau_{\rm MC}$ would entail a too large contribution from such
sources to the $170\,\mu$m counts (see Fig.~\ref{c175std}). In the
following we set $\tau_{\rm MC}\simeq 45$ at $1 \mu$m, a value
yielding a small, but non-negligible, contribution to the sub-mJy
$15\,\mu$m counts and a significant contribution to the
$170\,\mu$m counts below a few hundred mJy, somewhat improving the
quality of the fit in both cases (Figs.~\ref{c15std} and
\ref{c175std}).

The redshift distribution for $S_{15\mu{\rm m}}\geq 0.1\,$mJy
(Aussel et al. 1999; Elbaz et al. 2002; Franceschini et al. 2003)
is also satisfactorily reproduced (Fig.~\ref{nz15std01}) and that
for $S_{15\mu{\rm m}}\geq 1\,$mJy is also consistent with the
limited information currently available (Fig.~\ref{nz15std1};
Pozzi et al. 2003; Rowan-Robinson et al. 2003).

\begin{figure}[tbp]
\centering
\includegraphics[width=9truecm]{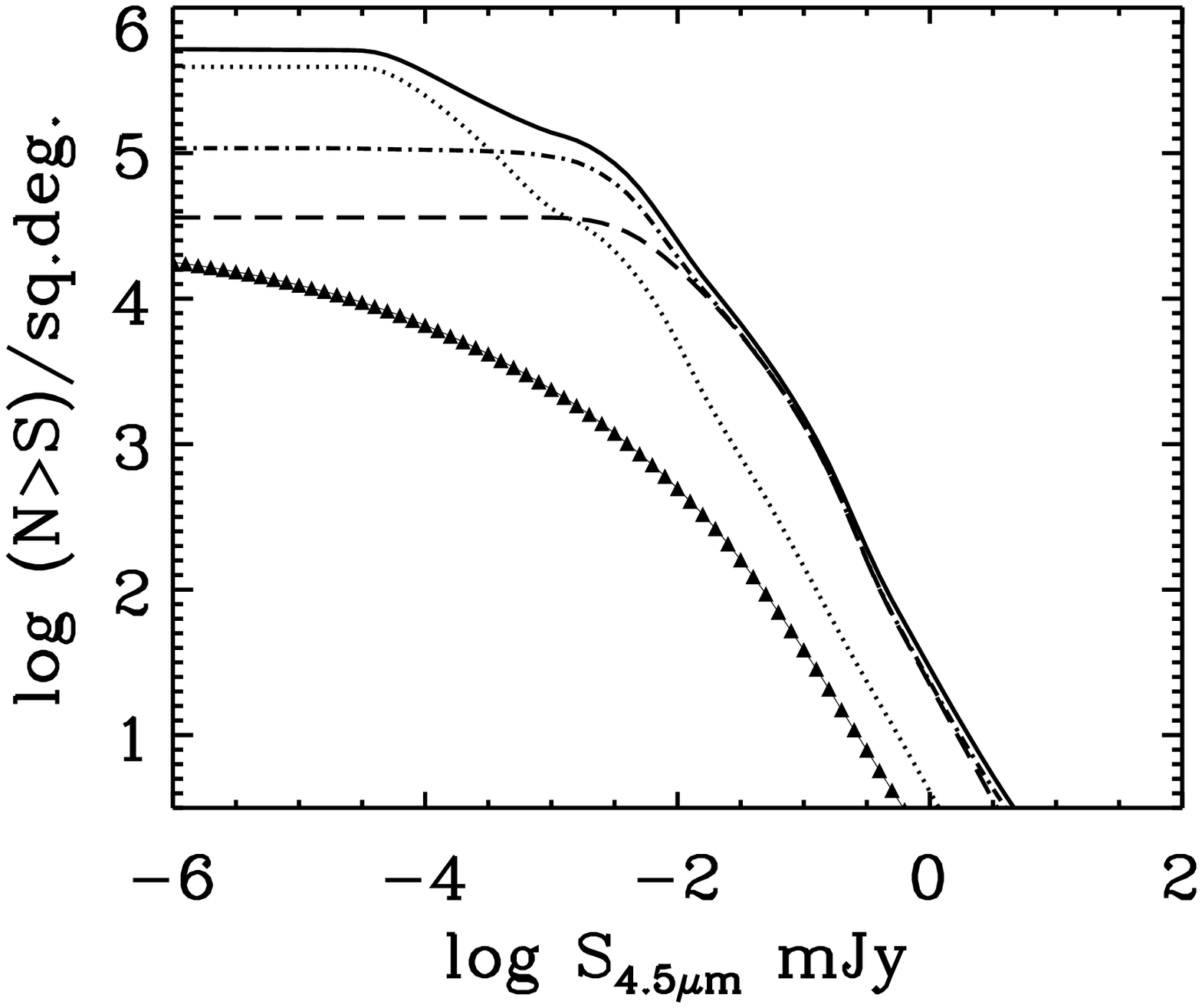}
\caption{Predicted $4.5 \mu$m IRAC source counts. The lines have
the same meaning as in Fig.~\protect{\ref{c3.6std}}.}
\label{c4.5std}
\end{figure}

\begin{figure}[tbp]
\centering
\includegraphics[width=9truecm]{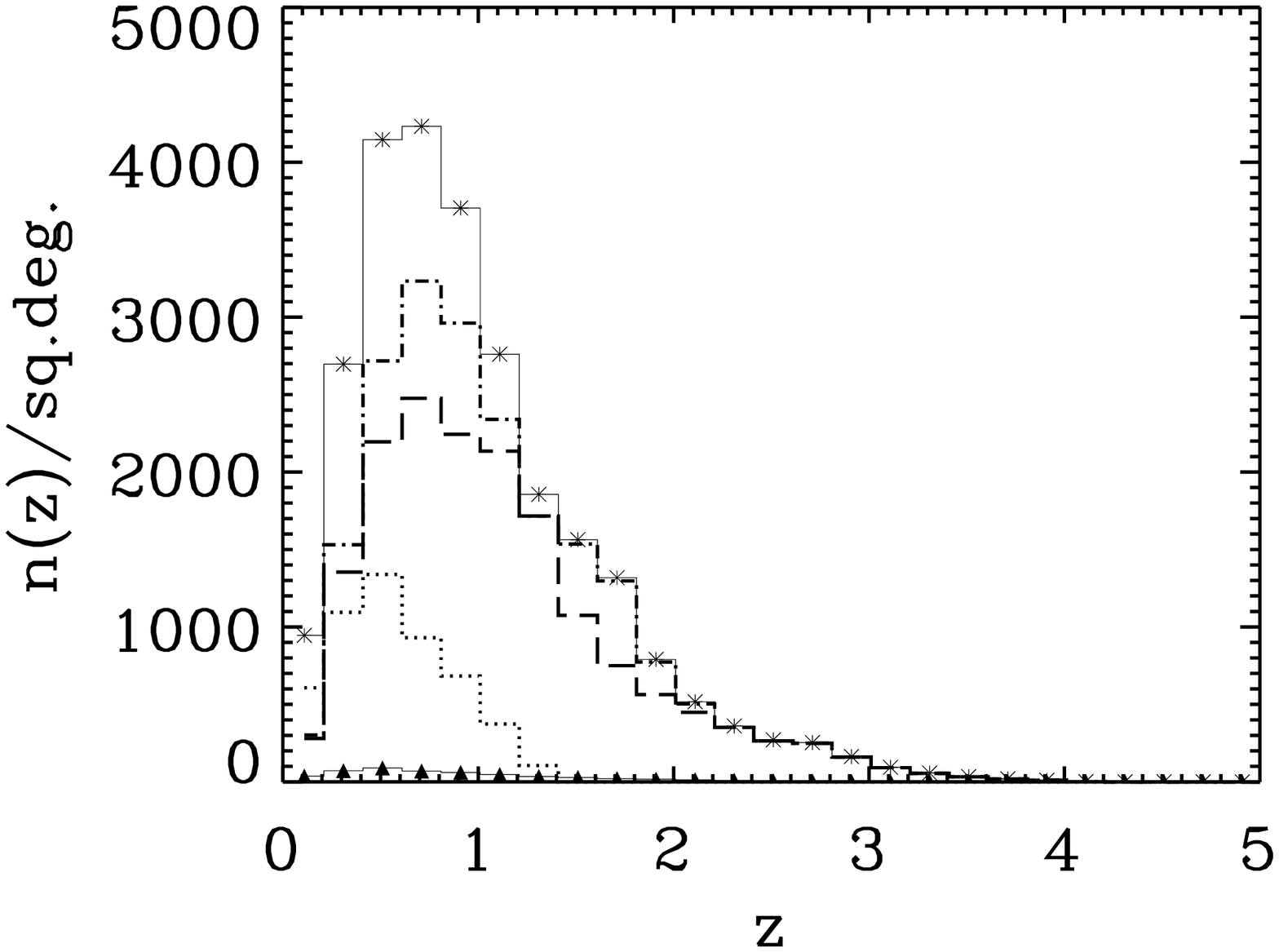}
\includegraphics[width=9truecm]{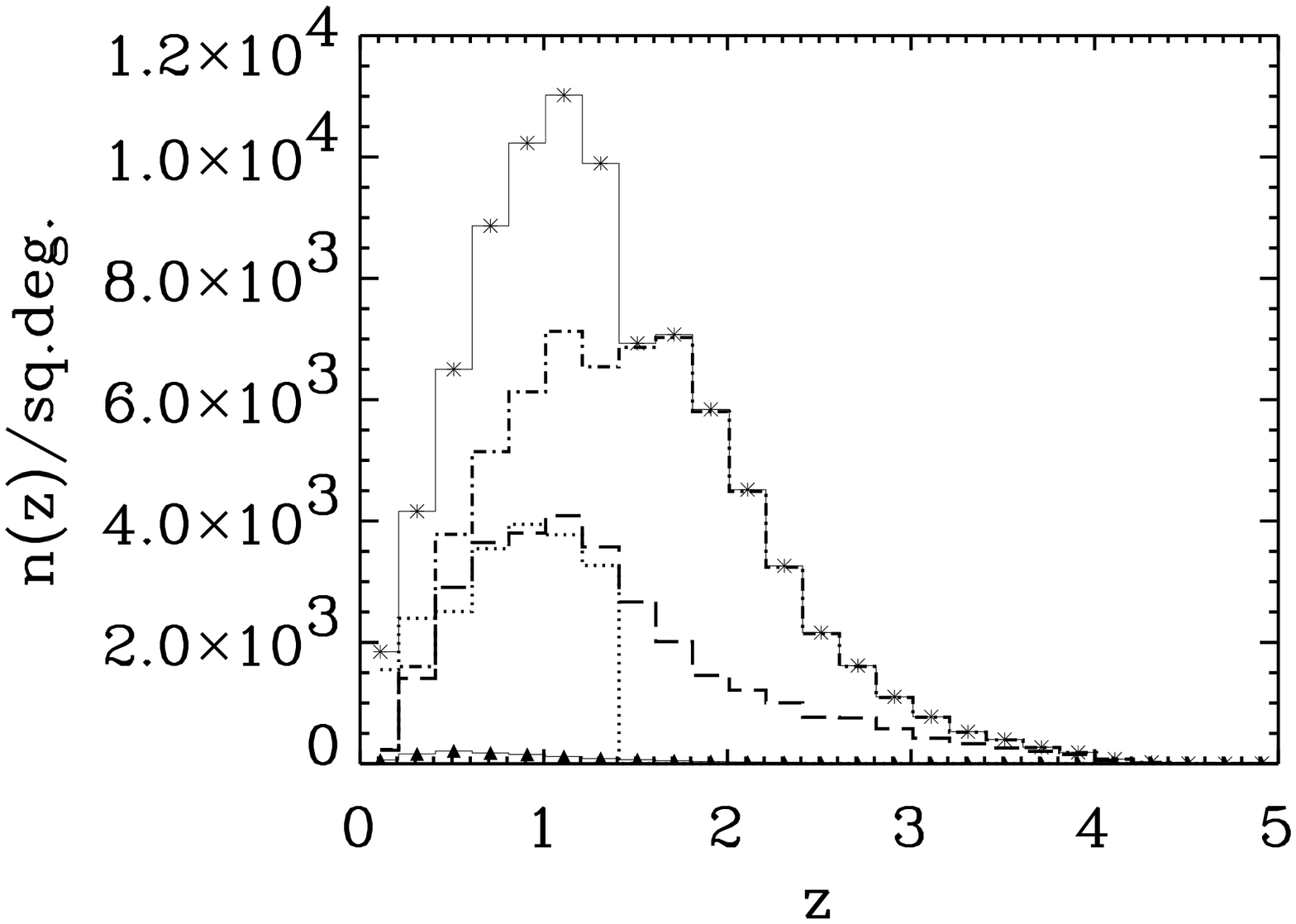}
\caption{Redshift distributions of sources with $S_{4.5\mu{\rm m}}
\ge 9.7\,\mu$Jy (upper panel) and $\ge 3 \mu\,$Jy, the SWIRE and
GOODS limits respectively. Dot-dashed line: spheroids (the
contribution of passively evolving spheroids is shown by the
long-dashed line); dotted line: late type galaxies; filled
triangles: AGNs. The solid histogram with asterisks is the sum of
all contributions. We expect to find $\simeq 2.5\times 10^4$
sources per square degree brighter than $9.7\,\mu$Jy, $\simeq
80$\% of which are spheroids (80\% of which passively evolving)
and $\simeq 8.7\times 10^4$ sources per square degree brighter
than $3\,\mu$Jy, $\simeq 74$\% of which are spheroids, almost
equally subdivided among the star-forming and passive evolution
phases. } \label{nz4.5std}
\end{figure}

Sajina et al. (2003) derived photometric redshifts for a
representative sample of 30 sources detected in the FIRBACK
$170\mu{\rm m}$ survey to a limiting flux density of
$S_{170\mu{\rm m}} \ge 135\,$mJy. The estimated redshift
distribution turns out to be bimodal, most sources being at $z\ll
1$, but with a clump of 5 sources at $z\sim 0.4$--1. The redshift
distribution predicted by our model has roughly the correct ratio
of low to high redshift sources, but the mean redshift of the
latter is somewhat higher than estimated by Sajina et al. (2003).
The redshift measurements, 90\% complete, of brighter
($S_{170\mu{\rm m}} \ge 200\,$mJy; $4\sigma$ detections) FIRBACK
sources by Patris et al. (2003) have shown that the vast majority
of them are moderate luminosity dusty galaxies at $z<0.3$, with a
rather cold SED. In Fig.~\ref{c175std} the redshift distribution
yielded by our model for S$\geq 223$mJy is compared with data by
Rowan-Robinson et al.\ (2003). As discussed below, the predicted
peak at $z>1$ due to star-forming galaxies, not seen in the data,
may be due to our rough treatment of their optically thick phase.
A slight decrease of the 170 to $850\,\mu$m flux density ratio
compared to that yielded by GRASIL or a slightly lower value of
$\tau_{MC}$ is enough to get rid of the $z>1$ peak.

\begin{figure}[tbp]
\centering
\includegraphics[width=9truecm]{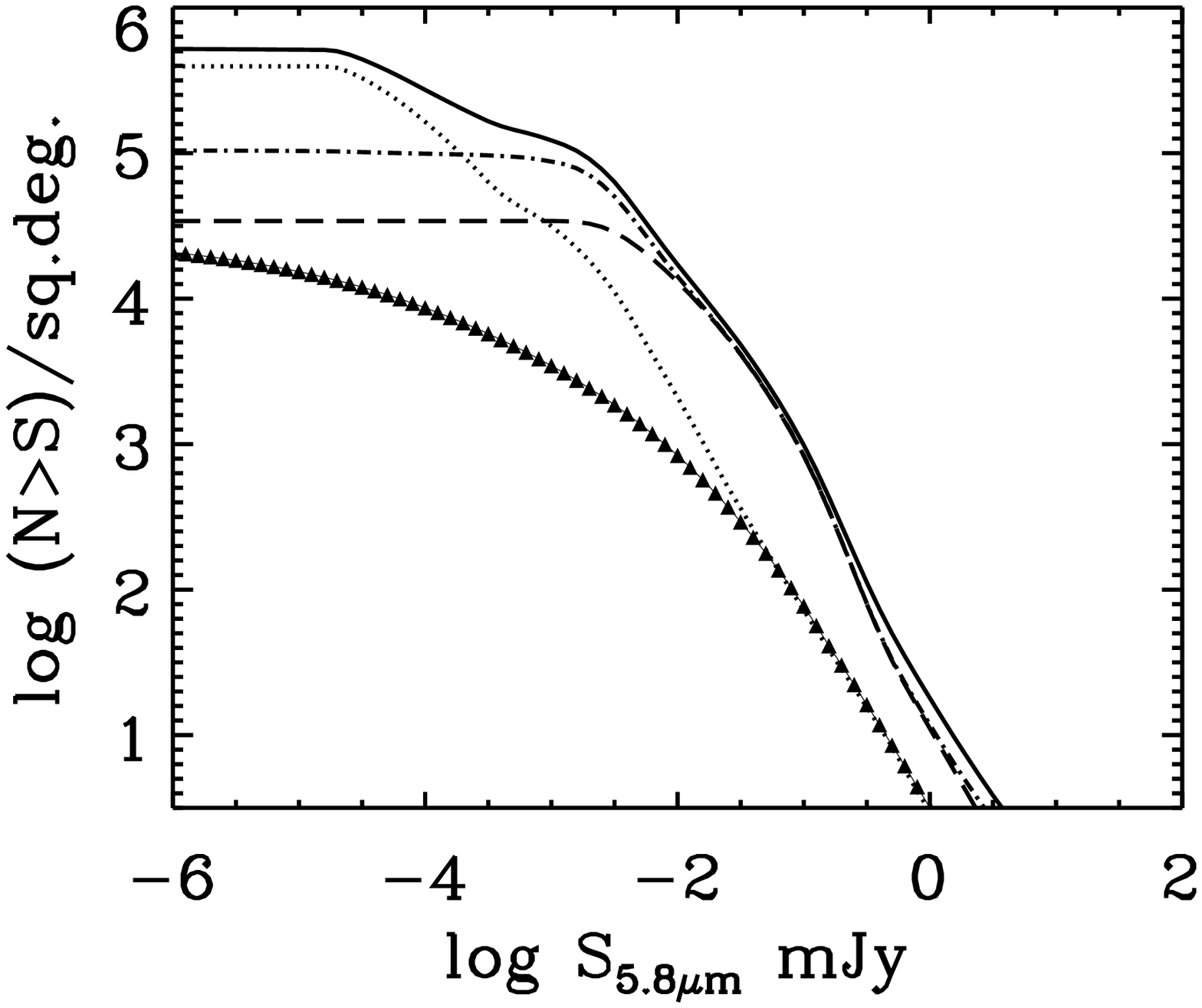}
\caption{Predicted $5.8 \mu$m IRAC source counts. The lines have
the same meaning as in Fig.~\protect{\ref{c3.6std}}.}
\label{c5.8std}
\end{figure}

\begin{figure}[tbp]
\centering
\includegraphics[width=9truecm]{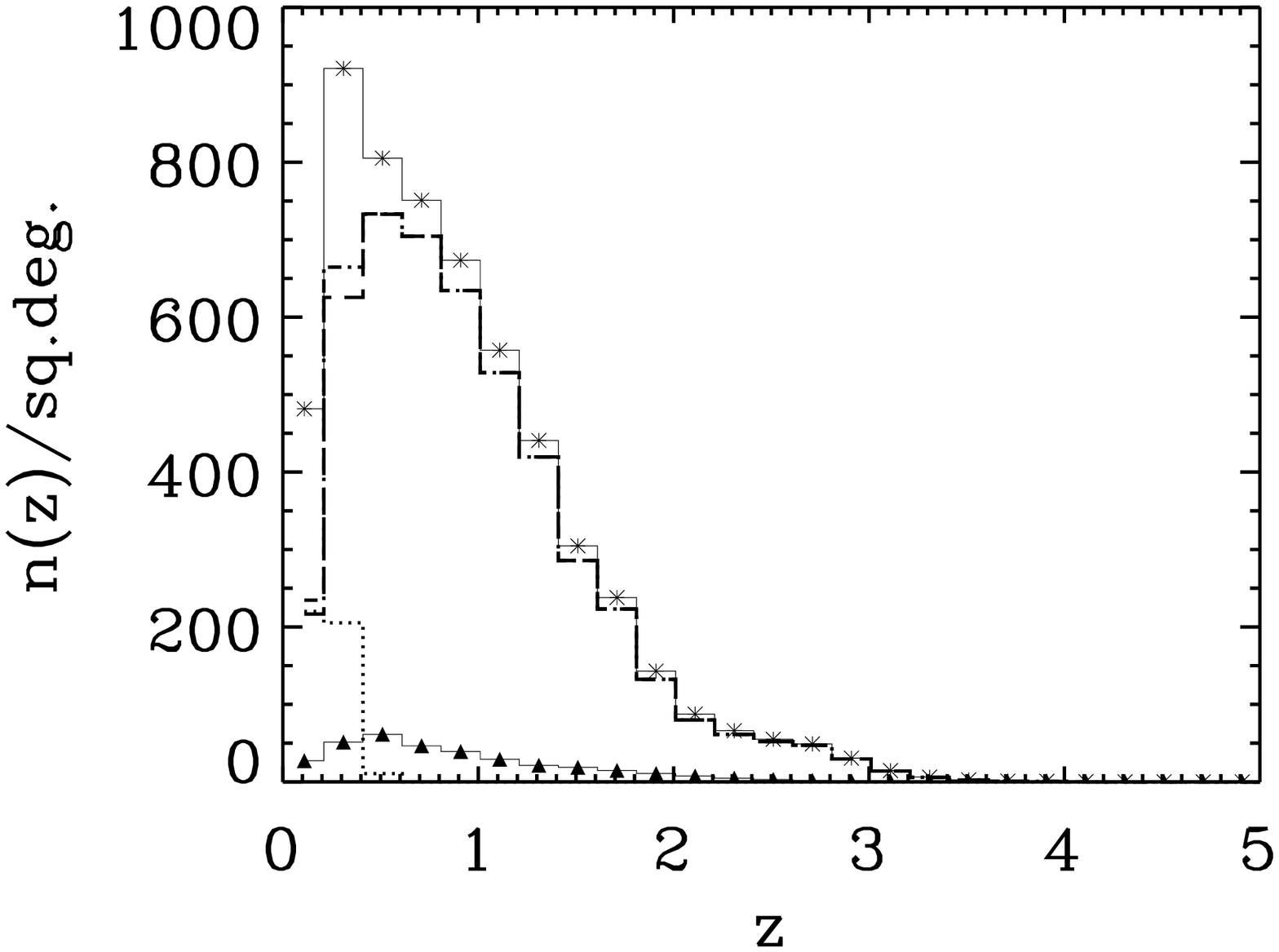}
\includegraphics[width=9truecm]{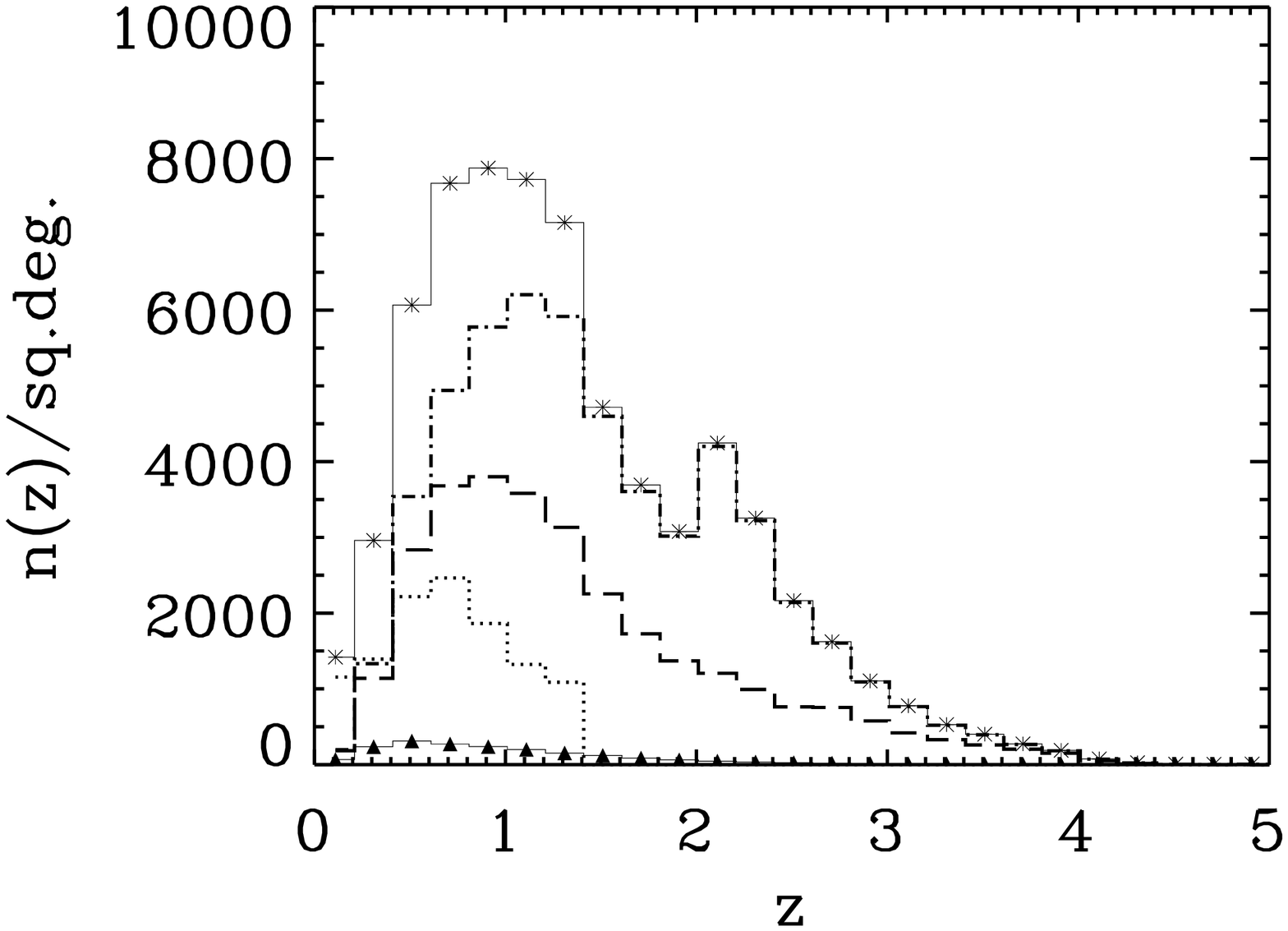}
\caption{Predicted redshift distributions of sources with
$S_{5.8\mu{\rm m}} \ge 27.5\,\mu$Jy (upper panel) and $\ge 3
\mu\,$Jy, the SWIRE and GOODS limits respectively. The lines have
the same meaning as in Fig.~\protect{\ref{nz4.5std}}. We expect
$\simeq 5.6\times 10^3$ sources per square degree brighter than
$27.5\,\mu$Jy, $\simeq 86$\% of which are spheroids (all passively
evolving) and $\simeq 6.5\times 10^4$ sources per square degree
brighter than $3\,\mu$Jy, $\simeq 80$\% of which are spheroids,
$\sim$ half of which are passively evolving.} \label{nz5.8std}
\end{figure}

The parameter $t_e$ is meant (cf. Sect.~\ref{sect:SED}) to control
the fraction of star-light, produced in star-forming spheroidal
galaxies, which is only reprocessed by cirrus dust, and is
therefore observable at optical wavelengths.

Since GRASIL assumes that the gas (and the dust) is in molecular
clouds (MCs) and in the cirrus, the mean optical depth to a star
outside its parent MC, $\tau_{\rm outside}$, is equal to
$\tau_{\rm cirrus}+\bar{N} \times 2\tau_{\rm MC}$, $\bar{N}$ being
the mean number of MCs along its line of sight, which is roughly
proportional to the size of the galaxy (or, for given
virialization redshift, to $M_{\rm vir}^{1/3}$). Thus, for small
enough galaxies the optical depth drops from $\tau_{\rm MC}$ to
$\tau_{\rm cirrus}$, which is $\ll 1$ at IR wavelengths, when the
stars move out of their parent MCs, after a time $t_e$. This same
condition is met in most cases when dealing with relatively normal
galaxies, even moderate starbursts (e.g. when reproducing the SED
of nearby galaxies), where the covering factor of MCs is $<<1$. On
the contrary, for large galaxies and for the really extreme
conditions met in forming spheroids according to our scenario,
$\tau_{\rm outside}$ can be substantially larger than $\tau_{\rm
MC}$. The present release of GRASIL, however, does not allow us to
deal accurately with the radiation transfer through the
distribution of MCs. On the other hand, such detailed treatment is
not crucial for our present purposes since whenever $\tau_{\rm
outside}\ge \tau_{\rm MC}$ essentially all the starlight is
reprocessed by dust. We have crudely approximated $\tau_{\rm
outside}$ as a step function: $\tau_{\rm outside}=\tau_{\rm
cirrus}$ for $M_{\rm vir} \le M_{\rm vir,crit}$ and $\tau_{\rm
outside}=\tau_{\rm cirrus}+ \tau_{\rm MC}$ for larger virial
masses. In order to take into account the dust heating by the
absorbed starlight, in practice we have kept the stars within
their MCs for the full duration of the starburst, for galaxies
with $M_{\rm vir} \ge M_{\rm vir,crit}$.

This treatment may overestimate the duration of the optically
thick phase, as well as the optical thickness, of the relatively
less massive galaxies (but still with $M_{\rm vir} \ge M_{\rm
vir,crit}$) which are responsible for the peak at $z>1$ in the
redshift distribution at $170\,\mu$m, predicted by the model but
not seen (Fig.~\ref{c175std}). We thus expect that, with a more
realistic treatment, we will get rid of such peak.

We find that, for the value of $\tau_{\rm MC}$ derived as
explained above, we can account for the K-band counts assuming for
$t_e$ a value between $\simeq 0.05$ and $0.4$ Gyr and setting
$M_{\rm vir,crit} \simeq 10^{12}\,M_\odot$. The results in
Fig.~\ref{ckstd} refer to $t_e=0.05\,$Gyr and illustrate the
relative contributions of star-forming and passively evolving
spheroids, as well as of starburst and spiral galaxies.

\begin{figure}[tbp]
\centering
\includegraphics[width=9truecm]{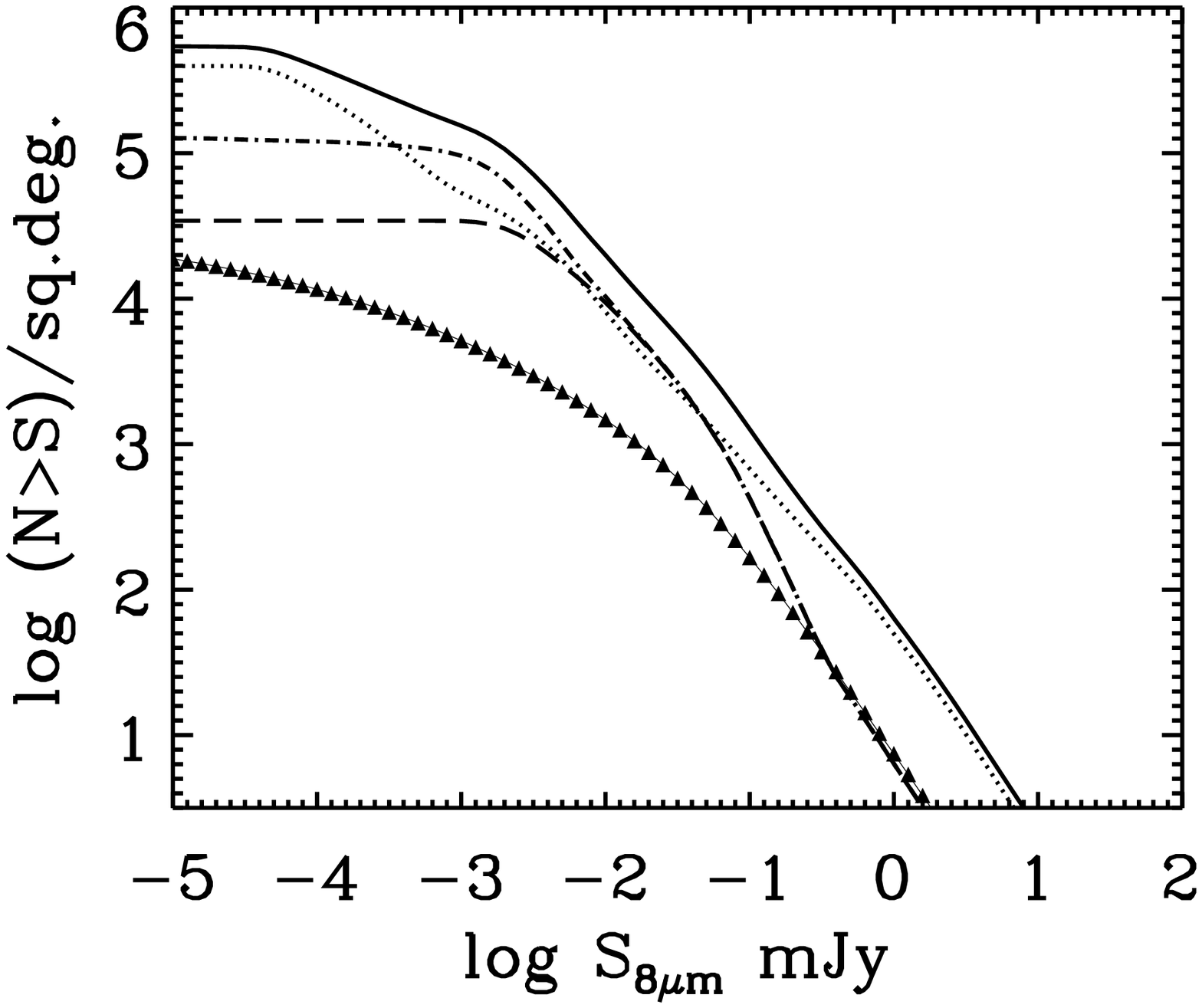}
\caption{Predicted $8 \mu$m IRAC source counts. The lines have the
same meaning as in Fig.~\protect{\ref{c3.6std}}.} \label{c8std}
\end{figure}

\begin{figure}[tbp]
\centering
\includegraphics[width=9truecm]{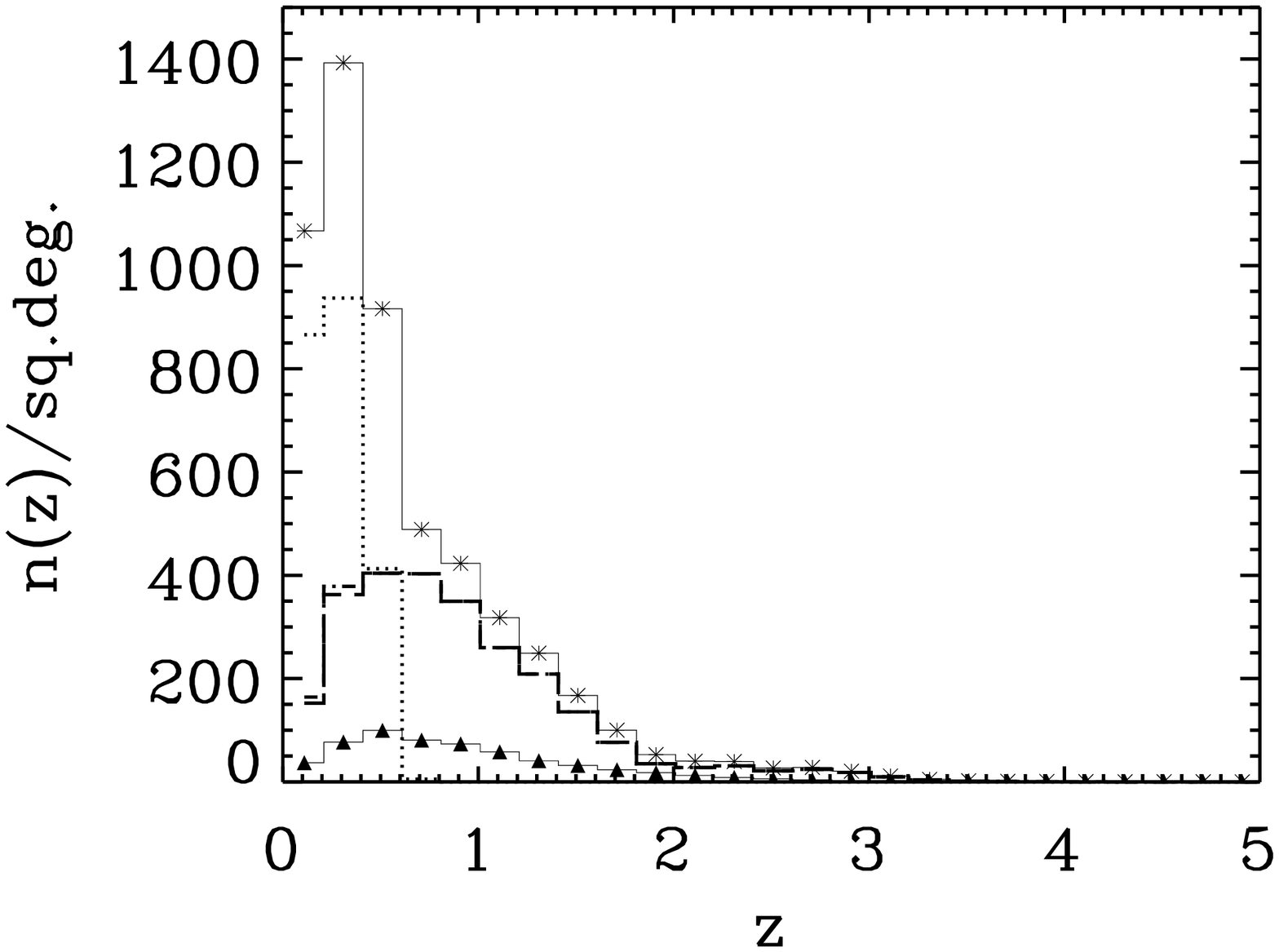}
\includegraphics[width=9truecm]{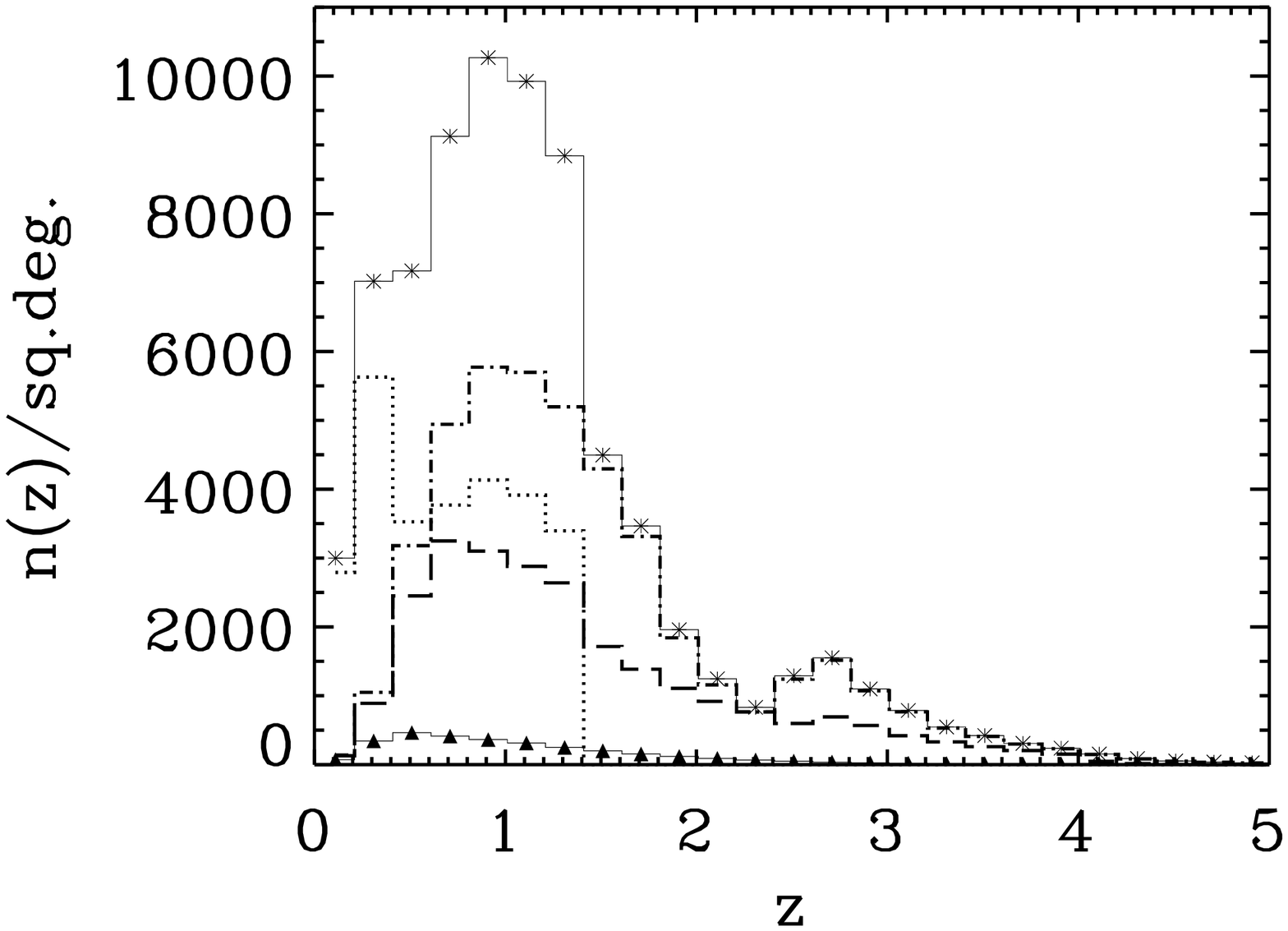}
\caption{Redshift distributions of sources with $S_{8\mu{\rm m}}
\ge 32.5\,\mu$Jy (upper panel) and $\ge 3 \mu\,$Jy, the SWIRE and
GOODS limits respectively. The lines have the same meaning as in
Fig.~\protect{\ref{nz4.5std}}. We expect $\simeq 5\times 10^3$
sources per square degree brighter than $32.5\,\mu$Jy, $\simeq
50$\% of which are spheroids (essentially all passively evolving)
and $\simeq 7.5\times 10^4$ sources per square degree brighter
than $3\,\mu$Jy, $\simeq 60$\% of which are spheroids, the
star-forming and passive evolution phases being almost equally
represented.} \label{c8std}
\end{figure}

\section{Comparison with the data}

\subsection{Counts and redshift distributions}
\label{sect:coured}

As discussed by Granato et al. (2001, 2004), in the present
framework the SCUBA-selected galaxies are interpreted as being
mostly star-forming spheroids ($M_{\rm vir} \geq 10^{11.6}\,
M_\odot$). This is borne out by Figs.~\ref{c850std} and
\ref{nz850std}. The values of GRASIL parameters derived above
yield a good fit to the $850\,\mu$m SCUBA counts, while
significantly higher/lower values of the optical depth of
star-forming spheroidal galaxies would lead to a substantial
over/under-prediction of the counts. The contributions of other
populations, such as normal late-type galaxies, radio sources, and
starburst galaxies are only important at flux density levels much
brighter or much fainter than those covered by SCUBA surveys.

The predicted redshift distribution of sources brighter than
$1\,$mJy at $850\,\mu$m (Fig.~\ref{nz850std}) has a narrow peak at
$z\sim 1$ mostly due to starburst galaxies, and a broad peak at
$z\sim 2$ with an extended tail towards higher redshifts, due to
star-forming spheroidal galaxies.

In Chapman et al. (2003), the spectroscopic redshifts of 10
representative SCUBA galaxies with $S_{850\mu{\rm m}}> 5\,$mJy
have been reported. The median redshift of their sample is $2.4$,
with a quartile range of $z=1.9-2.8$. The median and quartile
range for our model are $2.48$, $1.78-3.68$ and $2.33$,
$1.78-3.15$ respectively for the total (spheroids and late type
galaxies) and spheroids alone, in remarkably good agreement.

\begin{figure}[tbp]
\centering
\includegraphics[width=9truecm]{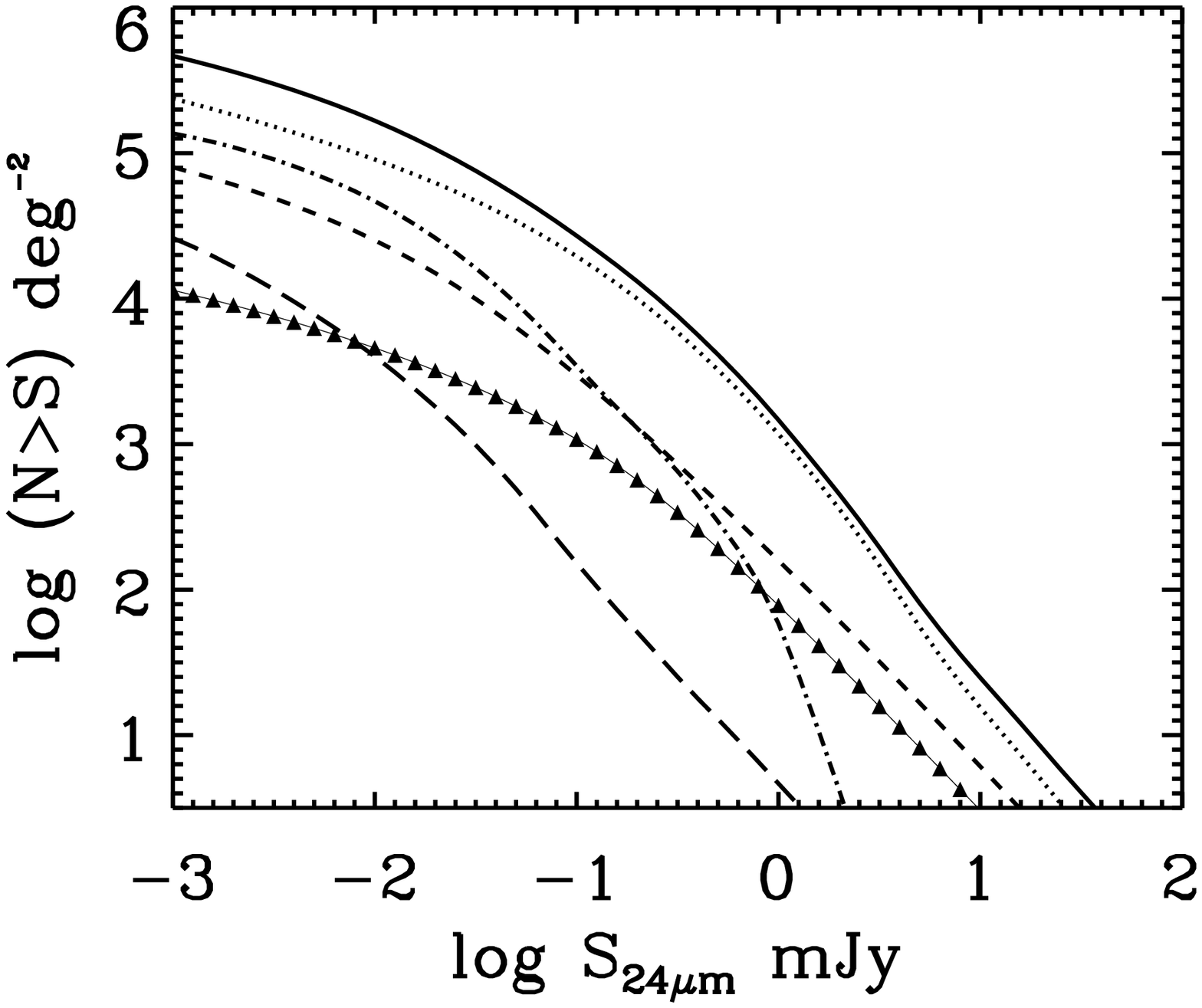}
\caption{Predicted contributions to the $24 \mu$m Spitzer/MIPS
source counts. Dot-dashed line: spheroids (all); long-dashed line:
passively evolving spheroids; dotted line: starbursts; dashed
line: spirals; filled triangles: AGNs.} \label{c24std}
\end{figure}

\begin{figure}[tbp]
\centering
\includegraphics[width=9truecm]{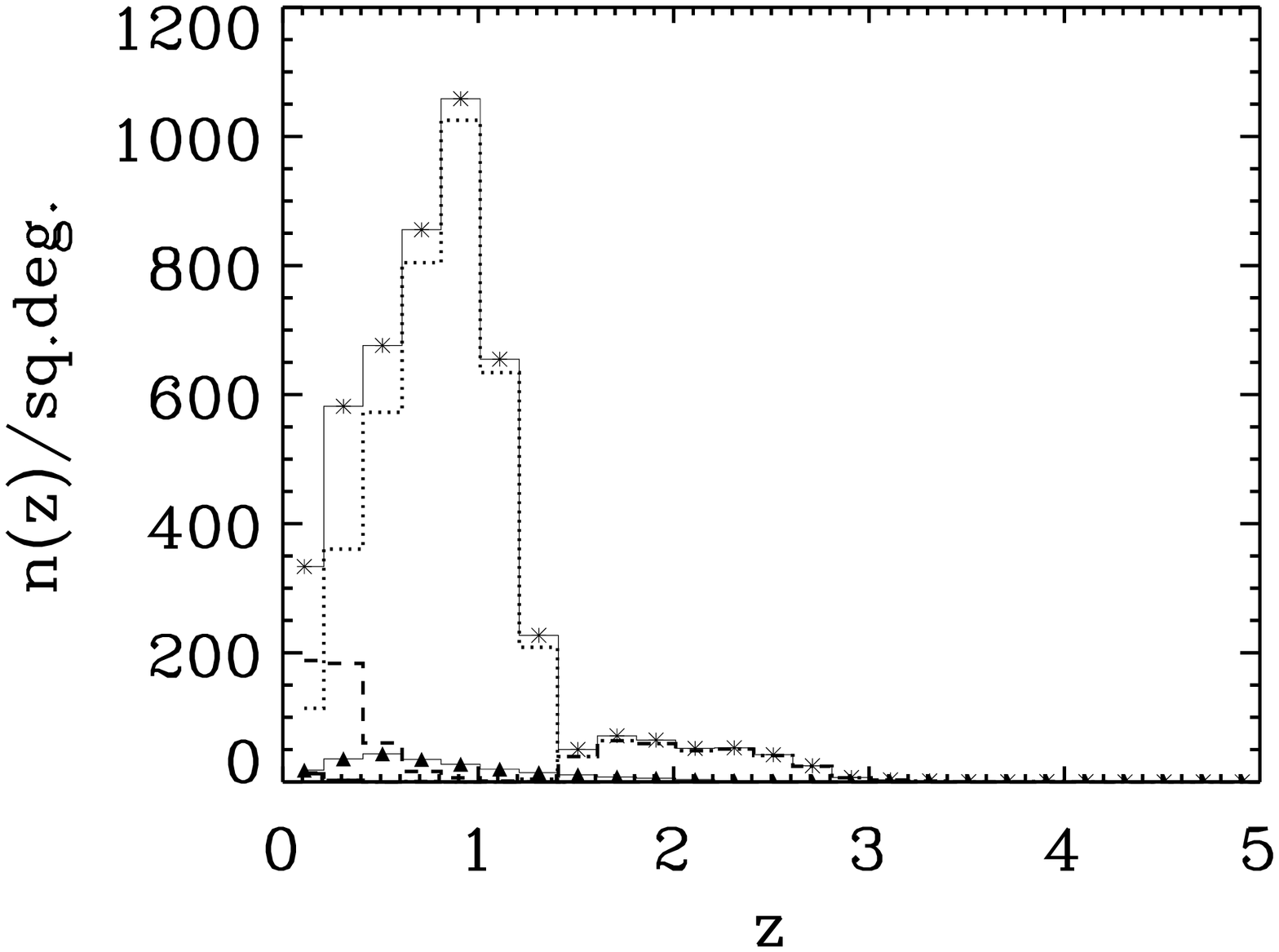}
\includegraphics[width=9truecm]{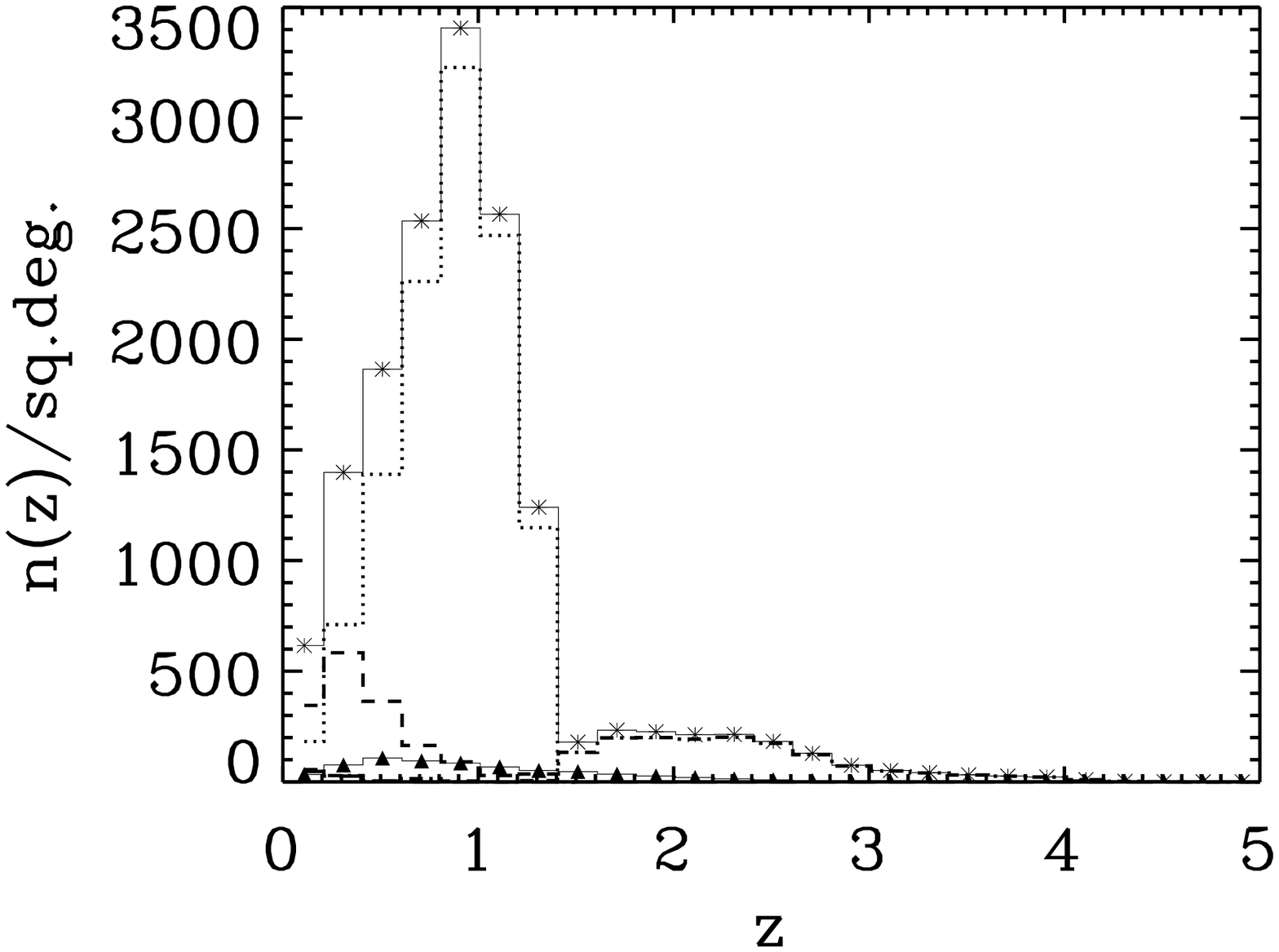}
\caption{Predicted redshift distributions for $S_{24 \mu{\rm m}}
\ge 0.45\,$mJy (the SWIRE flux density limit; upper panel) and
$S_{24 \mu{\rm m}} \ge 0.17\,$mJy (the estimated $5\sigma$
confusion limit). At these limits we will have, respectively, a
total of $\sim 4.7\times 10^3$ and $\sim 1.5\times 10^4$ sources
per square degree, $8 - 10$\% of which will be star forming
spheroids. Dot-dashed line: spheroids; dotted line: starbursts;
dashed line: spirals; filled triangles: AGNs.} \label{nz24std}
\end{figure}

While the sub-mm counts are informative on the star-formation
phase of the evolution of spheroidal galaxies, the K-band counts
in the range $14 \lsim {\rm K} \lsim 19$ (Fig.~\ref{ckstd}) are
dominated by the passive evolution phase. Fainter than $K \sim 19$
also star-forming spheroids begin to show-up and become
increasingly important at fainter magnitudes. The brighter (in
terms of apparent magnitude) such objects are those of lower
bolometric luminosity, which, in the GDS04 model, have a prolonged
star-formation phase extending to $z \lsim 1$ and whose stars are
kept within a high optical depth environment for a time $t_e =
0.05\,$Gyr (see Sect.~3). This is illustrated by
Fig.~\ref{nzk20std}, showing that, for $K\le 20$, the contribution
from star-forming spheroids peaks at $z$ somewhat smaller than 1.
Passively evolving spheroids essentially disappear at $K\ge 23$.
The observed high-$z$ tails of the distributions for $K\le 23$
(Fig.~\ref{nzk23std}) and for $K\le 24$ (Fig.~\ref{nzk24std}) are
thus mostly due to star-forming spheroids. At these faint
magnitudes, the $K$-band counts are dominated by late-type
galaxies, which account for the big peak in the range $0.2 \le z
\le 0.9$ (Figs.~\ref{nzk23std} and \ref{nzk24std}), as well as for
the low-$z$ shoulder at $K \le 20$ (Fig.~\ref{nzk20std}).

Note that the contribution of late type galaxies at faint flux
limits may be underestimated. In fact the Subaru Deep Field source
counts by Totani et al.\ (2001) (the triangles in
Fig.~\ref{ckstd}) are raw data. Maihara et al.\ (2001) provide the
counts corrected for incompleteness. However, as stated by the
authors, the correction is model-dependent. A mild density
evolution ($\sim (1+z)^{1.3}$) of dwarf galaxies would be
sufficient to fit the corrected faint $K$-band counts. But the
redshift distributions by Kashikawa et al.\ (2003)
(Figs.~\ref{nzk23std} and \ref{nzk24std}) refer to actually
observed sources, without any correction for incompleteness. We
have thus chosen to adopt the simple assumption of no evolution
for dwarves, which yields faint $K$-band counts consistent with
the uncorrected data.

Passively evolving spheroids also dominate the  $6.7 \mu$m ISOCAM
counts (Altieri et al. 1999; Flores et al. 1999; Oliver et al.
2002; Sato et al. 2003; Metcalfe et al. 2003; Fig.~\ref{c6.7std})
below $\sim 300\,\mu$Jy, while at brighter flux densities late
type galaxies take over. The model redshift distribution (lower
panel of Fig.~\ref{c6.7std}) is compatible with the (very limited)
redshift information (Flores et al. 1999, Sato et al.\ 2004).

\begin{figure}[tbp]
\centering
\includegraphics[width=9truecm]{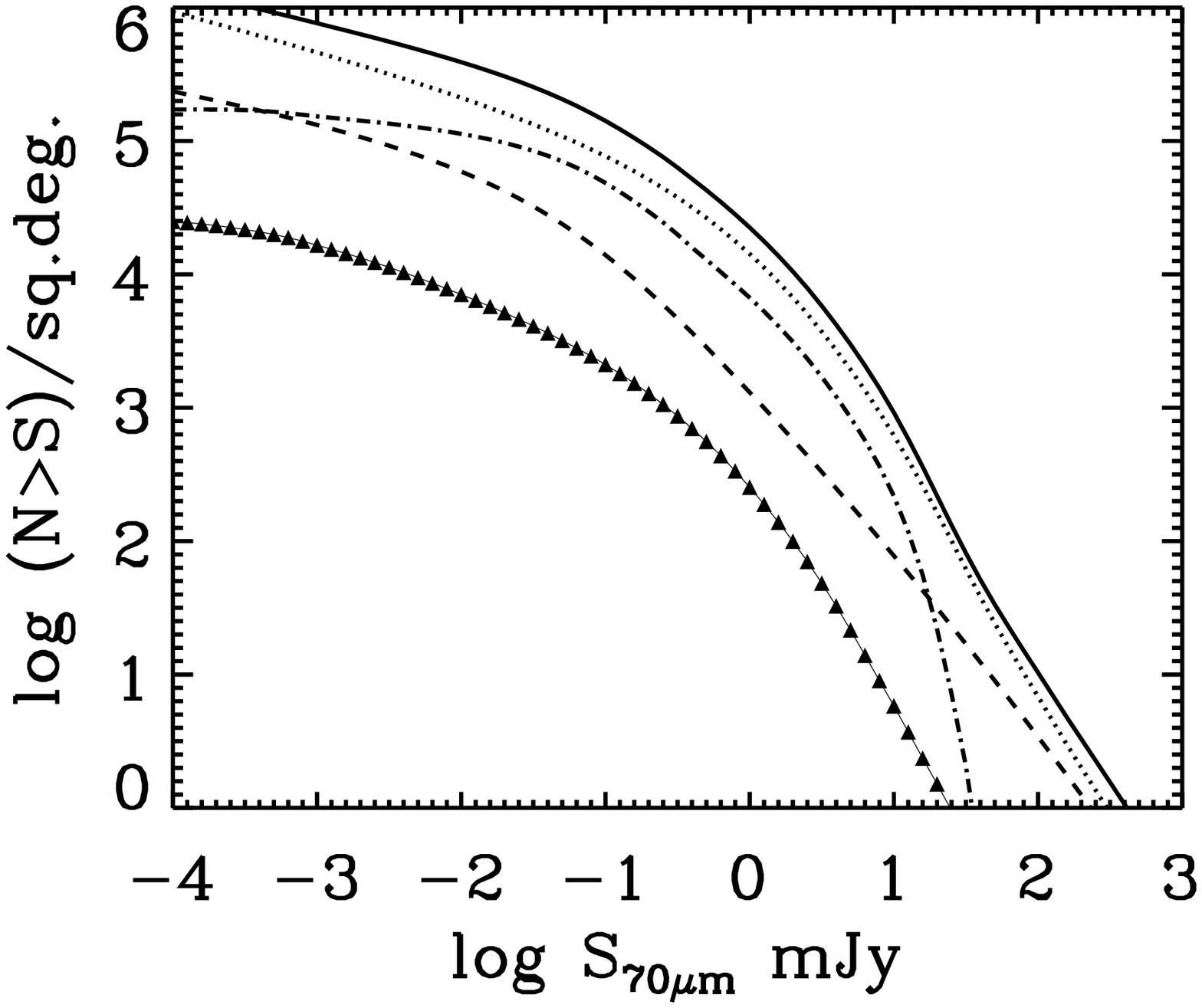}
\includegraphics[width=9truecm]{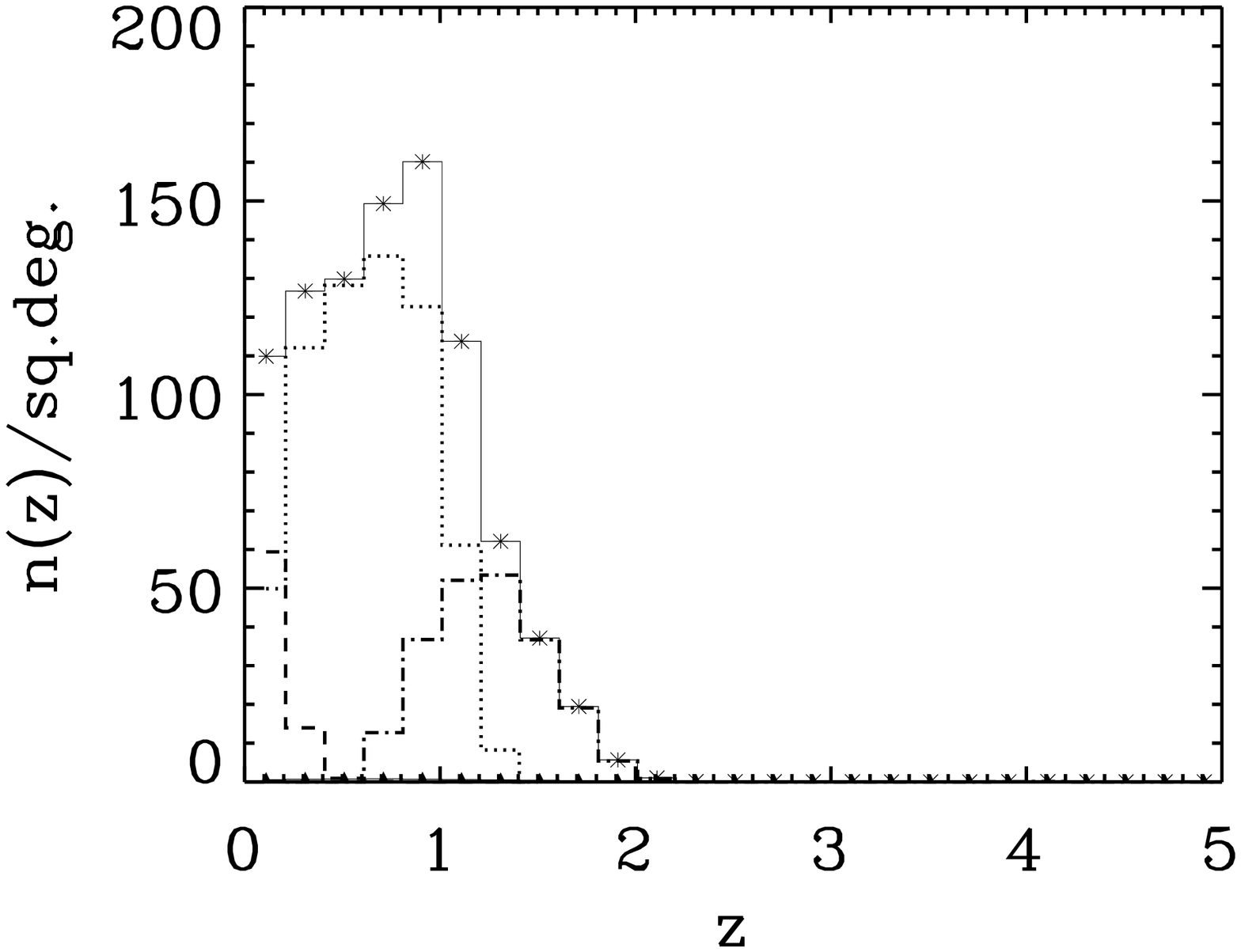}
\caption{Predicted $70 \mu$m source counts and redshift
distribution for $S_{70 \mu{\rm m}} \ge 10\,$mJy. Dot-dashed line:
(star-forming) spheroids; dotted: starbursts; dashed: spirals;
filled triangles: AGN. The thin solid histogram with asterisks in
the lower panel shows the sum of all contributions. We expect
$\sim 900$ sources per square degree brighter than 10 mJy; $\sim
23$\% of them are spheroids.} \label{c70std}
\end{figure}

\begin{figure}[tbp]
\centering
\includegraphics[width=9truecm]{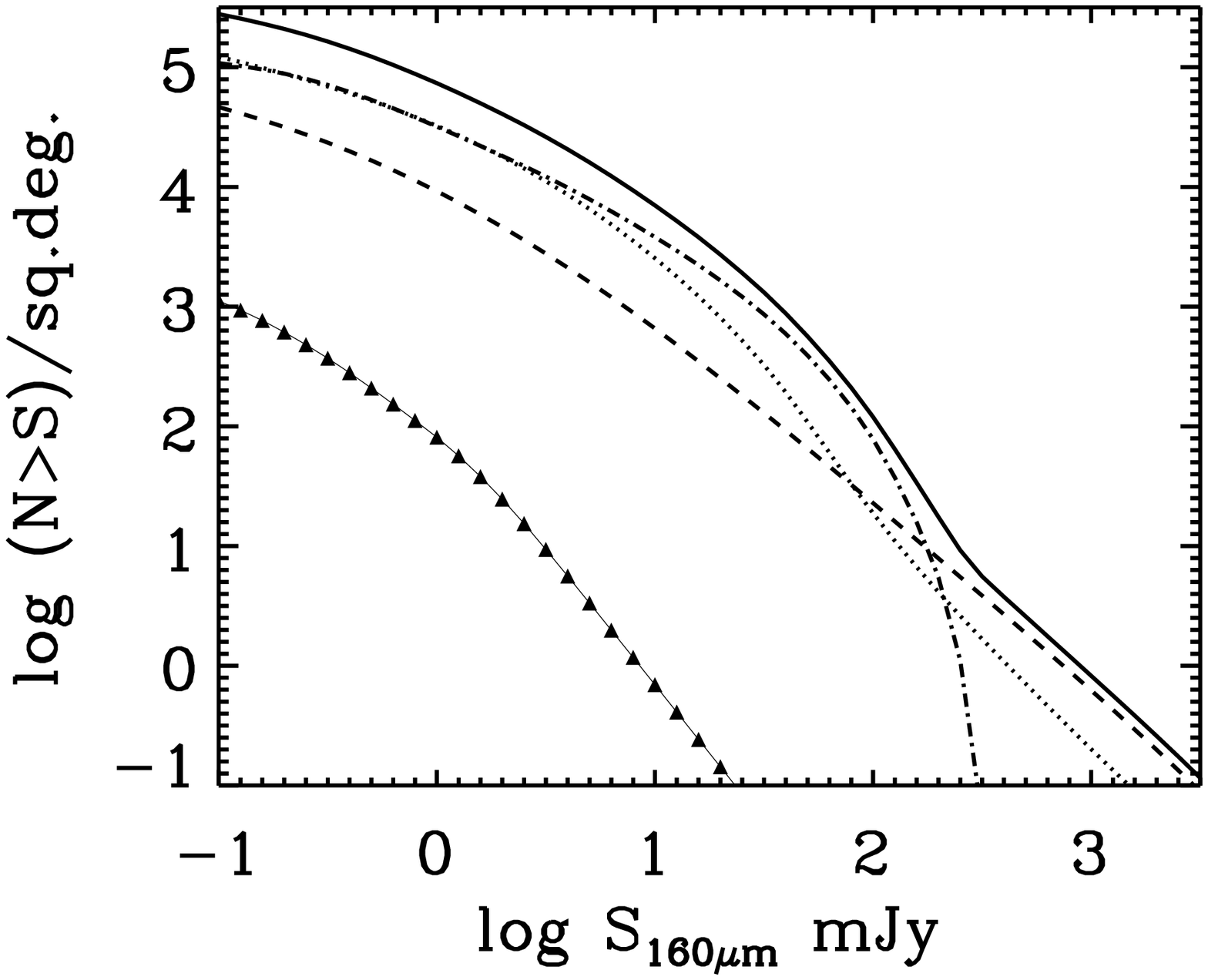}
\includegraphics[width=9truecm]{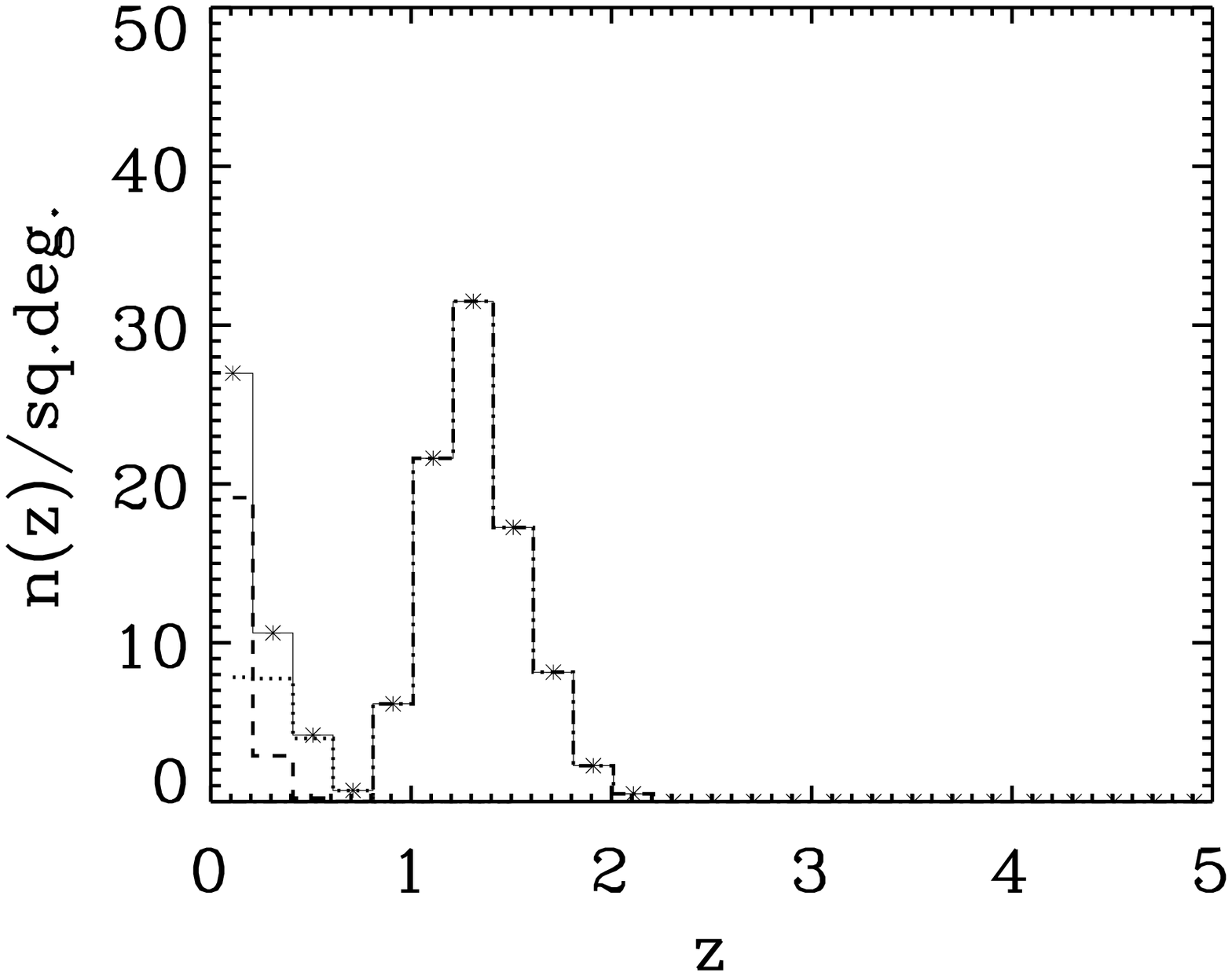}
\caption{Predicted $160 \mu$m source counts and redshift
distribution for $S_{160 \mu{\rm m}} \ge 96\,$mJy, the expected $5
\sigma$ confusion limit. Dot-dashed line: spheroids; dotted line:
starbursts; dashed line: spirals; filled triangles: AGN. The thin
solid histogram with asterisks in the lower panel is the sum of
the various contributions. We expect $\sim 130$ sources per square
degree brighter than 96 mJy; $\sim 65$\% of them are spheroids.}
\label{c160std}
\end{figure}

On the contrary, we interpret the observed IRAS $60\,\mu$m and
ISOPHOT $90\,\mu$m counts (Figs.~\ref{c60std} and \ref{c90std}) as
dominated by spiral and starburst galaxies, although star-forming
spheroidal galaxies may start contributing to the deepest
$90\,\mu$m point. Starburst galaxies dominate the $60\,\mu$m
counts, except at the brightest fluxes, while their contribution
to the ISOPHOT $90\,\mu$m counts is comparable to that of normal
late-type galaxies. The predicted $z$-distribution at the latter
wavelengths (lower panel of Fig.~\ref{c90std}) compares favourably
with the observational determination by Rowan-Robinson et al.
(2003).

GRASIL yields the SED of galaxies all the way to radio
frequencies, both for spheroids and for our templates for
late--type galaxies (see Bressan, Silva \& Granato 2002 for
details). Furthermore, we have adopted the Dunlop \& Peacock
(1990) models to take into account radio galaxies. The predicted
contributions of the various populations to the 1.4 GHz counts are
shown in Fig.~\ref{c1.4std}. In the flux density range $1\,\mu{\rm
Jy}\lsim S_{1.4{\rm GHz}} \lsim 30\,\mu{\rm Jy}$ we expect that
the counts are dominated by starburst and star-forming spheroidal
galaxies, with comparable contributions.

\subsection{Extremely Red Objects}

Extremely red objects (EROs), with $R-K>5$, have received special
attention in recent years (Daddi et al. 2000; Smith et al. 2002;
Roche et al. 2002, 2003; Cimatti et al. 2002, 2003; Takata et al.
2003; Yan \& Thompson 2003; Yan et al. 2004; Webb et al. 2004;
Moustakas et al. 2004), since their properties set crucial
constraints on the early evolutionary phases of massive spheroidal
galaxies. EROs are actually a mix of dusty star-forming and
evolved galaxies formed at high redshifts (Cimatti et al. 2002,
2003; Mohan et al. 2002).

We have worked out the counts and the redshift distribution of
both star-forming and passively evolving spheroidal galaxies
brighter than $K=19.2, 20, 20.3$ that, according to our model,
have ERO colours. Note that the observed redshift distributions
may be incomplete at high-$z$ (Yan et al. 2004); also, a mild star
formation activity, either due to residual gas or induced
occasionally by interactions, neglected by our simple model, could
decrease the number of galaxies with ERO colours, particularly at
high redshifts. On the whole, the agreement with observations,
which turn out to be very challenging for the other semi-analytic
models (e.g., Somerville et al.\ 2004) is quite good
(Fig.~\ref{ckstdero} and \ref{nzkstdero}).

\subsection{The IR background}
\label{sec:irb}

One further important constraint on the models comes from the IR
background. In Fig.~\ref{bgir} we compare the $1-1000\mu$m
background spectrum yielded by our model with the data collected
by Hauser \& Dwek (2001). The dotted line shows the summed
contributions of normal late-type and starburst galaxies. The leap
at $10\,\mu$m corresponds to the transition between the
evolutionary regime adopted in the MIR to mm region
(Sect.~\ref{sec:mirmm}) and the no-evolution one adopted in the
NIR (Sect.~\ref{sec:nir}). This sharp transition is clearly an
over-simplification and results, in particular, in an
underestimate of the contribution of late-type galaxies to the
background intensity. The underestimate is likely to be quite
large at wavelengths just below $10\,\mu$m.

In the wavelength range $10\mu{\rm m} \lsim \lambda \lsim
150\mu{\rm m}$ the background is dominated by starburst galaxies,
while at shorter wavelengths, spheroidal galaxies (mostly
passively evolving) take over. Longward of $\lambda \simeq
150\,\mu$m star-forming spheroids dominate if the starburst SED is
similar to that of M82 (lower panel). In the case of a cooler
starburst SED, like that of NGC6090 (upper panel), the
contribution of starburst galaxies to the sub-mm background may be
comparable to that of star-forming spheroidal galaxies. The actual
contribution of starburst galaxies is likely to be intermediate
since, as noted above, they show a variety of SEDs, and their
colours are correlated with luminosity.


\section{Predictions for Spitzer Space Telescope surveys}

We will focus on the two major extragalactic surveys carried out
with the Spitzer Space Telescope (formerly SIRTF): the SIRTF
Wide-area InfraRed Extragalactic survey (SWIRE; Lonsdale et al.
2003) and the Great Observatories Origins Deep Survey (GOODS;
Dickinson et al. 2003). SWIRE will survey 60--$65\,\hbox{deg}^2$
at high Galactic latitude in the four bands (3.6, 4.5, 5.6, and
$8\,\mu$m) of the InfraRed Array Camera (IRAC) and in the three
bands (24, 70, and $160\,\mu$m) of the Multiband Imaging
Photometer for SIRTF (MIPS). GOODS will provide deep images of
approximately $300\,\hbox{arcmin}^2$ and ultra-deep images of two
small fields in all IRAC bands, as well as deep exposures with
MIPS at $24\,\mu$m. The model counts and redshift distributions in
each of the IRAC and MIPS bands are shown in Figs.~\ref{c3.6std}
to \ref{c160std}.

The IRAC surveys will probe in depth the evolution of spheroidal
galaxies. The surveys to the SWIRE sensitivity limits (7.3, 9.7,
27.5, and 32.5 $\mu$Jy at 3.6, 4.5, 5.8, and 8 $\mu$m,
respectively; Lonsdale et al. 2003) will primarily test the
passive evolution phase, although star-forming spheroids will
begin to peep out particularly at the shorter wavelengths. GOODS
is expected to reach a limit of a $\sim 2$--$3 \mu$Jy in all the
IRAC bands. At this flux density limit, the majority of detected
sources are expected to be spheroidal galaxies, about half of
which in the active star-forming phase. Their redshift
distribution at $3.6\,\mu$m will peak in the range 0.6--1 or
around $z\simeq 1.5$ for the SWIRE and GOODS surveys,
respectively. The peak due to star-forming spheroids moves, at the
GOODS flux limit, to higher and higher redshifts with increasing
wavelengths, up to $z\simeq 2.7$ at $8\,\mu$m. The peak of the
overall distribution, for the GOODS survey, will keep at $z\simeq
1$, due to the contribution of late-type galaxies. We caution
that, as noted in Sect.~\ref{sect:coured}, the contribution of
late-- type galaxies is likely to be underestimated at the
shortest IRAC wavelengths because of our assumption of
no-evolution at $\lambda < 10\,\mu$m.

The GOODS survey in the $24 \mu$m  MIPS band is likely to be,
because of the limited spatial resolution, confusion limited.
Adopting Condon's (1974) approach to estimate the confusion noise,
$\sigma_{\rm conf}$, our model implies (neglecting the effect of
clustering, discussed by Negrello et al. 2004, as well as the
cirrus noise) a $5\sigma_{\rm conf}$ detection limits, $S_{\rm
lim,conf}$, of $0.17$ mJy, in good agreement with the earlier
estimates summarized in Table~3 of Lonsdale et al. (2003). These
estimates are well above the flux limit aimed at by the GOODS
survey (20--$80\,\mu$Jy; Dickinson et al. 2003), but a factor of
about 2.6 below the SWIRE sensitivity limit (0.45 mJy). We expect
that both the SWIRE and the GOODS surveys at this wavelength are
dominated by starburst galaxies, however with an 8--10\%
contribution from star-forming spheroids at redshifts in the range
1.5--3.

For the other MIPS bands we find $S_{\rm lim,conf}=$ 10 and 96 mJy
at 70 and 160 $\mu$m respectively, in good agreement with Xu et
al. (2003, model S3$+$E2) and well above the SWIRE sensitivities
($2.75$ and $17.5$ mJy, respectively). Thus we expect SWIRE to be
confusion limited at these wavelengths. The majority of sources
detected to the estimated confusion limits should be starburst
galaxies at $70\,\mu$m, while star-forming spheroids may take over
at $160\,\mu$m. However, as discussed in Sect.~\ref{sect:param},
our model is likely to somewhat overestimate the counts at the
latter wavelength.

\section{Conclusions}

Granato et al. (2001, 2004) have shown that the mutual feedback
between star-forming spheroidal galaxies and the active nuclei
growing in their cores can be a key ingredient towards overcoming
one of the main challenges facing the hierarchical clustering
scenario for galaxy formation, i.e. the fact that the densities of
massive high redshift galaxies detected by SCUBA and by deep
near-IR surveys are well above the predictions.

However, to take full advantage of the wealth of data on
extragalactic sources that are rapidly accumulating in the IR to
mm region to test evolutionary models and to assess their
parameters, we need to deal with complex and poorly understood
processes that determine the time-dependent SEDs of the various
populations of galaxies. Indeed, semi-analytic models must rely on
a large number of adjustable parameters. We have carried out a
detailed comparison of the physically grounded GDS04 model,
keeping their choice for the parameters controlling the
star-formation history, the chemical enrichment and the evolution
of dust and gas content of massive spheroidal galaxies. We are
therefore left with only two adjustable parameters, affecting
their near-IR to mm SED (see Sect.~\ref{sect:SED}) computed using
the code GRASIL that includes a full treatment of star-light
reprocessing by dust; as described in Sect.~\ref{sect:param},
their values are constrained mostly by $15\,\mu$m and $K$-band
counts. A simplified phenomenological approach
(Sect.~\ref{sec:other}) has been adopted to deal with the other
relevant galaxy populations (normal late-type and starburst
galaxies), and the contribution by AGN has been estimated by
coupling the cosmological evolution of AGN in the X-ray bands with
detailed SEDs (Sect.~\ref{sec:agn}).

The model predictions have then been tested against a broad
variety of observational data, including, in addition to the
$15\,\mu$m and $K$-band counts, the redshift distributions of
sources brighter than 0.1 and 1 mJy at $15\,\mu$m, the SCUBA
counts at $850\,\mu$m, the available (although still scanty) data
on the redshift distribution of sources brighter than 5 mJy at
$850\,\mu$m, the ISOPHOT 90 and $170\,\mu$m counts and the
corresponding redshift distributions, the IRAS $60\,\mu$m counts,
the radio 1.4 GHz counts, the ISOCAM $6.7\,\mu$m counts and
redshift distribution, the redshift distributions of galaxies to
the magnitude limits $K= 20$, 23, and 24, and the 1--$1000\,\mu$m
background spectrum. Specific predictions for the $K$-band counts
and redshift distributions of EROs have been worked out and
compared with data.

Encouraged by the good agreement of model predictions with all
these data sets, we have worked out detailed predictions for the
GOODS and SWIRE surveys with the Spitzer Space Telescope. In
agreement with previous estimates, we find that the GOODS deep
survey at $24\,\mu$m and the SWIRE surveys at 70 and $160\,\mu$m
are likely to be severely confusion limited. The GOODS surveys in
the IRAC channels (3.6 to $8\,\mu$m), reaching flux limits of a
few $\mu$Jy, are expected to resolve most of the background at
these wavelengths, to explore the full passive evolution phase of
spheroidal galaxies and most of their active star-forming phase,
detecting galaxies up to $z\simeq 4$ and beyond. A substantial
number of high $z$ star-forming spheroidal galaxies should also be
detected by the $24\,\mu$m SWIRE and GOODS surveys, while the 70
and $160\,\mu$m surveys will be particularly useful to study the
evolution of such galaxies in the range $1 \lsim z \lsim 2$.
However, starburst galaxies at $z \lsim 1$--1.5 are expected to be
the dominant population in MIPS channels, except, perhaps, at
$160\,\mu$m.

We plan to apply our model to make predictions for the forthcoming
surveys with Herschel, Planck, LMT and ALMA.


\begin{acknowledgements}

Work supported in part by MIUR (through a COFIN grant) and ASI. LS
and GLG acknowledge kind hospitality by INAOE where part of this
work was performed. We thank Andrea Cimatti for helpful
discussions and for providing the redshifts of EROs in advance of
publication. The LSGLG foundation is warmly acknowledged.

\end{acknowledgements}

\end{document}